\documentclass[a4paper]{aa}
\usepackage{graphicx}
\usepackage{txfonts}
\usepackage{color}
\usepackage{natbib}
\usepackage{lineno}
\usepackage{fix2col}
\usepackage{float}
\bibpunct{(}{)}{;}{a}{}{,} 

\def\aap{Astron. Astrophys.}

\def\ao{Applied Optics}

\def\apjl{Astrophys. J. Letters}

\def\apjs{Astrophys. J. Suppl.}

\begin{document}


\title{Inferring heat recirculation and albedo for exoplanetary atmospheres: Comparing optical phase curves and secondary eclipse data}
\titlerunning{Heat recirculation and albedo}

\author{P. von Paris\inst{1,2}    \and P. Gratier\inst{1,2} \and  P. Bord\'e\inst{1,2}  \and F. Selsis\inst{1,2} }

\institute{Univ. Bordeaux, LAB, UMR 5804, F-33270, Floirac, France \and CNRS, LAB, UMR 5804, F-33270, Floirac, France 
}

\abstract {Basic atmospheric properties such as albedo and heat redistribution between day and nightside have been inferred for a number of planets using observations of secondary eclipses and thermal phase curves. Optical phase curves have not yet been used to constrain these atmospheric properties consistently.}{We re-model previously published phase curves of CoRoT-1b, TrES-2b and HAT-P-7b and infer albedos and recirculation efficiencies. These are then compared to previous estimates based on secondary eclipse data.}{We use a physically consistent model to construct optical phase curves. This model takes Lambertian reflection, thermal emission, ellipsoidal variations and Doppler boosting into account. }{CoRoT-1b shows a non-negligible scattering albedo (0.11<A$_S$<0.3 at 95\,\% confidence) as well as small day-night temperature contrasts, indicative of moderate to high re-distribution of energy between dayside and nightside. These values are contrary to previous secondary eclipse and phase curve analyses. In the case of HAT-P-7b, model results suggest relatively high scattering albedo ($A_S \approx$0.3). This confirms previous phase curve analysis, however, it is in slight contradiction to values inferred from secondary eclipse data. For TrES-2b, both approaches yield very similar estimates of albedo and heat recirculation. Discrepancies between recirculation and albedo values as inferred from secondary eclipse and optical phase curve analyses might be interpreted as a hint that optical and IR observations probe different atmospheric layers, and hence temperatures. }{}


\maketitle

\section{Introduction}

\begin{figure*}
\begin{center}
\includegraphics[width=250pt]{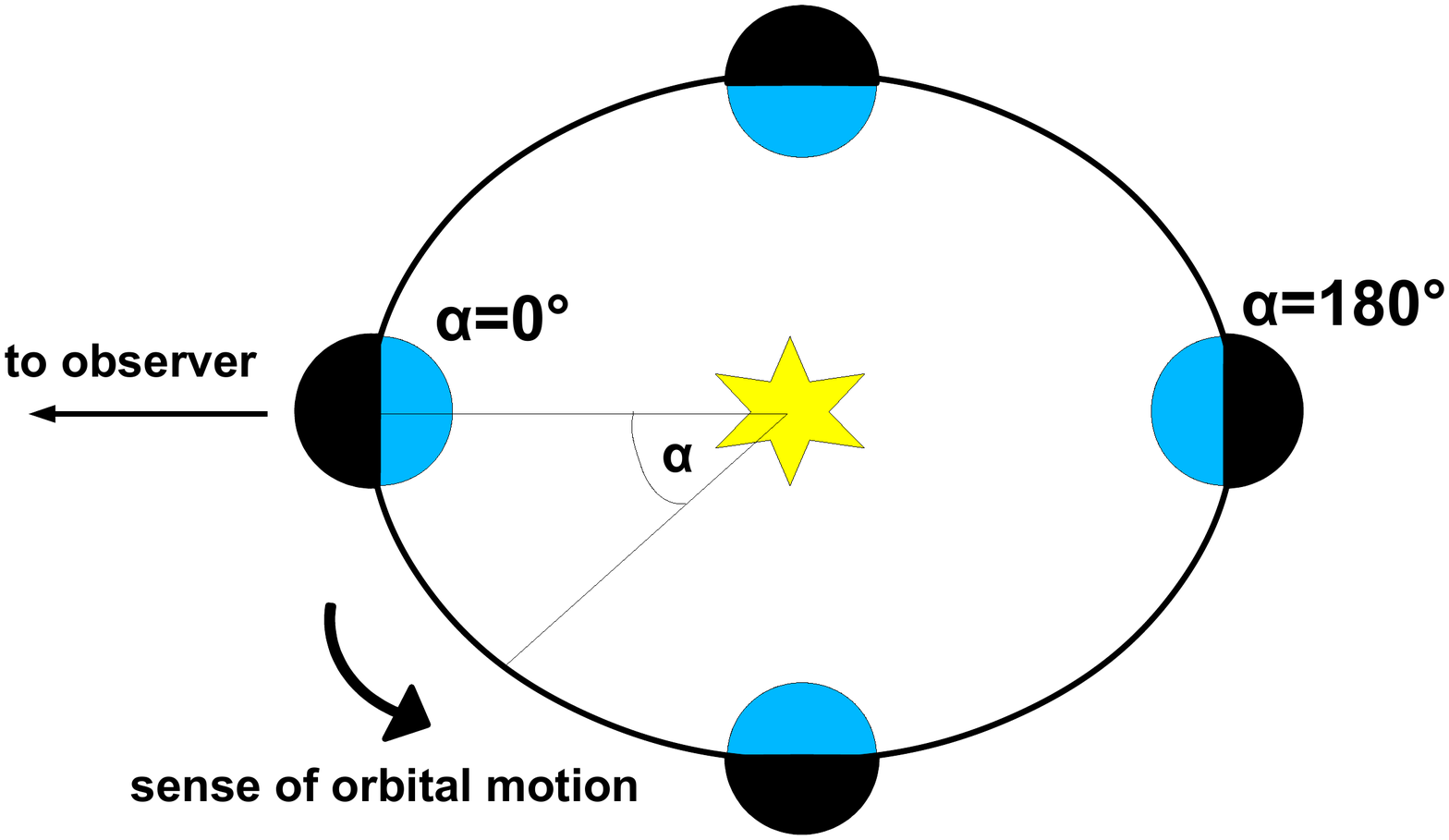}
\includegraphics[width=250pt]{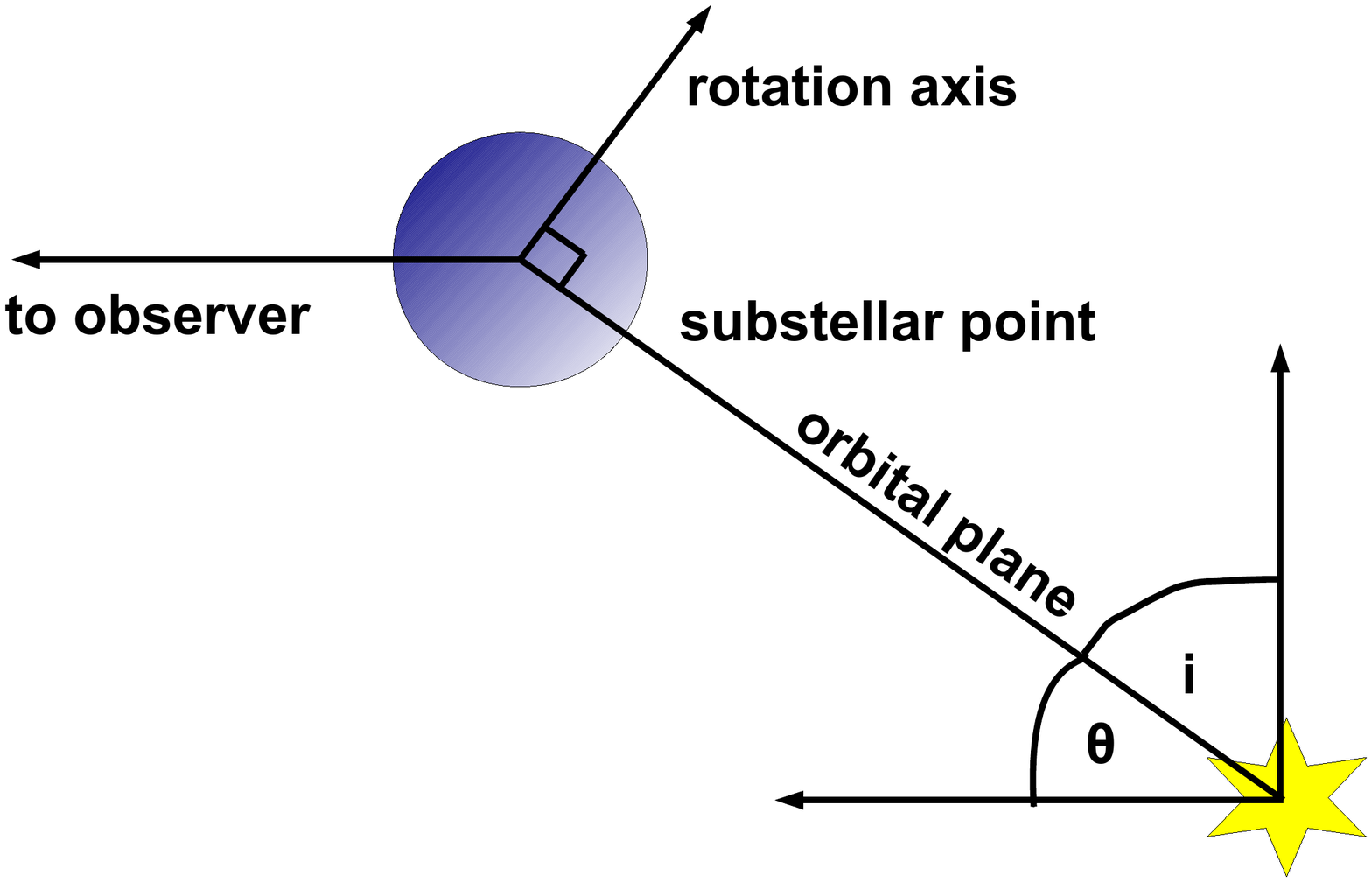}
\includegraphics[width=250pt]{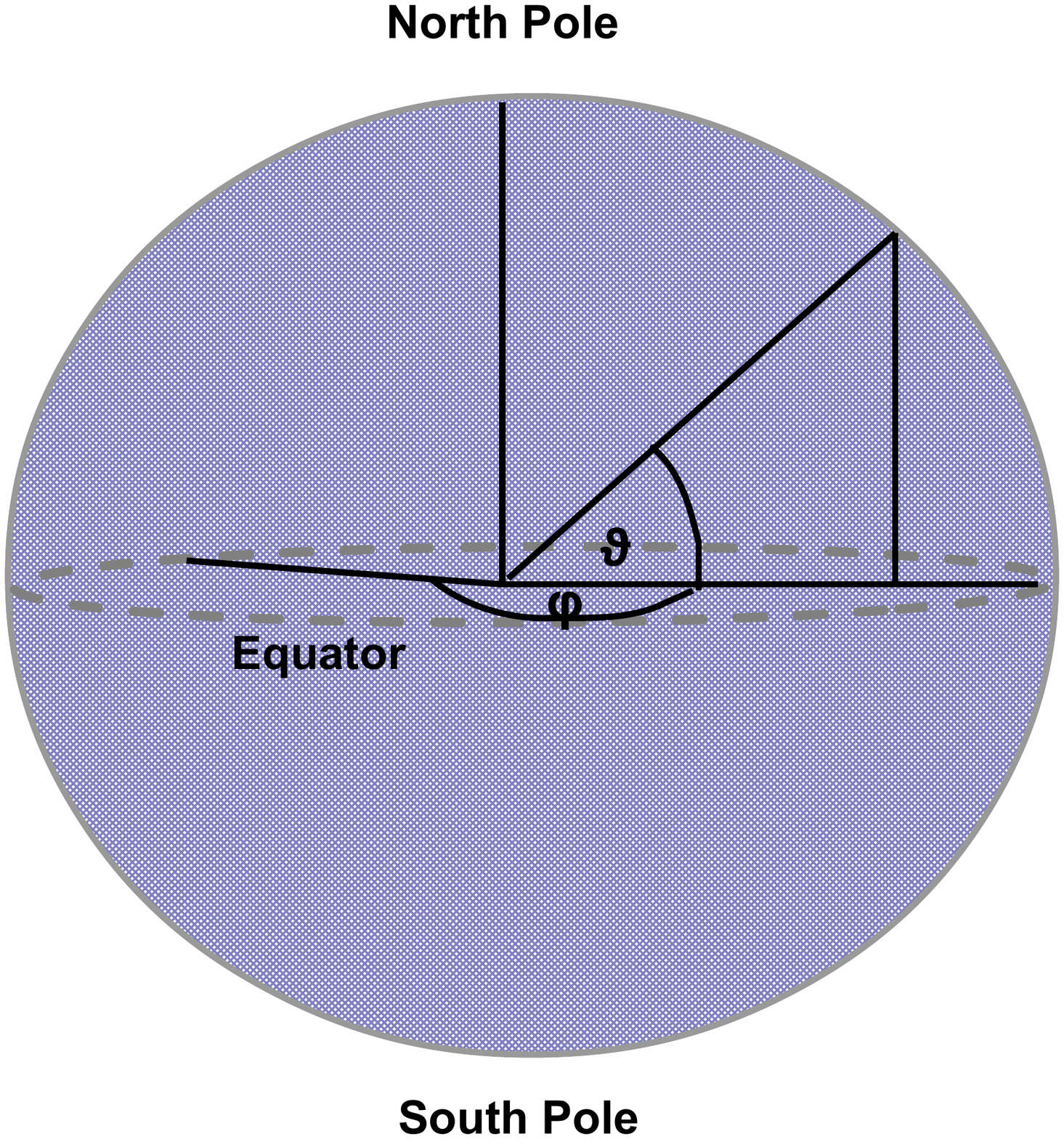}\\
\caption{Basic sketch of the geometry of the model.}
\label{illu}
\end{center}
\end{figure*}

In recent years, exoplanetary science has developed from a detection-centered astronomical science towards a characterization-centered planetary science. For basic properties of exoplanetary atmospheres such as albedo and heat redistribution from dayside to nightside, observational constraints are now available for a growing number of planets.

Radiative transfer and atmospheric modeling by, e.g., \citet{sudarsky2000} predicts very low optical albedos for cloud-free hot Jupiters because of strong absorption by alkali metals. When silicate clouds form high enough in the atmosphere, however, optical albedos could be significantly higher \citep{sudarsky2000}. The low optical albedos measured for a few hot Jupiters seem to confirm this (e.g., \citealp{rowe2006,rowe2008}), whereas high albedos inferred for other planets seem to indicate the potential presence of clouds (e.g., \citealp{quintana2013}, \citealp{demory2013inhomogen}, \citealp{esteves2013}).

Theoretically, the most basic circulation patterns of hot Jupiters are relatively well understood (e.g., \citealp{showman2002}, \citealp{madhusudhan2014}, \citealp{heng2015}). Tidally locked planets in close-in orbits always present the same hemisphere to the host star. In atmospheric models, the resulting strong irradiation contrasts and slow planetary rotations lead to inefficient recirculation throughout the IR photosphere of the planet (around pressures of 0.1-10\,mbar). The incident stellar energy is re-radiated before being circulated to the nightside. Consequently, strong temperature contrasts between dayside and nightside develop (e.g., \citealp{showman2002}, \citealp{parmentier2013}).When orbital distance or rotation period increases, atmospheric modeling predicts a transition to a circulation regime with much stronger recirculation and less pronounced day-night temperature differences (e.g., \citealp{showman2015}). Furthermore, in most 3D atmosphere models of hot Jupiters, a strong equatorial zonal jet appears, that then results in a displacement of the hottest point away from the substellar point (e.g., \citealp{showman2011}, \citealp{perez2013}, \citealp{showman2015}).

Observed thermal phase curves in the (near-)infrared (IR) of a few hot Jupiters seem to confirm these predictions (e.g., \citealp{knutson2007daynight_189733}, \citealp{knutson2009daynight_189733}, \citealp{stevenson2014}).  Assembling many different observations for a large number of hot Jupiters, \citet{cowan2011hot} and \citet{schwartz2015} point out a possibly emerging trend, in line with theoretical predictions (e.g., \citealp{perez2013}). Strongly irradiated planets show very inefficient recirculation, whereas for less irradiated planets, the whole range of recirculation is possible.

\citet{cowan2011hot} and \citet{schwartz2015} use published secondary eclipse data (i.e., an estimate of the planetary dayside flux) and thermal phase curves to perform their homogeneous analysis. For a few planets, optical phase curves are available, obtained with the CoRoT and Kepler satellites. These are not taken into account in \citet{schwartz2015}. However, for hot planets, even optical phase curves offer some constraints on thermal radiation, and therefore albedo and heat recirculation.

So far, published studies of optical phase curves could not be used for such an analysis. This is, in each case, because of one of the three following reasons.

Firstly, some models do not take thermal emission into account (e.g., \citealp{mazeh2010}, \citealp{barclay2012}, \citealp{esteves2013,esteves2015}). In such cases, attributing an observed phase curve asymmetry to an offset of the thermal hotspot is physically inconsistent (e.g., \citealp{esteves2015}). Secondly, a few models (e.g., \citealp{snellen2009}) only treat thermal emission and do not include scattered light. Therefore, inferring constraints on the scattering properties of the planet (as done in \citealp{snellen2009}) is equally inconsistent. Thirdly, models that take both thermal and scattering components into account (e.g., \citealp{mislis2012}, \citealp{faigler2013}, \citealp{placek2014}, \citealp{faigler2015}), do not use appropriate phase functions for these components. They assume, for instance, that thermal and scattered light have identical phase functions which is incoherent when assuming explicitly Lambertian scattering and blackbody thermal radiation (see below,  Appendix \ref{verification_model} and eqs. \ref{cellflux} and \ref{thermalflux}).

Therefore, we present a new model with a physically consistent treatment of both components. We apply our model to three hot Jupiters with well-characterized IR and optical measurements, namely CoRoT-1b, TrES-2b and HAT-P-7b, to compare to results from secondary eclipse analysis.

In Sect. \ref{formodel} and Sect. \ref{invmodel}, we describe the physical forward model, the inverse model and the Markov Chain Monte Carlo (MCMC) approach used in this work. Section \ref{setup} describes the model setup and planetary scenarios, Sect. \ref{results} presents the results, Sect. \ref{discuss} a discussion, and we conclude with Sect. \ref{summary}. The Appendices contain the fitting results, MCMC output, a short model verification as well as a discussion about Rayleigh scattering, non-Lambertian phase functions and stellar parameter uncertainties.

\section{Forward model}

\label{formodel}

\subsection{Geometry}

\label{geometry}

The basic geometry set-up of the star-planet system is shown in Fig. \ref{illu}. We adopt a coordinate system where the substellar point is fixed throughout the calculations at 0$^{\circ}$ latitude and 180$^{\circ}$ longitude. Therefore, by definition, local latitude $\vartheta$ is 0$^{\circ}$ at the equator (range -90$^{\circ}<\vartheta<90^{\circ}$) and local longitude $\varphi$ is 0$^{\circ}$ at local midnight (range $0^{\circ}<\varphi<360^{\circ}$). For planets with zero obliquity and a planetary rotation synchronized with the orbital period, our choice of coordinate system would establish a stationary map of the planet. For close-in giant planets, this is a reasonable assumption (but see, e.g., \citealp{arras2010} or \citealp{rauscher2014} for a discussion).

The planetary "surface" is divided into cells of 2.5$^{\circ}$x2.5$^{\circ}$ size (72 cells in latitude, 144 in longitude). We define here "surface" as the optical photosphere, i.e., where the planetary atmosphere becomes optically thick to visible radiation. It is generally identified with the surface of the sphere with radius $R_P$. The number of cells is reasonable for computational purposes and still allows for smooth light curves without noticeable effects of discretization. For each cell, the local surface normal \textbf{n}, stellar and observer directions (\textbf{s} and \textbf{o}, respectively) are calculated. A cell contributes to the reflected light if both \textbf{n}$\cdot$\textbf{s}$\geqslant0$ (i.e., dayside) and \textbf{n}$\cdot$\textbf{o}$\geqslant0$ (i.e., visible to the observer) are satisfied. The nightside is defined with \textbf{n}$\cdot$\textbf{s}$\leqslant0$, and correspondingly, the part of the nightside visible by the observer with \textbf{n}$\cdot$\textbf{s}$\leqslant0$ and \textbf{n}$\cdot$\textbf{o}$\geqslant0$ , simultaneously.

The subobserver latitude $\theta$ is given by the inclination $i$ of the orbital plane with respect to the observer:

\begin{equation}
\label{obslon}
\theta=90^{\circ}-i,
\end{equation}

i.e., an edge-on orbit has $i$=$90^{\circ}$, and a face-on orbit has $i$=0$^{\circ}$. The subobserver longitude $\phi$ as a function of time $t$ is given by the orbital phase $\alpha$ ($\alpha=0$ and $t=0$ at primary transit, or equivalently, inferior conjunction, see Fig. \ref{illu}):

\begin{equation}
\label{obslat}
\phi(t)=360^{\circ} - \alpha(t),
\end{equation}

with

\begin{equation}
\label{orbphase}
\alpha(t)=T(t)+\omega_P,
\end{equation}

where $T$ is the true anomaly and $\omega_P$ the argument of periastron. The true anomaly is calculated from the eccentric anomaly $E$:

\begin{equation}
\label{true}
\tan\left(\frac{T}{2}\right)=\sqrt{\frac{1+e}{1-e}}\tan\left(\frac{E}{2}\right),
\end{equation}

with $e$ eccentricity. $E$ is determined by numerically solving Kepler's equation:

\begin{equation}
\label{keplereq}
E-e\sin(E)=M,
\end{equation}
with $M$ mean anomaly.  The solution is obtained with a publicly available Fortran procedure\footnote{www.davidgsimpson.com/software/keplersoln\_f90.txt}, based on a method described in \citet{meeus1991}. The mean anomaly $M$ is given by

\begin{equation}
\label{meananomaly}
M=2\pi\cdot \left(x-\lfloor x \rfloor \right),
\end{equation}

where $x=\frac{t-t_{\rm{peri}}}{P_{\rm{orb}}}$, $P_{\rm{orb}}$ is the planetary orbital period, $t_{\rm{peri}}$ is the time of periastron passage and $\lfloor x \rfloor$ represents the floor function, i.e., the  greatest integer less than or equal to $x$. \footnote{In practice, we first calculate $M$, $T$ and $\alpha$ with $t_{\rm{peri}}=0$ and then interpolate such that $t=0$ occurs at $\alpha=0$.}

\subsection{Reflected light}

Stellar light incident on the planetary dayside is partly reflected back into space and towards the observer. The amount of reflected light reaching the observer depends, for instance, on the scattering properties of the planet (e.g., Rayleigh scattering produces a different phase function than Mie scattering, see, e.g., \citealp{madhusudhan2012} for a review). In this work, in the absence of any reliable information on the scattering properties, the Lambert approximation of diffuse scattering is used. In Appendix \ref{rayappend} we discuss the possible influence of Rayleigh scattering and other phase functions on the phase curve.

In the Lambert approximation, for each cell contributing to the observed flux \textbf{($\mathbf{n} \cdot \mathbf{o} \geqslant 0 $)}, the flux (in W\,m$^{-2}$) received by the observer from this cell\textbf{, $F_{\rm{r,o,cell}}$,}  is given by

\begin{equation}
\label{cellflux}
F_{\rm{r,o,cell}}=\cos z_s \cdot F_{\rm{\ast,p}} \cdot \frac{A_S}{\pi} \cdot  \cos z_o \cdot \frac{\Delta S}{d^2} ,
\end{equation}

with $z_s$ the local stellar zenith angle, $z_o$ the local observer zenith angle, $F_{\rm{\ast,p}}$ the stellar flux at the planet's orbit (in W m$^{-2}$), $\Delta S$ the surface element of the cell on the planet (in sr m$^2$), $d$ the observer-planet distance and $A_S$ the (potentially wavelength-dependent) planetary scattering albedo.  $A_S$ is assumed to be constant with time, hence neglecting any time-dependent processes such as cloud formation etc. Note that, since we use the Lambertian approximation in eq. \ref{cellflux}, the scattering albedo is related to the geometric albedo $A_G$ in a simple manner:

\begin{equation}
\label{ageo} A_G=\frac{2}{3}A_S.
\end{equation}

The surface element $\Delta S$ of the cell, as seen from the planet's center, is calculated as

\begin{equation}
\label{surfelement}
\Delta S=\Delta \Omega \cdot R_p^2, 
\end{equation}

with $R_P$ the planetary radius and $\Delta \Omega$ the solid angle (in sr) of the cell (angular extent 2.5$^{\circ}$x2.5$^{\circ}$, see above).

The stellar flux at the planet's orbit, $F_{\rm{\ast,p}}$, is calculated as follows:

\begin{equation}
\label{stellarflux}
F_{\rm{\ast,p}}=\pi \left(\frac{R_{\ast}}{r}\right)^2 \int_{\lambda_{\rm{low}}}^{\lambda_{\rm{high}}}I_{\rm{\ast,s}}q_I(\lambda)d\lambda,
\end{equation}

where $I_{\ast,s}$ is a stellar model intensity (in W\,m$^{-2}$\,sr$^{-1}$\,$\mu$m$^{-1}$) at the stellar surface, $q_I$ is the instrumental filter function, $\lambda_{\rm{low}}$ and $\lambda_{\rm{high}}$ define the wavelength interval of the bandpass, $r$ the star-planet distance and $R_{\ast}$ is the stellar radius\footnote{ Note that we calculated the star's solid angle, as seen from the planet, in eq. \ref{stellarflux} as $\pi \left(\frac{R_{\ast}}{r}\right)^2$. This assumes that $R_{\ast}\ll r$, a condition that is on the verge of breaking down for close-in planets such as the ones considered in this work. However, detailed modeling (not shown here), which took the large angular extent of the star (up to tens of degrees) into account, showed little to no influence on resulting phase curves. Therefore, we retain our simplifying assumption in the following.}. The numerical integration for eq. \ref{stellarflux} is done with a standard trapezoidal integration scheme using a roughly 1\,nm spectral resolution.

Stellar intensities  $I_{\ast,s}$ are obtained from the ATLAS stellar atmosphere grid \citep{castelli2004}\footnote{http://www.user.oats.inaf.it/castelli/grids.html}. Instrumental filter functions $T_l$ are taken, for CoRoT-1, from \citet{snellen2009} and, for TrES-2 and HAT-P-7, the Kepler Handbook\footnote{retrieved from http://keplergo.arc.nasa.gov/Instrumentation.shtml}. 

The time-dependent planet-star distance $r$ is calculated with

\begin{equation}
\label{distance}
r(t)=a\frac{1-e^2}{1+e\cos(T(t))},
\end{equation}

with $a$ the semi-major axis. 

The total reflected flux $F_{\rm{r,o}}$ received by the observer is the sum over all contributing cells:

\begin{equation}
\label{refflux}
F_{\rm{r,o}}=\sum_{ (\mathbf{n} \cdot \mathbf{s} \geqslant 0 )\wedge  (\mathbf{n} \cdot \mathbf{o} \geqslant 0 )} F_{\rm{r,o,cell}}.
\end{equation}

\subsection{Emitted light}

In addition to reflected starlight, the planet also emits thermal radiation that can contribute to the overall phase curve. In optical phase curves, this contribution would be negligible for long-period (and consequently colder) planets. However, for close-in hot Jupiters, the thermal component can become comparable to (or even dominate) the reflected component of the phase curve.

In this work, we make the specific assumption of two hemispheres (nightside and dayside) which radiate as a blackbody with uniform temperatures  $T_{\rm{night}}$ and $T_{\rm{day}}$, respectively. 

A widely used approach (for example, \citealp{snellen2009}, \citealp{alonso2009}, \citealp{mislis2012}, \citealp{schwartz2015}) is to relate these temperatures to the Bond albedo $A_B$  and the efficiency of heat re-distribution $\epsilon$. Temperatures are then calculated, based on \citet{cowan2011hot}: 

\begin{equation}
\label{daytemp}
 T_{\rm{day}}^4=T_{\ast}^4\frac{R_{\ast}^2}{a^2(1-e^2)^{0.5}}\cdot (1-A_B)(\frac{2}{3}-\frac{5}{12}\epsilon)=T_0^4\cdot (\frac{2}{3}-\frac{5}{12}\epsilon),
\end{equation}

and

\begin{equation}
\label{nighttemp}
 T_{\rm{night}}^4=T_{\ast}^4\frac{R_{\ast}^2}{a^2(1-e^2)^{0.5}}\cdot (1-A_B)\frac{\epsilon}{4}=T_0^4\cdot \frac{\epsilon}{4},
\end{equation}

where $T_{\ast}$ is the stellar effective temperature and the additional factor $(1-e^2)^{0.5}$ accounts for the mean flux received over an orbit \citep{williams2002}.

The parameter $\epsilon$ is a re-parameterization of the geometrical redistribution factor $f$. \citet{spiegel2010} define $f$ via the apparent dayside flux $F_d$ (i.e., close to secondary eclipse) and the total planetary luminosity $L_P$:

\begin{equation}
\label{f_def}
 L_P=\frac{1}{f}\cdot \pi R_P^2 \sigma T_{\rm{day}}^4=\frac{1}{f}F_d.
\end{equation}

For perfect heat recirculation, the entire planet is at a uniform temperature, and thus $f=\frac{1}{4}$, because both dayside and nightside hemispheres contribute to the planetary luminosity. At zero heat recirculation, the nightside emission is zero, and each point on the dayside hemisphere emits with its local radiative equilibrium temperature. Hence, as shown by \citet{spiegel2010} and \citet{cowan2011hot}, $f=\frac{2}{3}$.

Assuming radiative equilibrium, meaning that the total stellar flux $F_S$ intercepted by the planet equals its luminosity $L_P$, and approximating the star as a blackbody, one can re-arrange eq. \ref{f_def} to yield

\begin{equation}
\label{f_re}
f=\frac{F_d}{L_P}=\frac{F_d}{F_S}=\frac{T_{\rm{day}}^4}{T_0^4}.
\end{equation}

For $\epsilon$ to vary between 0 and 1 (corresponding to no recirculation and perfect redistribution, respectively) then simply requires a linear transformation of $f$, which leads directly to eq. \ref{daytemp}.

Physically, $\epsilon$ is determined by the atmospheric circulation and the strength of winds and jets that transport heat away from the illuminated hemisphere. For strongly irradiated planets, the radiative timescale is expected to be much shorter than the advective (dynamical) timescale, therefore large day-night temperature contrasts and small values of $\epsilon$ are expected. For less irradiated planets, a range of circulation regimes is possible (e.g., \citealp{showman2002}, \citealp{showman2011}, \citealp{perez2013}, \citealp{heng2015}, \citealp{showman2015}).

Another option in the model is to retain $T_{\rm{night}}$ and $T_{\rm{day}}$ as free parameters for the inverse modeling, an approach used by, e.g., \citet{placek2014}.

Since blackbody radiation is isotropic, the thermal flux $F_{\rm{t,o,cell}}$ received by the observer from each cell is given by

\begin{equation}
\label{thermalflux}
F_{\rm{t,o,cell}}(T_c)=\int_{\lambda_{\rm{low}}}^{\lambda_{\rm{high}}}B(T_c,\lambda)q_I(\lambda)d\lambda  \cdot \cos z_o \cdot  \frac{\Delta S}{d^2},
\end{equation}

with $B(T_c,\lambda)$ the blackbody intensity at the cell's temperature $T_c$ and $T_c=T_{\rm{night}}$ or $T_c=T_{\rm{day}}$, depending on the location of the cell. 

Again, the total emitted flux $F_{\rm{t,o}}$ is the sum over all cells which are visible for the observer (i.e., \textbf{n}$\cdot$\textbf{o}$\geqslant0$):

\begin{equation}
\label{emmflux}
F_{\rm{t,o}}=\sum_{ (\mathbf{n} \cdot \mathbf{s} \geqslant 0 )\wedge  (\mathbf{n} \cdot \mathbf{o} \geqslant 0 )} F_{\rm{t,o,cell}}(T_{\rm{day}})+\sum_{ (\mathbf{n} \cdot \mathbf{s} \leqslant 0 )\wedge  (\mathbf{n} \cdot \mathbf{o} \geqslant 0 )} F_{\rm{t,o,cell}}(T_{\rm{night}}).
\end{equation}

\begin{figure}[h]
\begin{center}
\includegraphics[width=250pt]{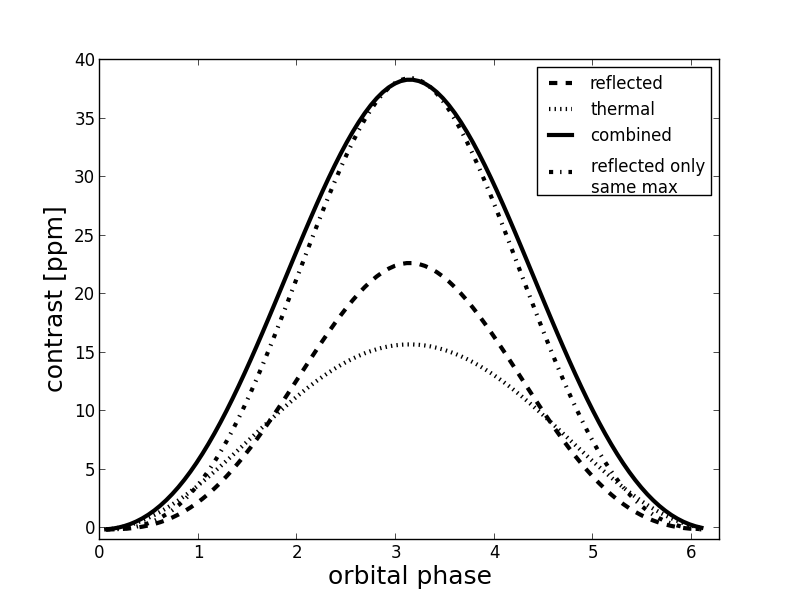}\\
\caption{Effect of thermal emission on the phase curve. See text for discussion.}
\label{planckeffect}
\end{center}
\end{figure}

Note that in our approach, thermal emission produces a different phase curve behavior compared to Lambert scattering of stellar radiation because of the additional factor $\cos z$ in eq. \ref{cellflux} compared to eq. \ref{thermalflux}.  Therefore it is potentially possible to disentangle reflected from emitted light. This effect is illustrated in Fig. \ref{planckeffect}.

Figure \ref{planckeffect} shows the reflective and the thermal contribution for a hot Jupiter planet in a 2-day orbit around a Sun-like star (500-1,000\,nm bandpass, $A_B$=$A_S$=0.15, $\epsilon$=0). In this example, the planetary mass is set to zero for illustration purposes. The reflected component is shown as a dashed line, the thermal component of the phase curve is the dotted line. Also shown are the combined phase curve (plain line) and a purely reflective phase curve (dot-dashed line) that has the same maximum amplitude as the combined phase curve. Due to the additional angle dependence of the reflected light ($\cos z_s$ in eq. \ref{cellflux}), the reflected phase curve is steeper than the thermal one and shows a more pronounced peak towards phase $\alpha=\pi$. Based on the slope, if the orbital period is short enough and the photometric precision good enough, it is possible to determine dayside and nightside temperatures, hence Bond albedo and heat redistribution, contrary to what is generally assumed in previous studies (e.g., \citealp{esteves2013,esteves2015}, \citealp{schwartz2015}) that did not incorporate thermal radiation consistently.

Of course, note that in reality, non-uniform distribution of temperatures (due to, e.g., chemical composition changes, \citealp{agundez2012}) will complicate the interpretation of thermal phase curves. However, because of the currently limited data, assuming uniform hemispheres is justifiable.

\subsection{Ellipsoidal variations and Doppler boosting}

In addition to the planetary contributions to the phase curve, our model also considers two modulations of the stellar light induced by the planet. These are the ellipsoidal variations (e.g., \citealp{pfahl2008}) and the Doppler boosting (e.g., \citealp{loeb2003}). 
In short, ellipsoidal variations are due to the tidal deformation of the star by the orbiting planet. As a consequence, the star represents a varying cross section to the observer, hence, the luminosity changes periodically, with a period half that of the orbital period of the companion. Doppler boosting is a consequence of the Doppler shift of the stellar spectrum induced by the stellar reflex motion. As the star orbits the barycentre of the star-planet system, the emitted stellar light shifts its wavelength, and thus periodically more or less light is emitted in the bandpass of the used instruments. 

We use the formalism developed in \citet{quintana2013} to calculate the respective contrasts:

\begin{equation}
\label{ellips}
C_{\rm{ell}}=-A_{\rm{ell}}\cdot \cos(2 \alpha),
\end{equation}

\begin{equation}
\label{doppler}
C_{\rm{dopp}}=A_{\rm{dopp}}\cdot \sin(\alpha),
\end{equation}

where $A_{\rm{ell}}$, $A_{\rm{dopp}}$ are the amplitudes of the ellipsoidal and Doppler variations. Note that we do not fit for a potential phase lag between tidal bulge on the star and the planet (as in, e.g., \citealp{barclay2012}). Furthermore, we do not incorporate other harmonics of the orbital period into our ellipsoidal variation term (contrary to what was as done, e.g., in \citealp{esteves2013}). $A_{\rm{ell}}$  is given by (see also eq. 2 in \citealp{quintana2013}):

\begin{equation}
\label{ellipsamp}
A_{\rm{ell}}=\alpha_{\rm{ell}}\frac{M_p}{M_{\ast}} \left(\frac{R_{\ast}}{a}\right)^3\sin^2(i),
\end{equation}

with $M_{\ast}$, $M_p$ the stellar and planetary mass, respectively. $\alpha_{\rm{ell}}$ is a parameter determined by the stellar limb ($u$) and gravity ($g$) darkening:

\begin{equation}
\label{darkening}
\alpha_{\rm{ell}}=0.15\cdot \frac{(15+u)(1+g)}{3-u}.
\end{equation}

The coefficients $u$ and $g$ depend on stellar characteristics such as effective temperature $T_{\ast}$, metallicity $\rm{[Fe}/\rm{H}]$ and surface gravity $\log g_s$ (determined from $M_{\ast}$ and $R_{\ast}$).  As in \citet{barclay2012}, \citet{quintana2013} or \citet{esteves2013}, we use pre-calculated tables for $u$ and $g$ to interpolate linearly in $T_{\ast}$, $\rm{[Fe}/\rm{H}]$  and  $\log g_s$. These tables are taken from model calculations presented in \citet{claret2011} for the Kepler and CoRoT bandpasses. We adopt their coefficients obtained with a microturbulence velocity of 2\,km\,s$^{-1}$ with ATLAS stellar models and, in the case of $u$, fitted with a linear least squares approach.

$A_{\rm{dopp}}$  is given by (see also eq. 5 in \citealp{quintana2013}):

\begin{equation}
\label{doppleramp}
A_{\rm{dopp}}=(3-\alpha_{\rm{dopp}})\frac{K}{c},
\end{equation}

where $c$ is the speed of light, $K$ the radial velocity semi-amplitude and $\alpha_{\rm{dopp}}$ a parameter which depends on the wavelength of observation $\lambda_{\rm{obs}}$ and the stellar effective temperature. As in \citet{quintana2013}, we use the approximate equations from \citet{loeb2003} for $\alpha_{\rm{dopp}}$:

\begin{equation}
\label{alphaloeb}
\alpha_{\rm{dopp}}=\frac{e^x(3-x)-3}{e^x-1},
\end{equation}

where $x=\frac{h\frac{c}{\lambda_{\rm{obs}}}}{k_BT_{\ast}}$ ($h$ Planck's constant, $k_B$ Boltzmann's constant). This approach of calculating $(3-\alpha_{\rm{dopp}})$ is somewhat different from the approach used in, e.g., \citet{esteves2013,esteves2015}. However, comparing results for TrES-2b (their work, 3.71, to 3.87 with our approach) and HAT-P-7b (3.41 to 3.59) indicates that both approaches yield values that are within 5\,\%. Hence mass determinations are expected to be comparable.

$K$ is determined as (with $G$ the gravitational constant)

\begin{equation}
\label{velocityamp}
K=\left(\frac{2\pi G}{P_{\rm{orb}}} \right)^{1/3} \frac{M_p}{M_{\ast}^{2/3}} \sin(i)\frac{1+e\cos \omega_P}{(1-e^2)^{0.5}}.
\end{equation}

\subsection{Asymmetries}

As demonstrated by, e.g., \citet{demory2013inhomogen} and \citet{esteves2015}, many exoplanets show asymmetries in their phase curves, with respect to secondary eclipse. This means that the maximum amplitude is reached before (after) secondary eclipse, meaning that the maximum brightness is shifted eastwards (westwards) of the substellar spot. In the case of a westwards shift, it has been interpreted as the presence of clouds that enhance the reflectivity of the "morning"\footnote{Note that for tidally-locked planets, such as assumed here, the star does not move across the celestial sphere for an observer on the planet. Therefore, "morning" (i.e., sunrise) and "evening" (i.e., sunset) as such don't exist. Rather, for eastward circulation, "morning" is defined as the terminator over which an air parcel would enter the dayside from the nightside. For illustration purposes, we will retain "morning" and "evening" throughout the text.}side. When the shift is eastwards, i.e., the "evening" side is brighter, previous studies attributed this to a shift in the hottest region of the atmosphere. Such a shift is associated with atmospheric circulation which transports heat away from the substellar point before it is re-radiated. Most 3D atmospheric models of hot Jupiters produce such an offset in thermal emission (e.g., \citealp{showman2002}, \citealp{showman2015}), and IR phase curves seem to confirm the theoretical predictions (e.g., \citealp{knutson2007daynight_189733}).

This change in brightness can be accounted for in the model in two ways. First, for the reflected-light component of the phase curve, we implement a simple dark-bright model, similar to the approach chosen in \citet{demory2013inhomogen}. Part of the planet has a scattering albedo $A_S$ (see eq. \ref{cellflux}), and part of the planet has a different scattering albedo $d_S\cdot A_S$, with $d_S$ being a free parameter such that $d_S\cdot A_S\leq1$. The extent in longitude of the latter part of the planet is controlled by two further parameters, namely $l_{\rm{start}}$ and $l_{\rm{end}}$. We do not consider any latitudinal variation of the scattering properties.

\begin{figure}[h]
\begin{center}
\includegraphics[width=120pt]{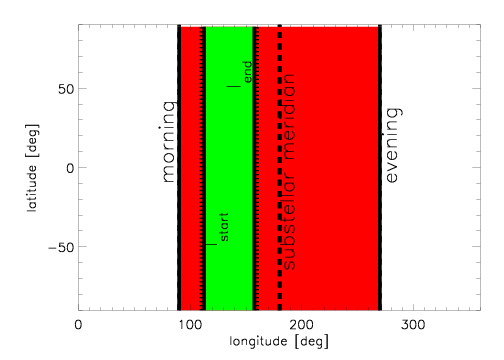}
\includegraphics[width=120pt]{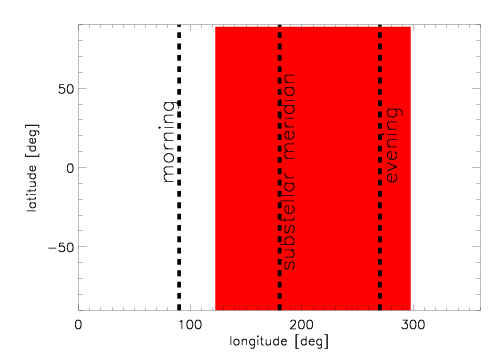}\\
\caption{Illustration of the asymmetric models. Left:  Reflective dayside (red) with bright "morning" (green). Right: Thermal offset modeled as a shifted dayside (red). See text for discussion.}
\label{asym}
\end{center}
\end{figure}

Second, for the thermal component of the phase curve, we consider a simple offset $\Theta_{\rm{d}}$ of the dayside such that the dayside has an extent in longitude between $\Theta_{\rm{d}}$ and  $180^{\circ}+\Theta_{\rm{d}}$. Figure \ref{asym} illustrates the asymmetric models.

\subsection{Phase curve}

The final time-dependent contrast $C$ between star and planet is then

\begin{equation}
\label{contrastref}
C(t)=\frac{F_R(t)+F_E(t)}{F_{\rm{\ast,o}}}+C_{\rm{ell}}+C_{\rm{dopp}}.
\end{equation}

The stellar flux at the observer, $F_{\rm{\ast,o}}$, is calculated analogous to $F_{\rm{\ast,p}}$ (see eq. \ref{stellarflux}):

\begin{equation}
\label{staratobserver}
F_{\rm{\ast,o}}=\pi \left(\frac{R_{\ast}}{d}\right)^2 \int_{\lambda_{\rm{low}}}^{\lambda_{\rm{high}}}I_{\rm{\ast,s}}q_I(\lambda)d\lambda.
\end{equation}

\subsection{Model parameter summary}

In total, our full physical model as described above contains up to 19 parameters, listed in Table \ref{parasummary}.

\begin{table}[h]
  \centering 
  \caption{Parameters of the forward model}\label{parasummary}
  \begin{tabular}{ll}
\hline
\hline
   Group& Parameters  \\
\hline
   Stellar (4)&  $T_{\ast}$, $R_{\ast}$, $M_{\ast}$, $\rm{[Fe}/\rm{H}]$ \\
   Orbital (4)&  $P_{\rm{orb}}$, $e$, $\omega_p$, $i$ \\
      Planetary (2)&  $R_p$, $M_p$ \\
Atmospheric (5)&  $A_S$, $A_B$, $\epsilon$, $T_d$, $T_n$\\
Asymmetries (4)&  $d_S$, $l_{\rm{start}}$, $l_{\rm{end}}$, $\Theta_{\rm{d}}$\\
\end{tabular}
\end{table}

These 19 parameters however are not independent. For instance, if $A_B$ and $\epsilon$ are fixed, $T_d$ and $T_n$ can be determined using eqs. \ref{daytemp} and \ref{nighttemp}.

\section{Inverse model}

\label{invmodel}

\begin{table*}
  \centering 
  \caption{Planetary scenarios for the forward model (C: CoRoT-1b, T: TrES-2b, H: HAT-P-7b)}\label{corot1mcmcsummary}
  \begin{tabular}{l|ll|cl}
\hline
\hline
   scenario& parameters & priors & planets &comments/constraints  \\
\hline
standard  & $\epsilon$, $A_S$, $M_P$ &$\epsilon$, $A_S$\hspace{0.7cm}uniform in [0,1]& C, T, H&$A_B$=$A_S$\\
  & &$M_P$ \hspace{0.85cm}uniform in [0.1,10] $M_{\rm{jup}}$&& $M_P$ fixed for C\\

standard + asy  & $\epsilon$, $A_S$, $M_P$,$d_S$, $l_{\rm{start}}$, $l_{\rm{end}}$ &$\epsilon$, $A_S$\hspace{0.7cm}uniform in [0,1]&T, H& $A_B$=$A_S$ \\
  & &$M_P$\hspace{0.95cm}uniform in [0.1,10] $M_{\rm{jup}}$&& \\
  & &$d_S$ \hspace{0.99cm}uniform in [0,10]&&$d_S\cdot A_S\leq 1$ \\
  & &$l_{\rm{start}}$, $l_{\rm{end}}$\hspace{0.22cm}uniform in [$90$,$270$]$^{\circ}$&&$l_{\rm{start}} \leq l_{\rm{end}}$ \\

  standard + off  & $\epsilon$, $A_S$, $M_P$, $\Theta_{\rm{d}}$ &$\epsilon$, $A_S$\hspace{0.7cm}uniform in [0,1]&T, H& $A_B$=$A_S$ \\
  & &$M_P$ \hspace{0.85cm}uniform in [0.1,10] $M_{\rm{jup}}$&& \\
  & &$\Theta_{\rm{d}}$\hspace{1.0cm}uniform in [0,$360$]$^{\circ}$&& \\

    standard + both  & $\epsilon$, $A_S$, $M_P$, $\Theta_{\rm{d}}$,$d_S$, $l_{\rm{start}}$, $l_{\rm{end}}$ &$\epsilon$, $A_S$\hspace{0.66cm}uniform in [0,1]&T, H& $A_B$=$A_S$ \\
  & &$M_P$\hspace{0.9cm}uniform in [0.1,10] $M_{\rm{jup}}$&& \\
  & &$\Theta_{\rm{d}}$\hspace{0.97cm}uniform in [0,$360$]$^{\circ}$&& \\
  & &$d_S$\hspace{1.0cm}uniform in [0,10]&&$d_S\cdot A_S\leq 1$ \\
  & &$l_{\rm{start}}$, $l_{\rm{end}}$\hspace{0.18cm}uniform in [$90$,$270$]$^{\circ}$&&$l_{\rm{start}} \leq l_{\rm{end}}$ \\

free A  & $\epsilon$, $A_B$, $A_S$, $M_P$ & $A_S$, $A_B$, $\epsilon$ uniform in [0,1]& C, T, H&$a_V \cdot A_S$<$A_B$<$a_V\cdot A_S$+(1-$a_V$)\\
  & &$M_P$\hspace{0.9cm}uniform in [0.1,10] $M_{\rm{jup}}$&& $M_P$ fixed for C\\

free T  & $T_d$, $T_n$, $A_S$, $M_P$ &$A_S$\hspace{1.0cm}uniform in [0,1]&C, T, H& \\
    &  &  $T_d$, $T_n$\hspace{0.5cm}uniform in [500, 3000] K& &$T_n \leq$$T_d$\\
  & &$M_P$\hspace{0.9cm}uniform in [0.1,10] $M_{\rm{jup}}$&& $M_P$ fixed for C\\
no scattering  & $\epsilon$, $A_B$ & $A_B$,$\epsilon$\hspace{0.75cm}uniform in [0,1]&C& $A_S$=0, \citet{snellen2009}\\
  \end{tabular}
\end{table*}

We use the Bayesian formalism to calculate posterior probability values $p(V_P | D)$ for the parameter vector $V_P$ in the model, given a set $D$ of observations. 

\begin{equation}
\label{bayes}
p(V_P | D)\propto p(D | V_P) \cdot p(V_P).
\end{equation}

The likelihood $p(D | V_P)$ is calculated assuming independent measurements and Gaussian errors for the individual data points. The priors $p(V_P) $ are taken to be uninformative over the entire parameter range allowed (for example, uniform over [0,1] for albedo and heat redistribution).

\subsection{MCMC algorithm}

To sample the full parameter space, we adopt a Markov Chain Monte Carlo (MCMC) approach. In this work, we use the emcee python package developed by \citet{foreman2013}, implementing an algorithm described in \citet{goodman2010}. emcee uses multiple chains (in this work, 500-1,000) to sample the parameter space.  The algorithm proposes, for each chain, new positions based on the position of the entire ensemble of chains. Compared to more traditional MCMC approaches, emcee converges quicker and is less likely to be dependent on initial conditions. Also, when using a high number of chains, the algorithm is less likely to become stuck in local minima since it is possible to eliminate chains from the ensemble (e.g., \citealp{hou2012}).

To ensure good convergence and avoid any contamination by initial conditions, the chains were run for 500-2,000 steps ($>$10-20 auto-correlation lengths for each parameter). The first few auto-correlation lengths were considered as burn-in and discarded for the calculation of parameter uncertainties. Convergence was checked by inspecting visually the evolution of the mean of the entire ensemble and calculating the Gelman-Rubin test. Initial positions were obtained with a random sample within the assumed prior to allow the sampler to start by exploring the entire parameter space.

\begin{figure}[h]
\begin{center}
\includegraphics[width=250pt]{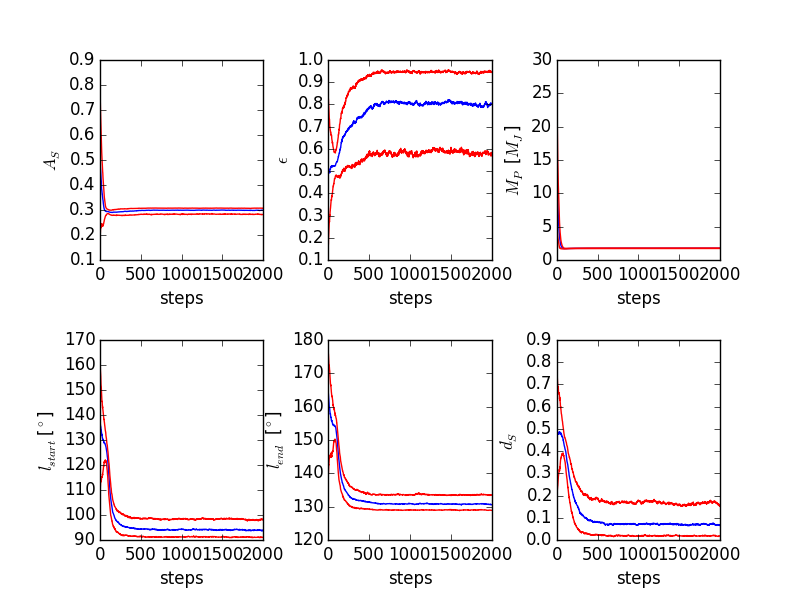}\\
\caption{Trace plots of model parameters for the HAT-P-7b "standard + asy" model (see Table \ref{corot1mcmcsummary}). Blue line traces the ensemble median, red lines correspond to the [0.16,0.84] percentiles (the dark red region in Fig. \ref{uncertain_illu} below).}
\label{asym_conv}
\end{center}
\end{figure}

\begin{figure}[h]
\begin{center}
\includegraphics[width=250pt]{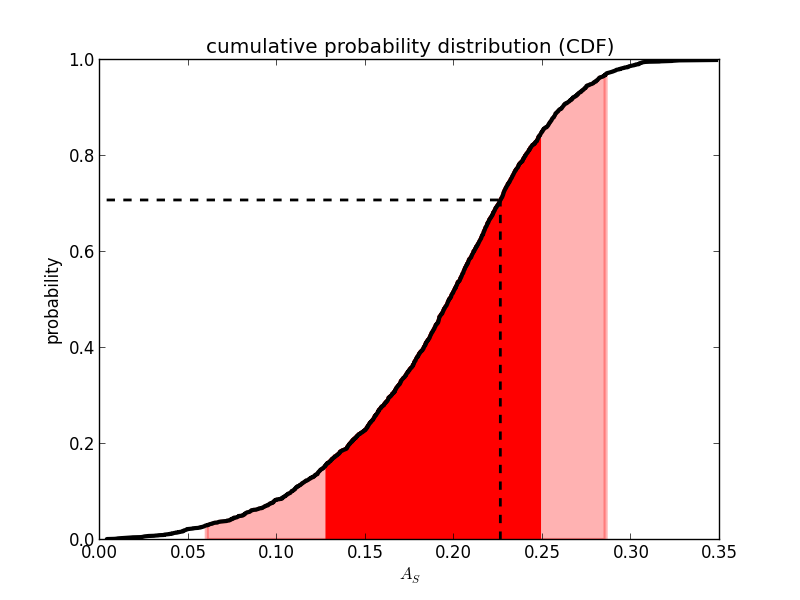}\\
\caption{Example of determination of uncertainty ranges via the cumulative probability distribution: Scattering albedo of CoRoT-1b in the "standard" scenario (see text for further details). 68\,\% credibility region in dark red,  95\,\% credibility region in light red. Dashed lines indicate maximum a-posteriori (MAP) value.}
\label{uncertain_illu}
\end{center}
\end{figure}

Uncertainty ranges are calculated by marginalizing over the posterior distribution, thinned by 60 steps, covering roughly 1-3 auto-correlation lengths of the particular parameter in question. We then determine 68\,\% and 95\,\% credibility regions as the [0.16,0.84] and [0.03,0.97] median-centered percentiles, respectively, of the cumulative probability distributions (CDF). If the parameter distribution were to be Gaussian, these credibility regions would correspond to the 1\, and 2\,$\sigma$ uncertainties, respectively. Figure \ref{uncertain_illu} illustrates our method to determine credibility regions and best-fit parameters. Best-fit parameters are determined as the maximum a-posteriori (MAP) set of parameters, i.e., the walker with the highest a-posteriori probability at the last step of the algorithm. Note that the MAP does not correspond to the median of the CDF in most cases. Furthermore, depending on the nature of the likelihood surface sampled by the chains (degeneracies, slope, etc.), the MAP, as determined from our sample, can occasionally even lie outside the [0.03,0.97] percentile (see Tables \ref{corot1mcmcresults}-\ref{hat7mcmcresults}, Figs. \ref{corot1triangle_2}-\ref{hat7triangle_3}).

\subsection{Goodness-of-fit criteria and model comparison}

We use three standard quantities (e.g., \citealp{feigelson2012}) to evaluate the best-fitting models obtained from the MCMC model: $\chi^2$, the reduced $\chi^2_{\rm{red}}$ and the Bayesian Information Criterion (BIC). These are evaluated for the MAP parameter set (not the median of the CDF).

$\chi^2$, the sum of the weighted squared residuals, is defined as

\begin{equation}
\label{chi2def}
\chi^2=\sum_{\rm{j=1}}^{N_D}\left(\frac{M_{V_P,j}-D_j}{\sigma_j}\right)^2,
\end{equation}

where $N_D$ is the number of data points, $\sigma$ the corresponding uncertainties and $M_{V_P}$ is the model prediction for the parameter vector $V_P$.

The reduced $\chi^2_{\rm{red}}$ is generally calculated by

\begin{equation}
\label{chi2reddef}
\chi^2_{\rm{red}}=\frac{\chi^2}{N_D-N_P},
\end{equation}

where $N_P$ is the number of parameters.  In order to be considered a good fit, $\chi^2_{\rm{red}}$ should be of the order of unity.

Note that by adding more parameters to the fitting model, the $\chi^2$ will decrease. Therefore, we use another criterion, the BIC (valid when $N_D\gg N_P$, which is the case for our calculations), defined as

\begin{equation}
\label{bicdef}
\rm{BIC}=\chi^2 + N_P \ln (N_D).
\end{equation}

The model that minimizes the BIC is taken to be the preferred model. The BIC penalizes overly complex models when complexity or sophistication is not warranted by the data, and also allows direct model comparison. For instance, the model probability ratio $p_M$ (i.e., the probability that the model is preferred over another one), can be expressed as

\begin{equation}
\label{bicdiff}
p_M=e^{-\frac{\Delta \rm{BIC}}{2}},
\end{equation}

where $\Delta \rm{BIC}=\rm{BIC}_{M1}-\rm{BIC}_{M0}$ is the difference between the model $M_1$ under consideration, and the model $M_0$ that minimizes the BIC.

The BIC is an approximation to the evidence (or marginal likelihood) $E_{\rm{model}}$ of a given model:

\begin{equation}
\label{evidencedef}
E_{\rm{model}}=\int p(D | V_P) \cdot p(V_P) dV_P.
\end{equation}

The calculation of the integral in eq. \ref{evidencedef} is in most case computationally very expensive. In emcee, this integral is estimated using a so-called thermodynamic integration, based on an algorithm proposed by \citet{goggans2004}. However, the calculation is very time-consuming (up to $\sim$50 times longer than the MCMC sampling), and does not, in our cases, provide any qualitatively additional information, compared to the BIC. Therefore, we only report the BIC and base our discussions on eq. \ref{bicdiff}.

\section{Model set-up}

\label{setup}

\subsection{Planets}

\subsubsection{CoRoT-1b}

CoRoT-1b is the first planet discovered from space \citep{barge2008} and the first planet with a detected optical phase curve \citep{snellen2009}. Secondary eclipses of CoRoT-1b have been detected in the optical \citep{alonso2009} and IR, both from ground (e.g., \citealp{rogers2009} and \citealp{gillon2009}) and from space (e.g., \citealp{deming2011}). These studies suggested that CoRoT-1b would be a very dark planet, with a low albedo ($A_G \lesssim$0.1) and inefficient heat redistribution ($\epsilon \lesssim$0.2), leading to high dayside temperatures of the order of 2,300\,K. 

We use the binned CoRoT red-channel phase curve data presented by \citet{snellen2009} (their Figure 1). All model parameters (stellar parameters, orbital period and inclination, planetary mass and radius) are taken from \citet{barge2008}, similarly to what was done in \citet{snellen2009}. The orbit is assumed to be circular. This is consistent with the analysis of secondary-eclipse timing (e.g., \citealp{rogers2009}, \citealp{alonso2009}, \citealp{deming2011}) that constrain the eccentricity to values of $e \lesssim$0.03. Given the data quality of the optical phase curve, the results are not sensitive to eccentricity. When allowing $e$ and $\omega_P$ to be fitted (simulations that are not shown here), constraints on either $A_B$ or $\epsilon$ did not change appreciably compared to the circular case.

\subsubsection{TrES-2b}

TrES-2b is a transiting hot Jupiter \citep{odonovan2006} discovered by the TrES survey. It was the first transiting exoplanet discovered in the Kepler field prior to Kepler's 2009 launch. Ground-based and Spitzer secondary eclipse photometry suggests moderately high temperatures around 1,500\,K (e.g., \citealp{odonovan2010}, \citealp{croll2010}). The optical phase curve has been detected in the Kepler data (e.g., \citealp{kipping2011}, \citealp{barclay2012}, \citealp{esteves2013}). Results indicate a very low geometric albedo ($A_G \leq 0.02$) and a rather efficient day-night redistribution of energy ($\epsilon$>0.5), since dayside and nightside temperatures are quite similar (around 1,300-1,500\,K, \citealp{esteves2013}). 

We use the binned Kepler phase curve data (\citealp{barclay2012}, their Figure 5). All model parameters (stellar parameters, orbital period and inclination, planetary radius) are taken from \citet{barclay2012}. The orbit is assumed to be circular, since secondary-eclipse timing provided constraints consistent with zero eccentricity (e.g., \citealp{odonovan2010}, \citealp{croll2010}) and optical phase curve analysis by \citet{esteves2015} did not find any significant eccentricity.

\subsubsection{HAT-P-7b}

As TrES-2b, HAT-P-7b is a very hot Jupiter discovered in the Kepler field \citep{pal2008} prior to the launch of the satellite. It was the first Kepler planet with a measured optical phase curve \citep{borucki2009}. Further analysis of the phase curve demonstrated a detection of ellipsoidal variations \citep{welsh2010}. \citet{christiansen2010} used Spitzer secondary eclipse data to infer maximum brightness temperatures of more than 3,000\,K. \citet{esteves2013} presented a new optical phase curve using full 3-year Kepler photometry. Measured phase curve amplitudes and secondary eclipse depths vary considerably between \citet{borucki2009}, \citet{welsh2010} and \citet{esteves2013}, leading to differences in inferred dayside and nightside temperatures of the order of 500\,K which are most likely a consequence of the extended data set in \citet{esteves2013}. Similar brightness temperature differences of 300-500\,K for the Spitzer secondary eclipse data have been found by \citet{cowan2011hot}, compared to \citet{christiansen2010}, which they attribute to the use of different stellar models. Based on calculated brightness temperatures and optical phase curves, \citet{christiansen2010} inferred a very inefficient heat redistribution ($\epsilon$ close to zero) between dayside and nightside and modest geometric albedos ($A_G$<0.1). \citet{esteves2013} found a geometric albedo of $A_G=0.18$ and a relatively homogenous temperature distribution, in contrast to \citet{christiansen2010}. \citet{schwartz2015} found moderate albedos, slightly lower than \citet{esteves2013} and moderate recirculation values, also in contrast to previous analysis \citep{christiansen2010}.

We use the binned Kepler phase curve data (\citealp{esteves2013}, their Figure 3) . However, model parameters are not taken from \citet{esteves2013} or \citet{esteves2015}. Instead, stellar parameters are taken from \citet{eylen2012}, and planetary parameters (inclination, radius) are taken from \citet{eylen2013}. We refer to Appendix \ref{phase_transit} for a discussion on our choice of parameters. Again, we assume a circular orbit, since detailed radial-velocity data is consistent with $e$=0 (e.g., \citealp{pal2008}, \citealp{winn2009}). Also, previous phase curve analysis did not find a hint of significant eccentricity \citep{esteves2015}

\subsection{Photometric fits}

\citet{snellen2009} analyze the phase curve of CoRoT-1b in terms of the normalized flux $F_{\rm{norm}}$, normalized to the primary, instead of secondary, eclipse (i.e., without transit, the flux at zero-phase would be unity). Therefore, we calculate the forward model as:

\begin{equation}
\label{corot1phase}
F_{\rm{norm}}(\alpha)=\frac{1+C(\alpha)}{1+C(0)}.
\end{equation}

Both \citet{barclay2012} and \citet{esteves2013} fit the phase curves in terms of variations in the photometric light curve, as in eq. \ref{contrastref}. They however introduce another, non-physical parameter, a zero-point flux offset $f_0$. This parameter is related to the data reduction and not a priori linked to any physical characteristics of the star-planet system. The light curve $L_C$ is then described with the following equation:

\begin{equation}
\label{f0phase}
L_C(\alpha)=C(\alpha)+f_0.
\end{equation}

We will use this equation for our analysis of TrES-2b and HAT-P-7b. Hence, $f_0$ is added to the tally of free parameters of the physical model (Table \ref{corot1mcmcsummary}).

\subsection{Planetary scenarios}

As summarized in Table \ref{corot1mcmcsummary}, we explore several different scenarios for the three planets. The main difference between these models lies in the treatment of the thermal component of the phase curve.

The first scenario ("standard") assumes scattering albedo $A_S$ and heat redistribution as fitting parameters and uses eqs. \ref{daytemp} and \ref{nighttemp} to calculate hemispheric temperatures. The Bond albedo $A_B$ is fixed to the scattering albedo ($A_S$=$A_B$). This allows our results to be compared directly to previous inferences of albedo and heat recirculation (e.g., \citealp{snellen2009}, \citealp{schwartz2015}) and is consistent with previous studies (e.g., \citealp{alonso2009}). The equality between Bond albedo and scattering albedo is motivated by the fact that a significant fraction $a_V$ of stellar light is emitted in the CoRoT red channel or the Kepler bandpass ($a_V\approx0.3$ for CoRoT-1).

The second scenario ("free albedo") relaxes this tight coupling between Bond albedo and scattering albedo. We allow both $A_B$ and $A_S$ to vary freely, but still calculate dayside and nightside temperatures from the Bond albedo, as in the first scenario. Based on energy conservation, however, we put some constraints on the Bond albedo:

\begin{equation}
\label{abondconstraint}
a_V \cdot A_S\leq A_B\leq a_V\cdot A_S+(1-a_V).
\end{equation}

This equation takes into account the contribution of the scattering albedo to the overall radiative budget. The lower limit of the Bond albedo ($A_{B,\rm{low}}=a_V \cdot A_S$) is a hard lower limit since it implies zero albedo outside the bandpass. Similarly, when assuming an albedo of unity outside the bandpass, we obtain the strict upper limit of $A_{B,\rm{high}}=a_V \cdot A_S+(1-a_V)$.

In a third approach ("free T"), we use $T_d$ and $T_n$ as fitting parameters (instead of Bond albedo and heat recirculation) and only impose $T_n \leq T_d$.

To compare directly to the phase-curve analysis by \citet{snellen2009}, we then use a fourth model approach for CoRoT-1b. We set $A_S$=0, i.e., the phase curve is produced by thermal emission only ("no scattering"). This was done because the physical model used in \citet{snellen2009} only accounts for thermal radiation (their eq. 1).

All scenarios for CoRoT-1b assume symmetric phase curves and do not fit for planetary mass because of the relatively low signal-to-noise ratio and reduced phase resolution of the binned phase curve. 

For TrES-2b and HAT-P-7b, however, the data quality is good enough that ellipsoidal variations are clearly seen. Therefore, we take the planetary mass to be a fit parameter. We also fitted the data with asymmetric standard models (see Table\ref{corot1mcmcsummary}), either with asymmetric scattering, a dayside offset or a combination of both.

\subsection{Energy balance}

The energy balance can also be expressed in terms of received and emitted flux. For this, we divide the planet in uniform hemispheres (see Sect. \ref{formodel}). The Bond albedo determines the overall received flux which must then be re-emitted on the day and night hemispheres.

\begin{equation}
\label{fluxbalance}
(1-A_B) \cdot F_{\rm{\ast,p}}=2\left (\sum_j^{N_{\rm{day}}}F_{j,\rm{day}}+\sum_j^{N_{\rm{day}}}F_{j,\rm{night}}\right ),
\end{equation}

where the sums contain the measured fluxes in the various wavelength bands (${N_{\rm{day}}}$ bands of the dayside spectrum, ${N_{\rm{night}}}$ of the nightside spectrum). Hence, under the assumption that outside the measured bands, no flux is emitted, we obtain an upper limit for the Bond albedo. This constraint is physically sound and does not rely on specific assumptions except radiative equilibrium and the hemispheric uniformity.

\section{Results}

\label{results}

\subsection{Convergence results}

The results of the convergence tests and the trace plots for all parameters are shown in Appendix \ref{convergence_res}. All simulations seem to have reached a stationary distribution and thus converged. Most parameters also pass the Gelman-Rubin (GR) test (generally, for most MCMC tools, a value of less than 1.1-1.2 is considered acceptable). Note, however, that the GR test is a test for good mixing of the ensemble. Thus, even when a stationary distribution is reached because of a large number of chains used in the calculation, parameters might fail the GR test (values larger than 1.2). This usually means that the inter-chain variance is large compared to the variance of individual chains (in our cases, sometimes orders of magnitude). For strongly non-linearly correlated parameters (e.g., in the "standard + off"  and "standard + both" scenarios of HAT-P-7b, see below), it will take a long time for a single chain to explore the entire permitted range. Thus, the GR statistic will be large, without necessarily impacting the convergence of the ensemble. This is clearly shown by the fact that the MCMC algorithm actually finds strong correlations.

A particularly good example of this are the "standard + asy" and "standard + both" scenarios for TrES-2b (see Figs. \ref{tres2_conv_2} and \ref{tres2_conv_3}, Table \ref{tres2gelman}). Since the parameters describing the albedo asymmetry are tightly anti-correlated, the MCMC algorithm finds basically two separated solutions (see also Fig. \ref{tres2triangle_3}), resulting in "bad" GR values. When restricting the calculations to one of the solutions (not shown), the GR test is passed by all parameters (values less than 1.17 in the "both" scenario, less than 1.09 in the "asymmetric" scenario).

\subsection{CoRoT-1b}

\begin{figure}[h]
\begin{center}
\includegraphics[width=250pt]{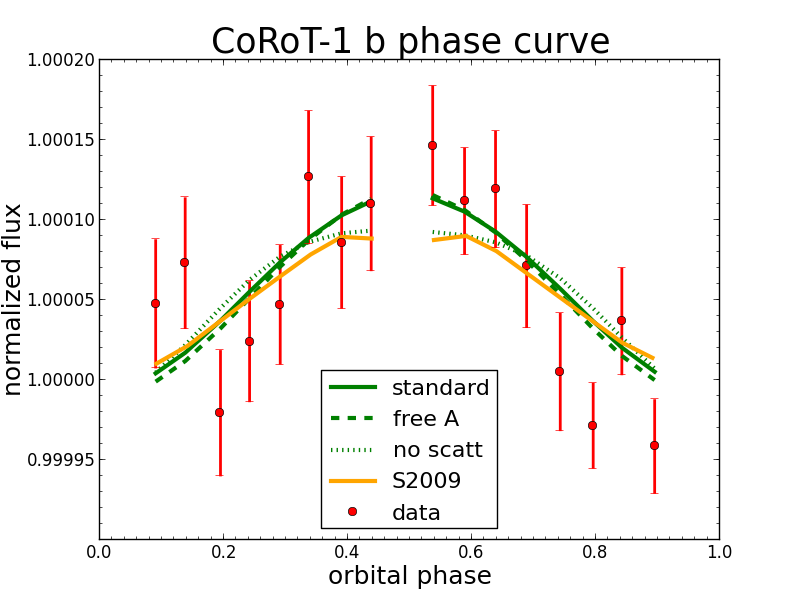}\\
\caption{CoRoT-1b red-channel phase curve: Comparison of best-fit models with data (red) and fit by \citet{snellen2009} (yellow). Primary transit and secondary eclipse not shown.}
\label{corot1b_phase}
\end{center}
\end{figure}

Figure \ref{corot1b_phase} shows the data and the best-fit model from \citet{snellen2009} as well as our best-fit models from the various planetary scenarios (Table \ref{corot1mcmcsummary}). Since the "free T" and the "free albedo" best-fit models are virtually identical, only the latter is shown, for clarity. Clearly, the standard and free-albedo best-fit models do not differ by much. It is also apparent that both the models presented in this work and the model of \citet{snellen2009} provide reasonable fits to the data.

Best-fit parameters as well as goodness-of-fit criteria and MCMC posterior parameter distributions are reported in Appendix \ref{mcmcresults}  (Table \ref{corot1mcmcresults}, Figs. \ref{corot1triangle_2} and \ref{corot1triangle_3}). Based on the BIC value, the data seem to slightly favor the standard scenario, however, all models in Table \ref{corot1mcmcresults} seem acceptable.

\begin{figure}[h]
\begin{center}
\includegraphics[width=250pt]{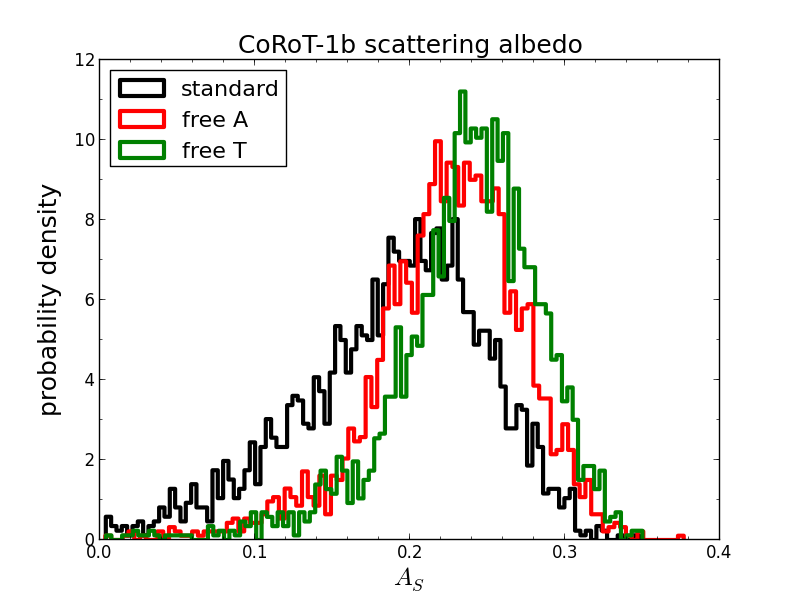}\\
\caption{Marginalized posterior distributions for CoRoT-1b scattering albedo in different models. $A_S$ is approximately independent of the thermal model (0.11<$A_S$<0.3 at 95\,\% confidence in the "free A" model).}
\label{corot1b_scatt}
\end{center}
\end{figure}

Furthermore, fit results suggest that the phase curve is dominated by scattering rather than by thermal emission. This can be  seen in Table \ref{corot1mcmcresults} from the fact that the inferred value of the scattering albedo $A_S$ is largely unaffected by the choice of the thermal model (0.11<$A_S$<0.3 at 95\,\% confidence in the "free A" model). The combined arithmetic mean of the scattering albedo in the three models ("standard", "free A", "free T") is $A_S$=0.22. The independence of $A_S$ of the thermal model is illustrated in Fig. \ref{corot1b_scatt}. 

\begin{figure}[h]
\begin{center}
\includegraphics[width=120pt]{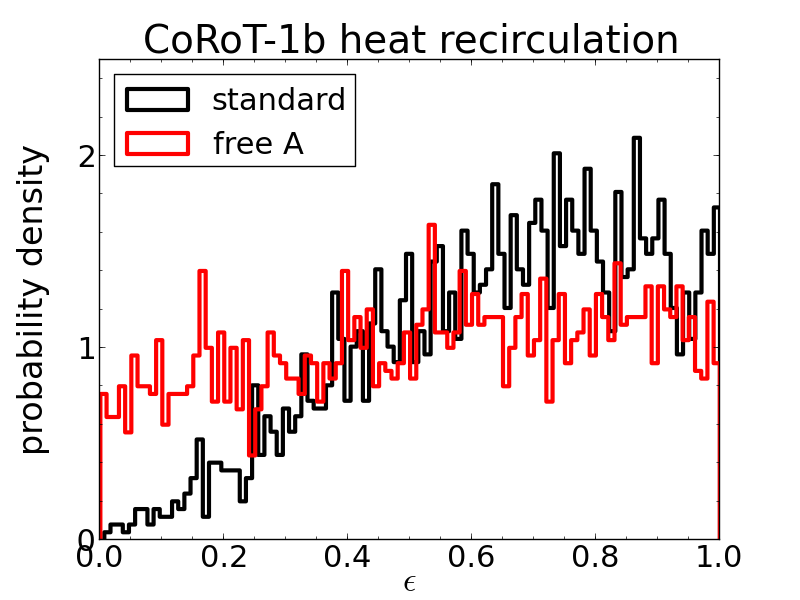}
\includegraphics[width=120pt]{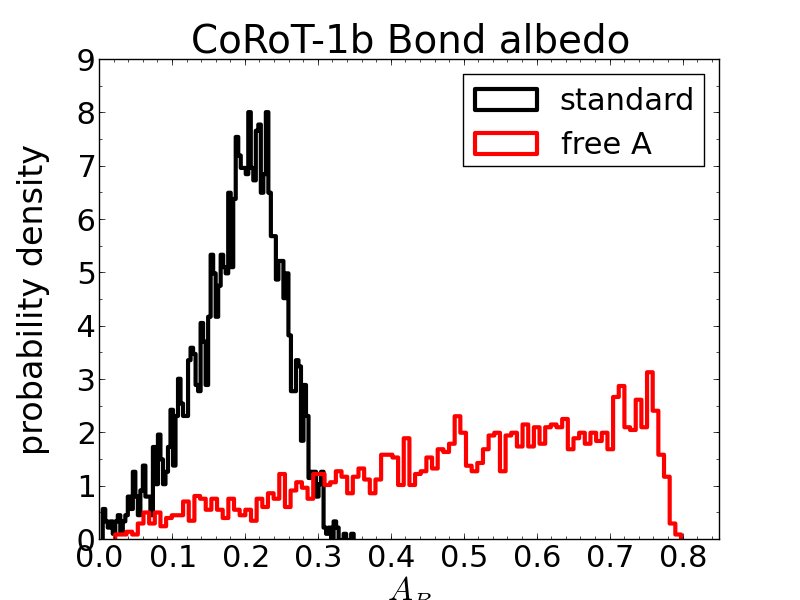}\\
\caption{Marginalized posterior distributions for CoRoT-1b $\epsilon$ (left) and $A_B$ (right) in different models. Constraints on $A_B$ are weak. $\epsilon$ is unconstrained and depends on the thermal model.}
\label{corot1b_bond}
\end{center}
\end{figure}

Constraints for Bond albedo are weak, and heat recirculation is essentially unconstrained. In addition, constraints on $\epsilon$ show some dependence on the choice of the thermal model employed (see Fig. \ref{corot1b_bond}). This suggests that indeed the optical phase curve does not contain much information on the thermal component, hence is dominated by reflected starlight rather than thermal emission. 

Independent circumstantial evidence for this can be drawn from the observed, flat transmission spectrum of CoRoT-1b \citep{schlawin2014} which can be interpreted as a hint for the presence of clouds (or, at least, a reflecting layer in the upper atmosphere).

\begin{figure}[h]
\begin{center}
\includegraphics[width=250pt]{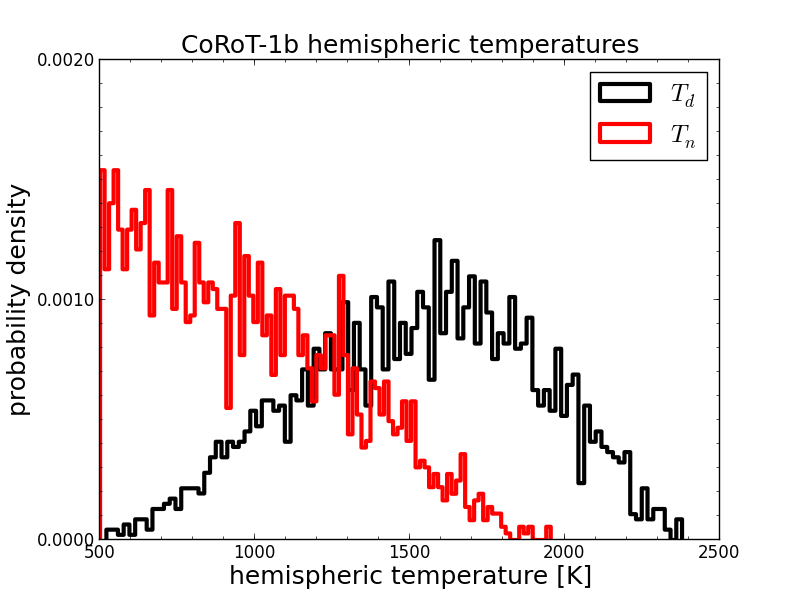}\\
\caption{Marginalized posterior distributions for CoRoT-1b dayside and nightside temperatures in the "free T" model.}
\label{corot1b_temp}
\end{center}
\end{figure}

Figure \ref{corot1b_temp} shows the constraints on dayside and nightside temperatures in the "free T" model. Inferred dayside temperatures are much lower than the IR brightness temperatures derived from secondary-eclipse measurements. Essentially, results are consistent with zero phase curve contribution from thermal emission, in accordance with the "standard" and "free A" models. Also, in accordance with \citet{snellen2009}, we find that the nightside emission is consistent with zero.

Our results are somewhat contrary to the results from the IR secondary-eclipse measurements (\citealp{schwartz2015}) and conclusions of the analysis of the optical phase curve by \citet{snellen2009}. Both studies used the same formalism as our "standard" scenario (i.e., using eqs. \ref{daytemp} and \ref{nighttemp}), and they conclude that CoRoT-1b is probably a low-albedo planet with inefficient heat recirculation. Note, however, that the phase curve model used by \citet{snellen2009} only takes thermal radiation into account, hence the scattering albedo is, by default, zero. Our results suggest significant scattering and, depending on the thermal model, at least some recirculation.

\begin{figure}[h]
\begin{center}
\includegraphics[width=250pt]{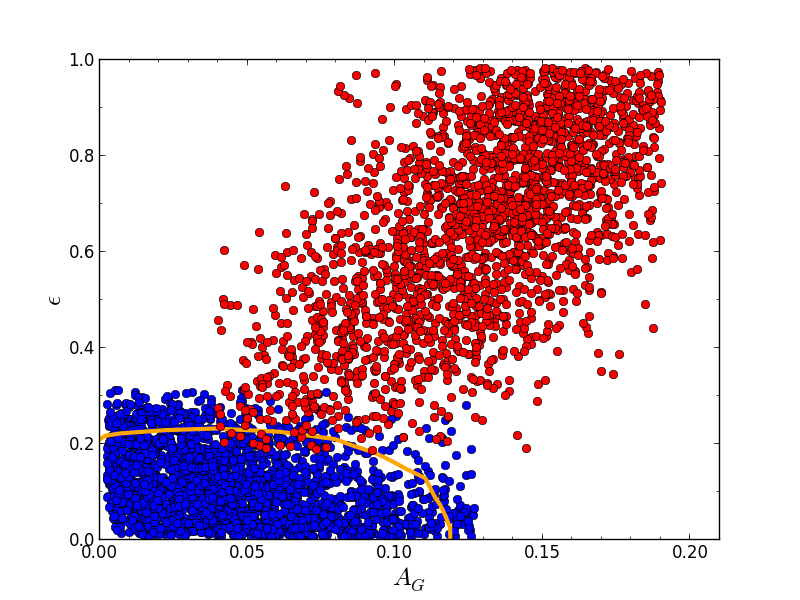}\\
\caption{Joint credibility regions of recirculation and geometric albedo (see eq. \ref{ageo}) for CoRoT-1b in the "standard" (red dots) and "no scattering" (blue dots) scenarios. Orange contour: 1\,$\sigma$ uncertainty region in \citet{schwartz2015}. Our "standard" model strongly disagrees with previous work.}
\label{corot1b_circulation}
\end{center}
\end{figure}

These contrasting results are illustrated in Fig. \ref{corot1b_circulation}. For this, we calculated the joint credibility regions in the two-dimensional $A_G$-$\epsilon$ space such that points simultaneously lie in the 95\,\% credibility regions of each parameter. The joint credibility region thus corresponds to an approximately 90\,\% probability. Figure \ref{corot1b_circulation} shows these credibility regions for both the "standard" and the "no scattering" scenario. It is clear that the 1\,$\sigma$ uncertainty region of \citet{schwartz2015} and our joint credibility region barely overlap in the "standard" case. By contrast, the "no scattering" case yields approximately the same constraints as the analysis by \citet{snellen2009} and \citet{schwartz2015}.

However, the no-scattering case is equivalent to imposing a strong prior on the scattering albedo ($A_S$=0) that seems rather ad-hoc. Therefore, on physical grounds, we prefer our standard model over the no-scattering case, even though goodness-of-fit criteria could not be used to decide formally which model to prefer, given that the $\Delta$BIC is small (see Table \ref{corot1mcmcresults}).

Despite the lack of information on the thermal emission of CoRoT-1b from its optical phase curve, we can however still put some constraints on the Bond albedo using the estimated value of $A_S$ and eq. \ref{abondconstraint}. Fit results for the free-T scenario (see Table \ref{corot1mcmcresults}) translate, at 95\,\% confidence, to 0.03<$A_B$<0.82, or, when using the best-fit values, 0.06<$A_B$<0.8, with $a_V \approx 0.26$ which is consistent with results from the "free A" scenario.

When using the reported IR brightness temperatures of CoRoT-1b (\citealp{rogers2009}, \citealp{deming2011}) and our optical brightness temperatures (${N_{\rm{day}}}$=4, ${N_{\rm{night}}}$=1 in eq. \ref{fluxbalance}), we obtain $A_B$<0.85, consistent with results from eq. \ref{abondconstraint}. This is not a strong constraint, since the spectral coverage is not large. However, it is a mostly model-independent result. Especially, nightside emission measurements are missing, except for our optical phase curve analysis (resulting in a non-detection since the nightside emission is consistent with zero at the level of the measurement errors).

\subsection{TrES-2b}

\begin{figure}[h]
\begin{center}
\includegraphics[width=250pt]{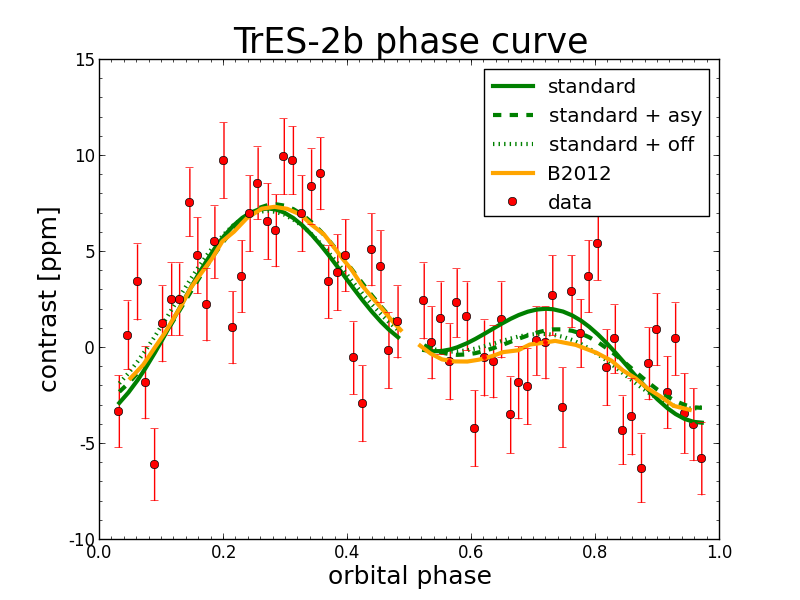}\\
\caption{TrES-2b phase curve: Comparison of best-fit models with data (red) and fit by \citet{barclay2012} (orange). Primary transit and secondary eclipse not shown.}
\label{tres2b_phase}
\end{center}
\end{figure}

Figure \ref{tres2b_phase} shows the various best-fit models of the different MCMC scenarios. It is apparent that the main difference between symmetric and asymmetric models is in the second peak where asymmetric models provide a better fit to the data. For clarity, the "free A" and "free T" scenarios are not shown. In Table \ref{tres2mcmcresults}, we state best-fit parameters as well as 95\,\% credibility regions for the parameters. Parameter posterior distributions are shown in Figs. \ref{tres2triangle_1}-\ref{tres2triangle_3} in the Appendix.

\begin{figure}[h]
\begin{center}
\includegraphics[width=250pt]{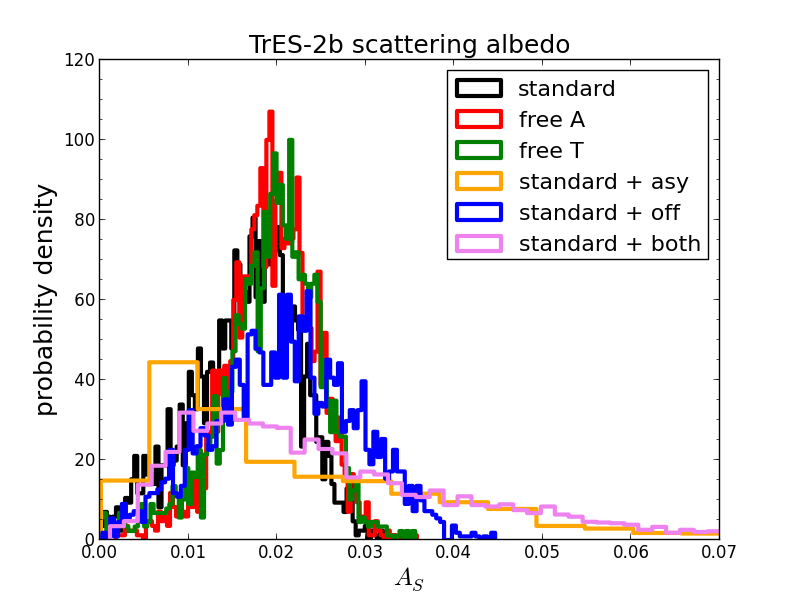}\\
\caption{Marginalized posterior distributions for TrES-2b $A_S$ in different models. TrES-2b is a dark planet in all models ($A_S$<0.03 at 95\,\% confidence).}
\label{tres2b_alb}
\end{center}
\end{figure}

Results presented in Fig. \ref{tres2b_alb} confirm the very dark nature of TrES-2b, with scattering albedos $A_S$<0.03 at 95\,\% credibility, as already inferred by previous authors (e.g., \citealp{kipping2011}, \citealp{barclay2012}, \citealp{esteves2013,esteves2015}).

\begin{figure}[h]
\begin{center}
\includegraphics[width=250pt]{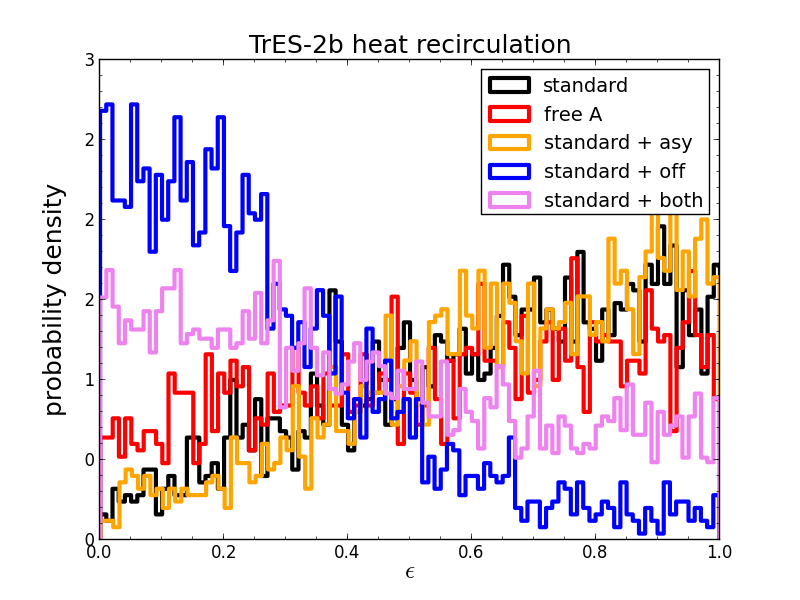}\\
\caption{Marginalized posterior distributions for TrES-2b $\epsilon$ in different models. $\epsilon$ depends strongly on the chosen model.}
\label{tres2b_eps}
\end{center}
\end{figure}

Furthermore, as shown in Fig. \ref{tres2b_eps}, the value of $\epsilon$ depends strongly on the choice of the planetary scenario (asymmetric vs. symmetric), as is the case for CoRoT-1b. For the asymmetric models, this is immediately obvious. With $\epsilon$ close to unity, there will be no strong contrast between dayside and nightside, hence no asymmetry can be produced in the "standard + off" scenario. In order to produce a noticeable effect on the phase curve, inefficient heat recirculation is required ($\epsilon$ close to zero). As can be seen in Fig. \ref{tres2b_eps}, the distributions for $\epsilon$ in the "standard + off" and "standard + both" scenarios are close to each other, which suggests that the preferred mechanism to produce an  asymmetric phase curve is probably thermal radiation, not reflected light (i.e., $\epsilon$ is probably small).

\begin{figure}[h]
\begin{center}
\includegraphics[width=250pt]{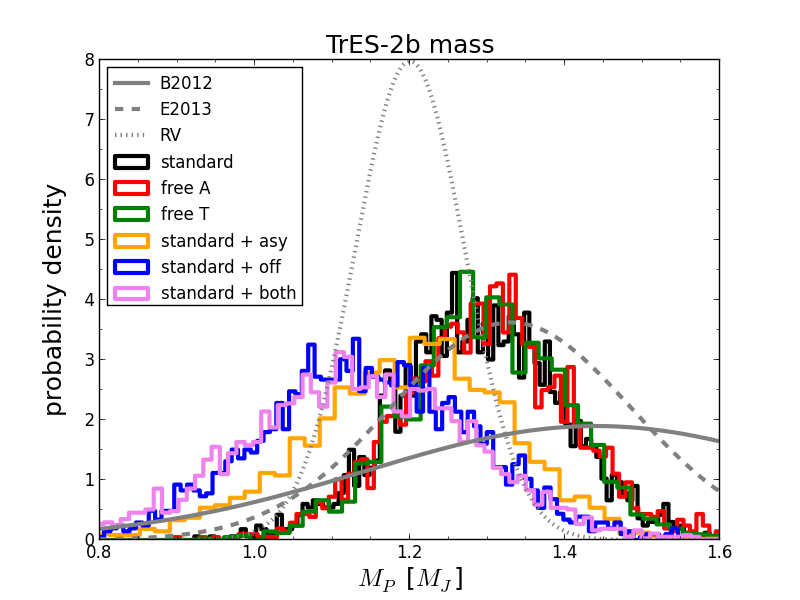}\\
\caption{Marginalized posterior distributions for TrES-2b $M_P$ in different models. Mass from previous studies (including RV measurements) in gray.}
\label{tres2b_mass}
\end{center}
\end{figure}

Figure \ref{tres2b_mass} shows the marginalized posterior distributions for the inferred planetary mass. Also shown are results of previous phase curve modeling by \citet{barclay2012} and \citet{esteves2013} as well as RV mass determinations. It is obvious that the precision of the photometric mass is worse than that of the RV data. However, our mass values are consistent with previous studies.

It is clear that our model provides a relatively good fit to the observed phase curve (Fig. \ref{tres2b_phase}, $\chi^2_{\rm{red}}\approx$2-2.2, see Table \ref{tres2mcmcresults}). Compared to the best-fit model by \citet{barclay2012}, our symmetric models consistently calculate a noticeably higher photometric contrast post-eclipse. However, as already noted by \citet{esteves2015}, this is simply because the model of \citet{barclay2012} allows for a separate, independent fitting of the beaming and ellipsoidal amplitudes that are adjusted to compensate for the apparent decrease in the phase curve (see also discussion in \citealp{faigler2015}). This is also the main reason why beaming and ellipsoidal masses do not agree with each other in \citet{barclay2012} or \citet{esteves2013}.

When allowing for asymmetric phase curves, the fit becomes slightly better.  In terms of the respective BIC values (see Table \ref{tres2mcmcresults}), the "standard + off" scenario is slightly favored, although a $\Delta \rm{BIC} \approx 3.5$ is not enough to detect firmly an asymmetry. Our tentative detection is therefore not in contradiction with conclusions of \citet{esteves2015} who state that symmetric models are favored for TrES-2b.

Figure \ref{tres2triangle_2} (right panel) shows some interesting correlations between $\epsilon$, $A_S$ and $\Theta_D$. For an increasing albedo, the offset of the dayside also increases. This is because, with increasing contribution of scattered light to the phase curve, the offset must become more pronounced to affect the phase curve and produce a visible asymmetry. Furthermore, for low albedos (e.g., high temperatures and low scattering contribution), inefficient heat recirculation is required to produce a phase curve at all. Upon increasing the scattering albedo, higher values of $\epsilon$ are allowed, but that reaches a maximum. Beyond this maximum, scattered light will dominate the phase curve, and again, $\epsilon$ must decrease to produce a significant thermal asymmetry (i.e., large day-night temperature differences).

Figure \ref{tres2triangle_3} (left panel) illustrates a degeneracy in the "standard + asy" scenario, between $A_S$, $d_S$, $l_{\rm{start}}$ and $l_{\rm{end}}$. Since the asymmetric phase curve requires a lower post-eclipse amplitude to fit the data, the "evening" side must be brighter than the "morning" side. This can be achieved in two ways: either high $A_S$ and correspondingly $d_S$<1 and $l_{\rm{start}}$ and $l_{\rm{end}}$ delimiting part of the "morning" side, or low $A_S$ and correspondingly $d_S$>1 and $l_{\rm{start}}$ and $l_{\rm{end}}$ delimiting part of the evening side. However, note that from a physical standpoint, it is unclear how the albedo could be higher on the evening side. Mostly, it is assumed that clouds are responsible for the scattering. These are supposed to dissipate over time while circulating over the dayside hemisphere, therefore post-eclipse maxima are not generally attributed to clouds (e.g., \citealp{demory2013inhomogen}, \citealp{esteves2015}). Another possibility for post-eclipse maxima being due to albedo changes would be the photodissociation of absorbers such as TiO or VO. In atomic form, the absorption would be much less efficient, hence the planetary albedo would increase. Investigating this possibility is, however, beyond the scope of this work.

\begin{figure}[h]
\begin{center}
\includegraphics[width=250pt]{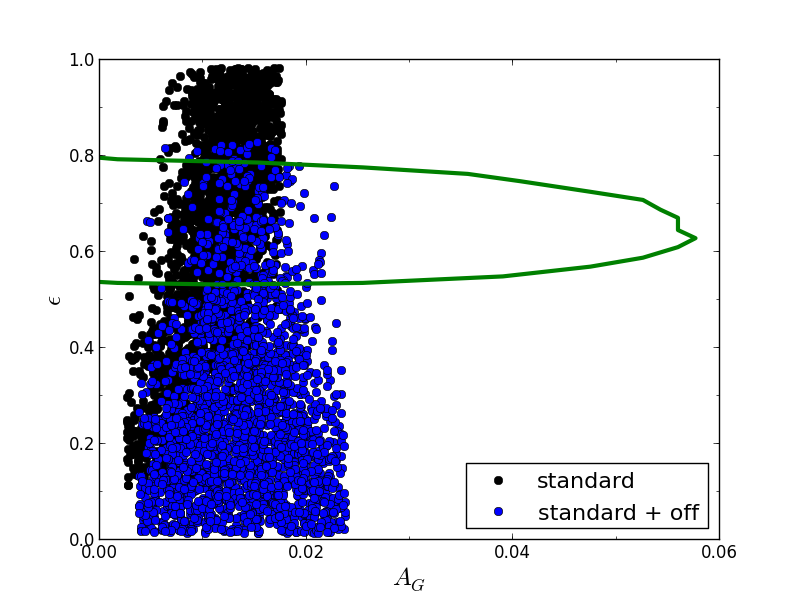}\\
\caption{Joint credibility regions of recirculation and geometric albedo for TrES-2b in the "standard" (black dots) and "standard + off" (blue dots) scenarios. Green contour: 1\,$\sigma$ uncertainty region in \citet{schwartz2015}. Both this and previous work are consistent with each other.}
\label{tres2b_circulation}
\end{center}
\end{figure}

Figure \ref{tres2b_circulation} shows the constraints on recirculation and geometric albedo as inferred from our "standard" and "standard + off" scenarios, compared to results by \citet{schwartz2015}. As for CoRoT-1b, we use joint credibility regions to illustrate our inferred range in the $A_G$-$\epsilon$ plane. Both our results and the results of \citet{schwartz2015} are consistent with each other and certainly agree better than for CoRoT-1b (see Fig. \ref{corot1b_circulation}). However, note that the 1\,$\sigma$ region of \citet{schwartz2015} contains roughly 30\,\% of the points of the "standard" model, but only about 8\,\% of the points of the "standard + off" model.

Similarly to CoRoT-1b, we use the calculated $A_S$ values to put constraints on the overall Bond albedo of TrES-2b (eqs. \ref{abondconstraint} and \ref{fluxbalance}). Using $a_V \approx 0.4$, we obtain, based on the optical phase curve, $A_B$<0.6. Putting together the dayside brightness temperature measurements from IRAC and Ks bands (\citealp{odonovan2010}, \citealp{croll2010}), we then derive $A_B$<0.68. These constraints are somewhat tighter than the ones derived for CoRoT-1b because the spectral coverage is larger and TrES-2b is a cooler planet.

\subsection{HAT-P-7b}

\begin{figure}[h]
\begin{center}
\includegraphics[width=250pt]{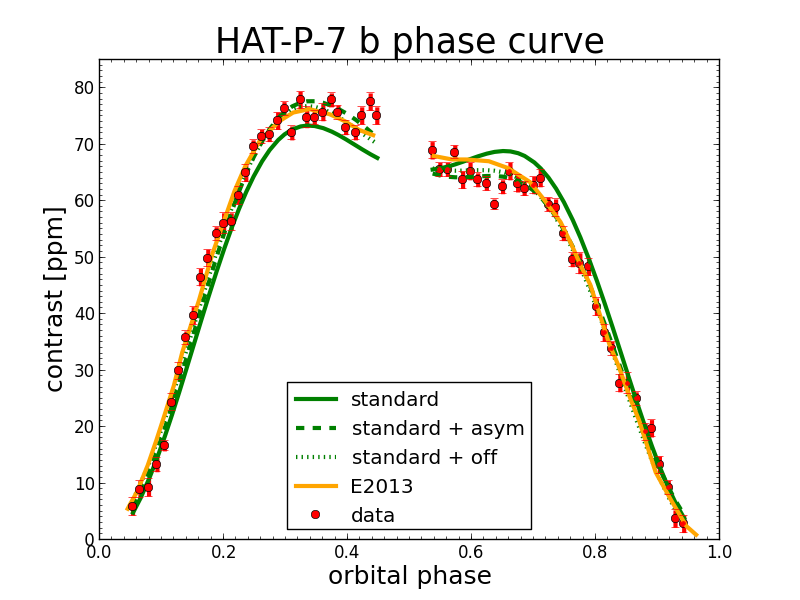}\\
\caption{HAT-P-7b phase curve: Comparison of best-fit models with data (red) and fit by \citet{esteves2013} (orange). Primary transit and secondary eclipse not shown.}
\label{hat7b_phase}
\end{center}
\end{figure}

Figure \ref{hat7b_phase} shows our best-fit models of the different MCMC scenarios. In contrast to TrES-2b, the phase curve of HAT-P-7b is dominated by reflected light, rather than by the ellipsoidal variations (even though these are still clearly visible). Again, for clarity, the "free A" and "free T" scenarios are not shown. In Table \ref{hat7mcmcresults}, we state best-fit parameters as well as 95\,\% credibility regions for the parameters. Parameter posterior distributions are shown in Figs. \ref{hat7triangle_1}-\ref{hat7triangle_3} in the Appendix.

\begin{figure}[h]
\begin{center}
\includegraphics[width=250pt]{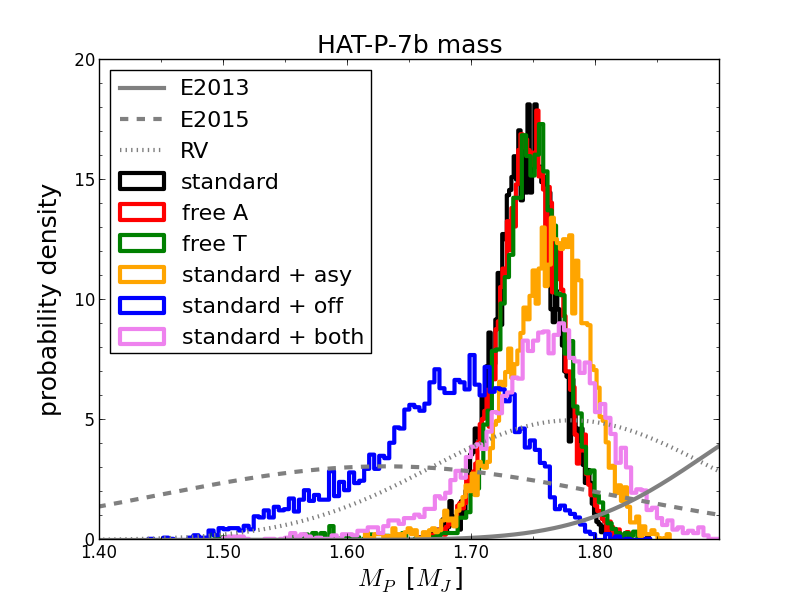}\\
\caption{Marginalized posterior distributions for HAT-P-7b $M_P$ in different models. Mass from previous studies (including RV measurements) in gray.}
\label{hat7_mass}
\end{center}
\end{figure}

Figure \ref{hat7_mass} shows the marginalized posterior distributions for the inferred planetary mass. Also shown are results of previous phase curve modeling by \citet{esteves2013} and \citet{esteves2015} as well as RV mass determinations. 
Again, as for TrES-2b, our estimated mass values are consistent with previous studies, and the determined planetary mass is not greatly affected by the choice of the phase curve model. Note that the formal uncertainties on planetary mass are somewhat smaller in our work than the RV uncertainties. This is mainly due to the excellent photometric quality of the phase curve and the fact that we fix stellar parameters, i.e., the stellar mass does not contribute to the final uncertainty on mass estimates. Note also the strong disagreement between mass estimates from \citet{esteves2013} and \citet{esteves2015} (plain and dashed gray lines in Fig. \ref{hat7_mass}, respectively). This is because the former uses separate beaming and ellipsoidal amplitudes as fitting parameters, while the latter uses planetary mass as a fitting parameter, as we do here. The beaming amplitude is adjusted to account for the asymmetry of the phase curve, thus planetary mass estimates from beaming and ellipsoidal amplitudes do not agree (4.2 compared to 1.6 $M_J$, see Table 5 in \citealp{esteves2013}). Hence, it is clearly demonstrated that a separate fitting of both amplitudes can potentially lead to incorrect mass estimates.

\begin{figure}[h]
\begin{center}
\includegraphics[width=250pt]{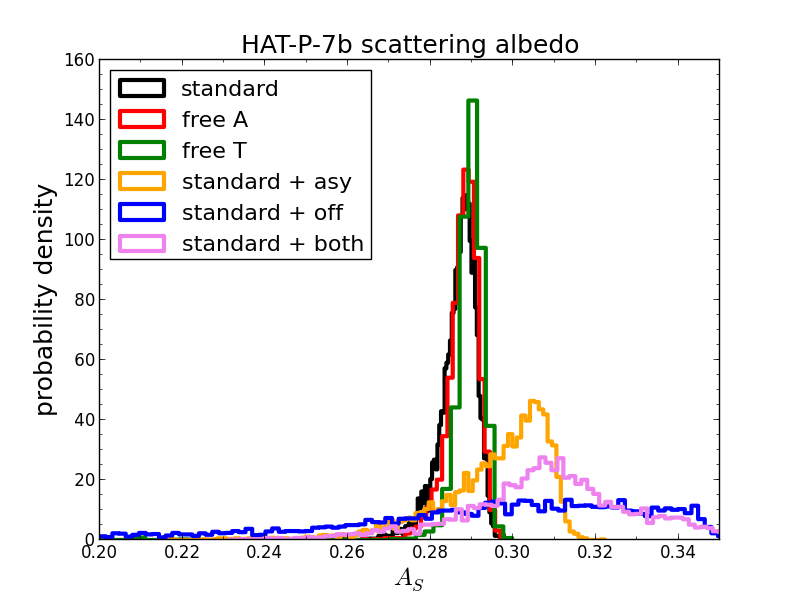}\\
\caption{Marginalized posterior distributions for HAT-P-7b $A_S$ in different models. $A_S$ is mostly independent of the adopted model. 0.26<$A_S$<0.34 at 95\,\% confidence.}
\label{hat7_alb}
\end{center}
\end{figure}

Figure \ref{hat7_alb} shows the inferred scattering albedo for the different scenarios. The values are broadly consistent with the values from \citet{esteves2015} who find a geometric albedo of $A_G \approx$ 0.2, close to our values (recall $A_G$=$\frac{2}{3} A_S$, eq. \ref{ageo}). The fact that $A_S$ is mostly independent of the specific planetary scenario suggests that the estimated value of $A_S$ (0.26<$A_S$<0.34 at 95\,\% confidence) is robust. The arithmetic mean for the combined scenarios is $A_S$=0.28.

\begin{figure}[h]
\begin{center}
\includegraphics[width=250pt]{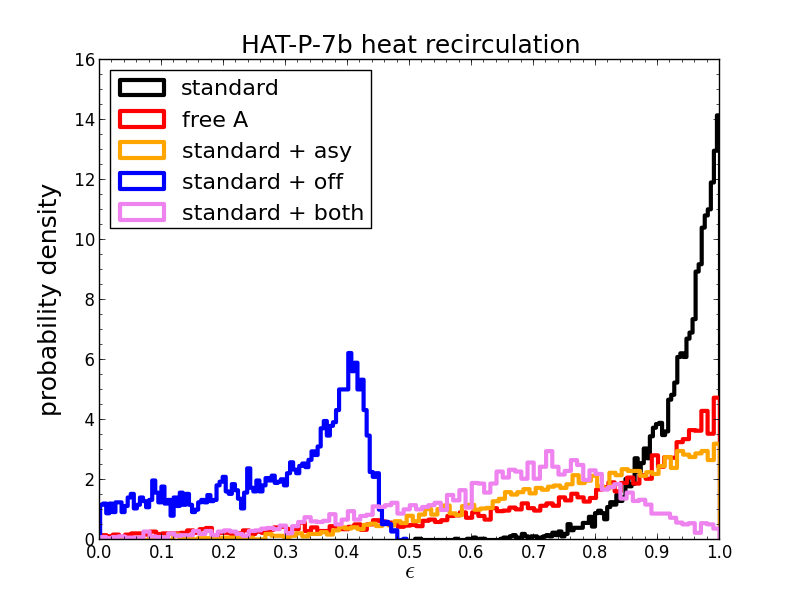}\\
\caption{Marginalized posterior distributions for HAT-P-7b $\epsilon$ in different models. $\epsilon$ is strongly dependent on the adopted planetary scenario.}
\label{hat7_eps}
\end{center}
\end{figure}

As before, $\epsilon$ depends on the choice of the thermal model, as illustrated in Fig. \ref{hat7_eps}. Similar to what has been found for TrES-2b, the "standard + off" scenario requires a very different distribution for $\epsilon$ in order to produce a thermal offset in the phase curve. The fact that, for HAT-P-7b, the distribution of the "standard + both" scenario is closer to the "standard + asy" distribution, suggests that for HAT-P-7b, the asymmetry in the phase curve is better explained by scattered light than by thermal emission. This is supported by the BIC values (see Table \ref{hat7mcmcresults}). Both the "standard +asy" and the "standard + both" scenarios are strongly favored by a probability of about 10$^5$ compared to the "standard + off" scenario ($\Delta$BIC$\approx$20). This result indicates that the preferred model explanation for the asymmetry is reflected light, rather than a thermal offset (but see above for a discussion of the physical problems of this solution). Our results quite clearly suggest asymmetric models rather than symmetric ones ($\Delta$ BIC>400), again confirming previous phase curve analysis \citep{esteves2015}.

The $\chi^2_{\rm{red}}$ values in Table \ref{hat7mcmcresults} are somewhat high (3.7 for the preferred model). Usually, such a high value might indicate that either the model is not capturing correctly the physical behavior of the system, or that the errors are under-estimated. However, the $\chi^2_{\rm{red}}$ is dominated by a few points (7 outliers contribute 50\% of the total $\chi^2_{\rm{red}}$), and one particular point contributes around 15\%. Since this is found for all fit scenarios (i.e., always the same outliers), this suggests that the error bars are indeed under-estimated. When removing the apparently systematic outliers, the $\chi^2_{\rm{red}}$ is reduced to less than 2. This in turn suggests a relatively good fit.

Figure \ref{hat7triangle_2} (right panel) shows the correlations for the "standard + off" scenario, as discussed above for TrES-2b. These correlations are much stronger and cleaner in this case, since the signal-to-noise ratio of the phase curve is much better for HAT-P-7b.

\begin{figure}[h]
\begin{center}
\includegraphics[width=250pt]{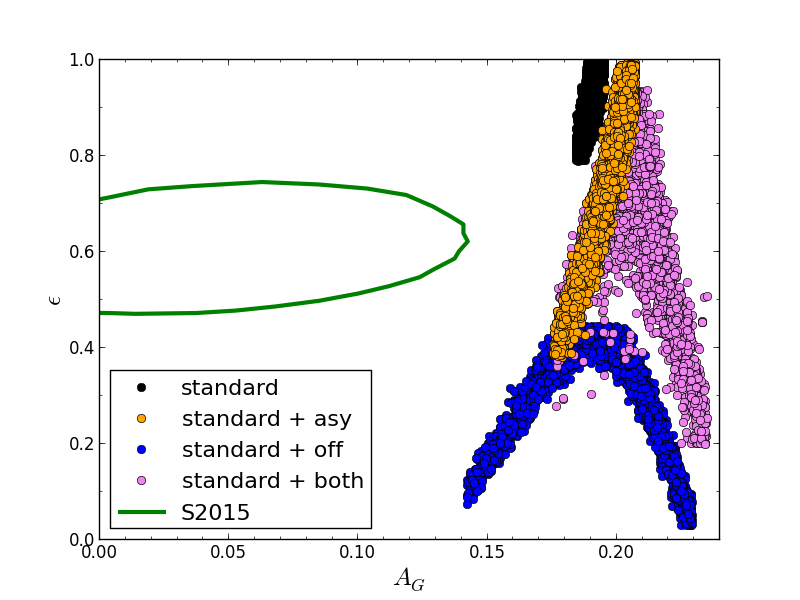}\\
\caption{Joint credibility regions of recirculation and geometric albedo for HAT-P-7b in different scenarios. Green contour: 1\,$\sigma$ uncertainty region in \citet{schwartz2015}. Both our and previous work strongly disagree on the inferred $\epsilon$ and $A_G$ values.}
\label{hat7_circulation}
\end{center}
\end{figure}

Figure \ref{hat7_circulation} shows the constraints on recirculation and geometric albedo as inferred from our scenarios, compared to \citet{schwartz2015}. As above, we use joint credibility regions to illustrate our inferred range in the $A_G$-$\epsilon$ plane. Note that the "standard+asy" model results and the low-albedo part of the "standard+both" model overlap (see also Fig. \ref{hat7triangle_3}). It is clearly seen that our results and the results of \citet{schwartz2015} strongly disagree, as was the case for CoRoT-1b. However, because of the good data quality of the HAT-P-7b phase curve, the disagreement is stronger than for CoRoT-1b. There is no overlap between our 90\,\% joint credibility regions and the 1\,$\sigma$ uncertainty region of \citet{schwartz2015}. This is because of the very well-constrained scattering albedo, which is nearly independent of the thermal and asymmetry model that was chosen (see also Fig. \ref{hat7_alb}).

We use the calculated $A_S$ values and measured IR dayside emission spectrum to constrain $A_B$ for HAT-P-7b (eqs. \ref{abondconstraint} and \ref{fluxbalance}). As for TrES-2b, we have $a_V \approx 0.4$. Therefore, based on the optical phase curve, we find 0.11<$A_B$<0.72. The observed Spitzer spectrum translates into $A_B$<0.87. These constraints are very loose because HAT-P-7b is a rather hot planet, compared to TrES-2b. Near-IR measurements covering the 1-3\,$\mu$m range (close to the Wien peak of the thermal radiation) would be highly desirable to further constrain the energy balance and add more constraints on the Bond albedo.

\section{Discussion}

\label{discuss}

For Solar System objects, phase curves have been incredibly useful to determine the scattering properties of atmospheric particles (size distribution, vertical location and extent, composition). Ground-based data as well as spacecraft observations (e.g., Venus Express, Voyager 1 and 2, Pioneer 10 and 11) have been used to investigate, e.g., Venus (e.g., \citealp{arking1968}, \citealp{garcia2014}, \citealp{petrova2015}), Titan (e.g., \citealp{rages1983}), Jupiter (e.g., \citealp{tomasko1978}, \citealp{smith1984}), Saturn (e.g., \citealp{tomasko1984}) or Uranus (\citealp{rages1991}, \citealp{pryor1997}). 

Large differences are found in the broadband phase curves of, e.g., Jupiter and Saturn (e.g., \citealp{dyudina2005}) or Mars, Mercury and Venus (e.g., \citealp{mallama2009}). These are of course attributable to differences in cloud structure and composition, the absence or presence of an atmosphere, topographic surface features or the amount of dust-covered or bare regolith, to name but a few factors influencing the phase curves.

Sophisticated radiative transfer models in combination with cloud and aerosol models are needed to interpret these observations correctly and retrieve scattering properties.

In comparison, exoplanet studies suffer from the incredibly crude data available at present (in terms of signal-to-noise ratio, spectral resolution or spectral coverage). Even though recent progress has been astonishing, we do not expect exoplanet data to approach Solar-System quality in the near future. Hence, interpretation of exoplanet observations does not require models of comparable complexity yet, although more complex models, which take into account, for example, cloud formation or temperature gradients, have recently been published (e.g., \citealp{webber2015}, \citealp{hu2015}) and applied to, e.g., the well-characterized phase curve of Kepler-7b. However, most studies, including this work, rely on simpler models and make strong assumptions to infer planetary and atmospheric properties.

For example, the model used in this work relies on the following two assumptions, in line with previous studies (e.g., \citealp{snellen2009}, \citealp{cowan2011hot}, \citealp{schwartz2015}): 

\begin{itemize}
  \item Day and night hemispheres are assumed to be respectively described by a single, uniform temperature, without any longitudinal or latitudinal gradients. In reality, this is unlikely to be true. Secondary-eclipse mapping of the hot Jupiter HD189733b has already demonstrated that the brightness distribution is far from uniform (e.g., \citealp{dewit2012}). This can be interpreted as a non-uniform temperature distribution. IR phase curves also clearly show temperature gradients (e.g., \citealp{knutson2007daynight_189733,knutson2009daynight_189733}, \citealp{crossfield2010}). In the hypothetical no-recirculation limit ($\epsilon=0$), non-uniformity effects might produce a thermal beaming dominated by the sub-stellar point that could potentially enhance planetary emitted radiation (e.g., \citealp{selsis2011}, \citealp{schwartz2015}). However, given the relatively low signal-to-noise ratios, the number of effects contributing to the optical phase curve and the large bandpass of Kepler and CoRoT, the optical phase curve is not expected to be very sensitive to the temperature distribution.
  \item The observed brightness temperature in a given spectral bandpass equals the bolometric equilibrium temperature hence constrains the energy budget of the atmosphere and can be related to the Bond albedo. This is a fundamental assumption that is unlikely to hold once better spectral resolution becomes available. As suggested by, e.g., \citet{barclay2012}, the photospheres for optical and IR observations (as well as for day- and nightside emission) are probably located at different pressures. Hence, the observations would probe different temperatures and dynamical regimes. Depending on pressure, circulation and temperature regimes can be quite different (e.g., \citealp{parmentier2013}, \citealp{agundez2014}, \citealp{showman2015}). Hence, observed brightness temperatures in either spectral domain would not be necessarily related to the bolometric equilibrium temperature.

  \end{itemize}

It is possible that these assumptions are not violated (or at least, not strongly) for many exoplanets and that they more or less hold. Our results, however, imply that optical and IR data lead to different conclusions for the same objects (in two out of three cases) when applying these assumptions. Therefore, it seems that they are too strong and overly simplified. It is a subject of future research to reconcile this finding with the current data quality, which does not necessarily warrant complex models or a level of sophistication much higher than the models presented here or in previous work.

We point out, however, that in the case for CoRoT-1b, both our model results and the results by \citet{schwartz2015} are marginally compatible, since their 1\,$\sigma$ uncertainty regions and our 90\,\% credibility regions slightly overlap. Therefore, a re-analysis of the CoRoT-1b phase curve with the newly released, improved data pipeline might reduce the photometric uncertainties and provide a more decisive answer to resolve the apparent contradiction between phase curve and secondary eclipse analyses.

\section{Conclusions}

\label{summary}

We have presented a simple, yet physically consistent, model of optical phase curves for exoplanets. It includes Lambertian scattering, thermal emission (under the assumption of uniform hemispheric temperatures), ellipsoidal variations and Doppler boosting. It can account for asymmetric phase curves by longitudinally asymmetric scattering albedos and an offset of thermal radiation compared to the sub-stellar point.

This model has been used to re-analyze published phase-curve data of CoRoT-1b, TrES-2b and HAT-P-7b. Results are then compared to an analysis of secondary-eclipse data of these planets by \citet{schwartz2015}.

We have shown that for CoRoT-1b and HAT-P-7b, inferred albedo and heat recirculation values from optical phase curves are different compared to previously published results. For TrES-2b, both methods yield similar results.

We find that CoRoT-1b has a rather higher scattering albedo than previously found. We find 0.11<$A_S$<0.3 at 95\,\% confidence, which is in slight contrast with previous analyses, which found $A_S$<0.15 (\citealp{snellen2009}, \citealp{schwartz2015}). Also, full phase curve analysis favors a strong redistribution of stellar incident energy to the nightside, contrary to previous studies, which suggested a very inefficient recirculation  (\citealp{snellen2009}, \citealp{schwartz2015}). These contradictions are mainly because previous optical phase curve analysis of CoRoT-1b by \citet{snellen2009} considered only thermal emission.

In line with previous studies on the optical phase curve of HAT-P-7b (e.g., \citealp{esteves2013,esteves2015}), we find an appreciable albedo ($A_S$$\approx$0.3),  slightly higher than inferred from secondary eclipse data. In contrast to previous studies based on secondary eclipse data \citep{schwartz2015}, the analysis of the optical phase curve favors moderate to efficient heat recirculation. Asymmetric models are found to best fit the observed phase curve.
 
These differences between secondary eclipse and optical phase curve analyses occur most likely because optical and IR observations probe different atmospheric layers. Furthermore, our results suggest that some of the assumptions made (specifically, that observed brightness temperatures constrain the energy budget) are probably too strong and should be relaxed.

Future work will aim, among others, at re-analyzing further planets with published optical phase curves and reconciling different observations in the optical and the IR for CoRoT-1b.

\begin{acknowledgements}

This study has received financial support from the French State in the frame of the "Investments for the future" Programme IdEx Bordeaux, reference ANR-10-IDEX-03-02. P. G. acknowledges support from the ERC Starting Grant (3DICE, grant agreement 336474). We thank the anonymous referee for a very constructive and positive feedback. Computer time for this study was provided in parts by the computing facilities MCIA (M\'esocentre de Calcul Intensif Aquitain) of the Universit\'e de Bordeaux and of the Universit\'e de Pau et des Pays de l'Adour.

\end{acknowledgements}

\bibliographystyle{aa}
\bibliography{literatur_idex}

\appendix

\section{Model verification}

\label{verification_model}

We verify that the implementation of model equations leads to the correct limits for reflected and emitted light.

Equations \ref{lambert_formula} and \ref{z_formula} show the phase function for a standard Lambertian sphere. These equations have been used in most studies of optical phase curves so far.

\begin{equation}
\label{lambert_formula}
 \Phi (z)=\frac{1}{\pi}\left(\sin (z)+(\pi-z)\cos(z)\right).
\end{equation}

\begin{equation}
\label{z_formula}
 \cos (z) = -\sin(i) \cos (\alpha_0).
\end{equation}

Note that $\alpha_0$ is 0 at opposition and $\pi$ at primary transit, contrary to $\alpha$ (see eq. \ref{obslat}).

For thermal radiation, the phase function is described by the illuminated fraction $L$ of the planetary disk, i.e., the dayside:

\begin{equation}
\label{illu_formula}
\Phi (z) = \frac{1}{2}(1- \cos (z)).
\end{equation}

\begin{figure}[h]
\begin{center}
\includegraphics[width=250pt]{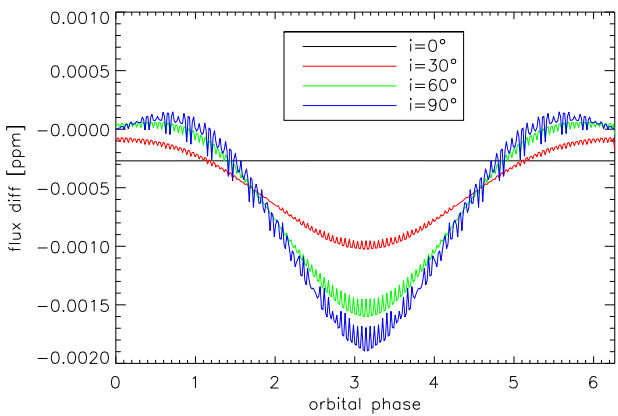}\\
\caption{Model test: Reflected flux compared to exact Lambertian sphere.}
\label{fluxverif}
\end{center}
\end{figure}

In Fig. \ref{fluxverif}, we show the difference between the exact formulation (eqs. \ref{lambert_formula} and \ref{z_formula}) and our model for the reflected component. The considered case is a hot Jupiter in a 10-day orbit around a Sun-like star, at varying orbital inclinations. The amplitude of the signal is of the order of a few ppm (10$^{-6}$). The difference is about three orders of magnitude less, which clearly indicates that the model correctly incorporates Lambertian scattering. The high-frequency structure in the residuals is due to the spatial discretization of the numerical model and has no effect on physical results.

\begin{figure}[h]
\begin{center}
\includegraphics[width=250pt]{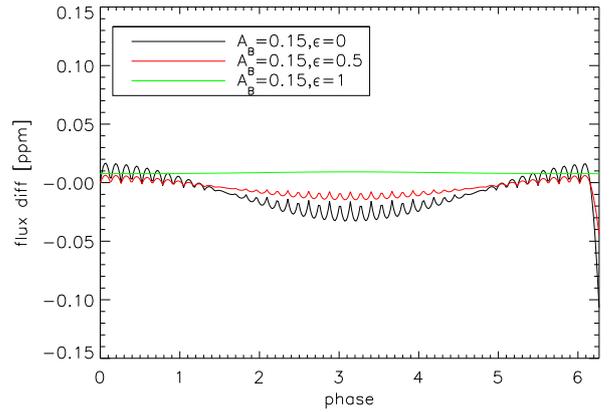}\\
\caption{Model test: Emitted flux compared to exact solution.}
\label{planckverif}
\end{center}
\end{figure}

Figure \ref{planckverif} shows the comparison of our model to the exact solution for thermal radiation. In this case, we show a hot Jupiter in a 2-day orbit around a Sun-like star ($A_B$=0.15, $\epsilon$=0), in order to get an appreciable signal. The difference at peak amplitude is about 0.01\,ppm for a total amplitude of $\approx$1.6\,ppm. This amounts to an error of less than 1\,\%, which we deem acceptable. Again, the high-frequency structure in the residuals is due to the spatial discretization of the numerical model and the time resolution.

\section{Phase curves of transiting planets}

\label{phase_transit}

If the data is of high enough quality, in terms of signal-to-noise ratio or time resolution, more and more parameters can be added to the fit. However, when analyzing phase curves of transiting planets, some parameters can be related to one another self-consistently, by the shape of the primary transit. For instance, the transit depth of the primary transit directly yields the radius ratio $k_r$ between planet and star:

\begin{equation}
\label{radius_ratio}
k_r=\frac{R_p}{R_{\ast}}.
\end{equation}

Furthermore, for circular orbits, transit duration and transit shape can be related to the orbital inclination $i$ and the projected star-planet separation $k_p$ in units of stellar radii:

\begin{equation}
\label{semi_ratio}
k_p=\frac{a}{R_{\ast}}.
\end{equation}

Assuming $M_P<<M_{\ast}$, we can write Kepler's 3rd law as follows:

\begin{equation}
\label{kepler3}
\frac{P_{\rm{orb}}^2}{a^3}=\frac{4\pi^2}{GM_{\ast}}.
\end{equation}

Since the orbital period $P_{\rm{orb}}$ of transiting planets is usually known to within a few minutes or better, and $k_p$ and $k_r$ are mostly determined to an accuracy of better than 1\,\%, it is possible to calculate the stellar mass and the planetary radius, given a stellar radius. Equation \ref{kepler3} yields, in this case, an analytic relation between stellar radius and stellar mass.

\begin{figure}[h]
\begin{center}
\includegraphics[width=250pt]{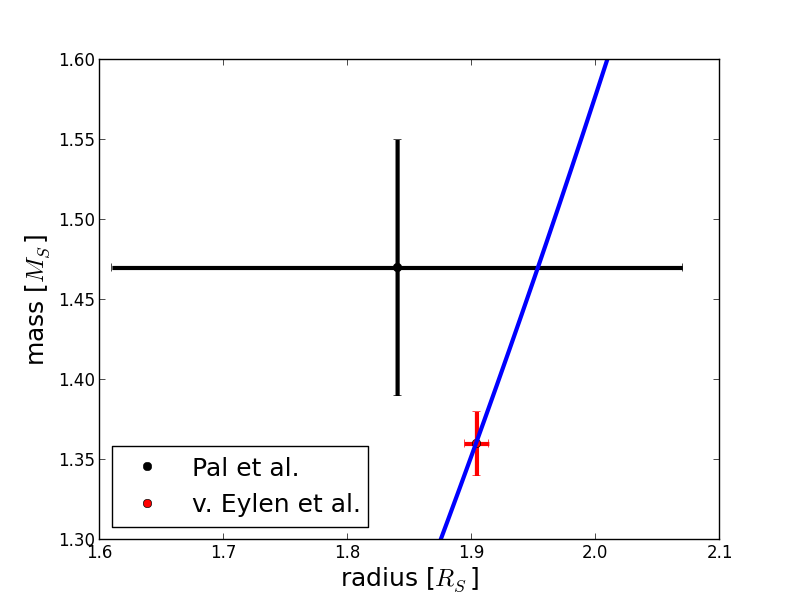}\\
\caption{Consistency between reported radius and mass determinations for the star HAT-P-7, by using the determined orbital period $P_{\rm{orb}}$=2.204\,days, $a/R_{\ast}$ values from Table \ref{hattransittable}, blue line is eq. \ref{kepler3}}
\label{hat7b_cons}
\end{center}
\end{figure}

Such a relation is shown in Fig. \ref{hat7b_cons} (blue line) for HAT-P-7, using a period of $P_{\rm{orb}}$=2.204\,days \citep{pal2008} and $k_p$=4.1512 \citep{esteves2013}. Also shown are stellar parameters taken from \citet{pal2008} who used high-resolution spectroscopy and \citet{eylen2012} who used asteroseismology (see Tables \ref{hatstellartable} and \ref{hattransittable} for a compilation).

\begin{table}[h]
  \caption{Stellar parameters for HAT-P-7. In case of asymmetric uncertainties in the original publication, the larger one is stated.}\label{hatstellartable}
\begin{tabular}{llcc}
\hline
\hline
study&  $M_{\ast}$[$M_{S}$]  &  $R_{\ast}$[$R_{S}$] \\
\hline
\citet{pal2008}                            & 1 .47$\pm$0.08    & 1.84$\pm$0.23 \\
\citet{eylen2012}                        & 1.361$\pm$0.021& 1.904$\pm$0.01\\
\end{tabular}
\end{table}

\begin{table}[h]
  \caption{Planetary parameters for HAT-P-7b. In case of asymmetric uncertainties in the original publication, the larger one is stated.}\label{hattransittable}
\begin{tabular}{l|cc}
\hline
\hline
study&  $k_r$ &  $k_p$ \\
\hline
\citet{pal2008}            &    0.0763$\pm$0.001           & 4.35$\pm$0.38        \\
\citet{esteves2013}     &  0.07749$\pm$0.000013  & 4.1512$\pm$0.0026   \\
\citet{eylen2013}         & 0.077462$\pm$0.000034 & 4.1547$\pm$0.0042 
\end{tabular}
\end{table}

It is clearly seen that fixing stellar parameters at $R_{\ast}$=1.84\,$R_{S}$ and $M_{\ast}$=1.47\,$M_{S}$, as done by \citet{esteves2013}, results in inconsistent system parameters. These then introduce a significant error in estimating the planetary mass from the ellipsoidal variations.

To illustrate the effects on planetary mass estimates, we performed inverse modeling of the HAT-P-7b phase curve, adopting the "standard" scenario from Table \ref{corot1mcmcsummary}, i.e., fitting for mass, albedo ($A_B$=$A_S$) and heat recirculation. Consistent models use the stellar parameters from \citet{eylen2012}, i.e., $R_{\ast}$=1.90\,$R_{S}$ and $M_{\ast}$=1.36\,$M_{S}$, whereas the inconsistent models use $R_{\ast}$=1.84\,$R_{S}$ and $M_{\ast}$=1.47\,$M_{S}$, as done in \citet{esteves2013}.

\begin{figure}[h]
\begin{center}
\includegraphics[width=250pt]{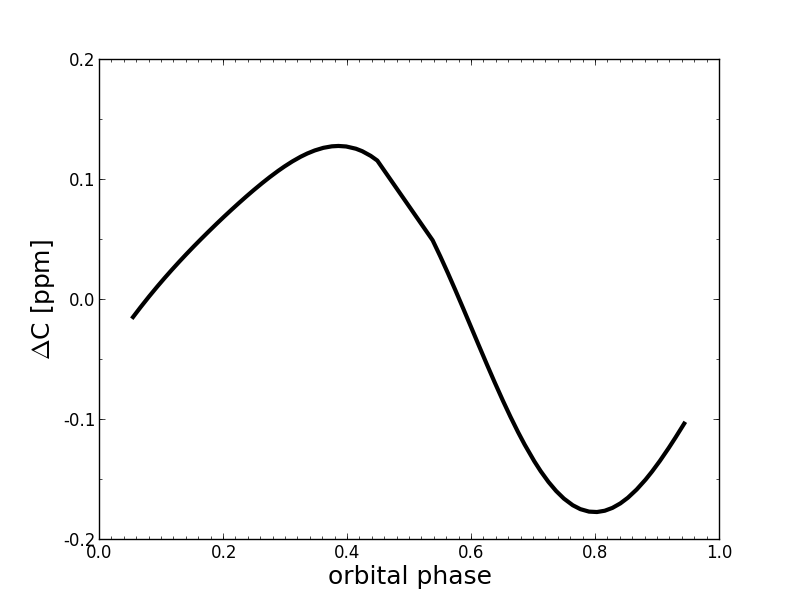}\\
\caption{Residuals between consistent (stellar mass and radius from \citealp{eylen2012}) and inconsistent (stellar mass and radius from \citealp{pal2008}) best-fit models of the HAT-P-7b optical phase curve. }
\label{hat7b_inconsistent}
\end{center}
\end{figure}

Figure \ref{hat7b_inconsistent} shows the residuals $\Delta C=C_{\rm{con}}-C_{\rm{incon}}$ of the best-fit models. As is clearly seen, both scenarios result in virtually identical fits, which only differ by about 0.2\,ppm (compared to the roughly 75\,ppm amplitude of the phase curve). Furthermore, both best-fit models result in similar $\chi^2_{\rm{red}}$ values of 10.8 and 10.97, respectively.

\begin{figure}[h]
\begin{center}
\includegraphics[width=250pt]{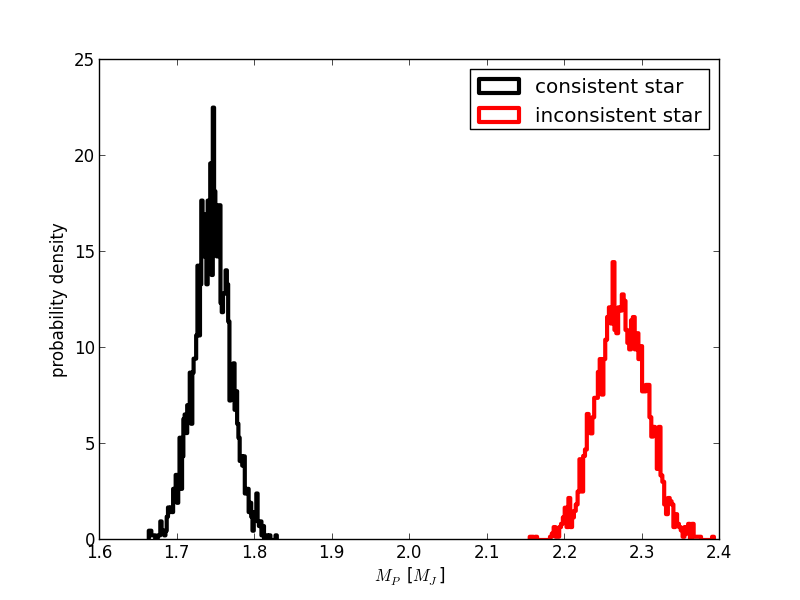}
\caption{Marginalized posterior distributions for HAT-P-7b $M_P$, for different adopted stellar parameters in the standard model. }
\label{tricons}
\end{center}
\end{figure}

 All parameters except planetary mass are not affected by the choice of stellar parameters. Figure \ref{tricons} shows the marginalized posterior distributions for the planetary mass, for both sets of stellar parameters. It is clear that the estimated planetary mass varies by as much as 30\,\%. In case of inconsistent stellar parameters, the planetary mass is severely over-estimated, compared to RV results.
Therefore, we chose the stellar parameters stated in \citet{eylen2012} since these are consistent with parameters deduced from primary transit analysis.

\section{Optional phase function choices}

\label{rayappend}

\subsection{Empirical Solar System phase functions}

A few previous studies (e.g., \citealp{collier2002}, \citealp{kane2010longperiod}) used an empirically derived phase function instead of the Lambert phase function in eq. \ref{lambert_formula}. This phase function was obtained from a fit to optical observations of Venus and Jupiter.

\begin{equation}\label{delta_m_emp}
   \Delta m (\alpha_0)=0.09\frac{\alpha_0}{100^{\circ}}+2.39\left (\frac{\alpha_0}{100^{\circ}}\right)^2-0.65\left(\frac{\alpha_0}{100^{\circ}}\right)^3.
\end{equation}

\begin{equation}\label{emp_phase}
 \Phi (\alpha_0)=10^{-0.4\cdot \Delta m (\alpha_0)}.
\end{equation}

\begin{figure}[h]
\begin{center}
\includegraphics[width=250pt]{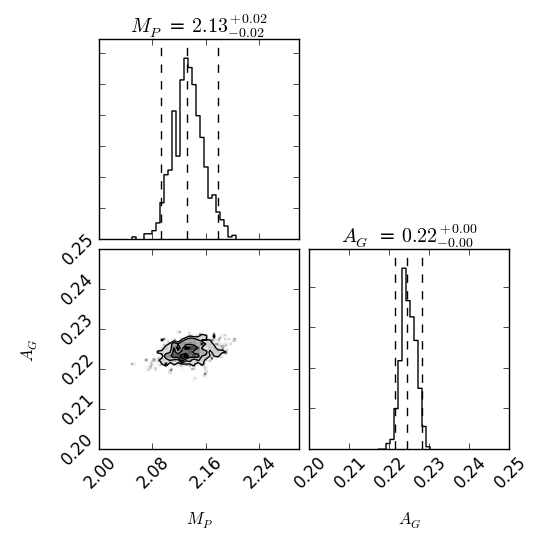}\\
\caption{Constraints on geometric albedo and planetary mass, as derived from the standard model using the empirical phase function of eq. \ref{emp_phase}.}
\label{hat7b_emp}
\end{center}
\end{figure}

When fitting the phase curve of HAT-P-7b with the empirical phase function, we obtain a geometric albedo of about $A_G\approx$0.2, consistent with previous estimates using the Lambertian approximation. However, as shown in Fig. \ref{hat7b_emp}, the estimated planetary mass is far larger. Even when changing from a uniform prior to a Gaussian prior based on RV measurements (1.78$\pm$0.08, \citealp{esteves2015}), the fitted mass is greatly over-estimated. Hence, our results suggest that eq. \ref{emp_phase} as a particular choice of phase function is probably not correct for HAT-P-7b. Possible reasons to explain this include, e.g., the higher temperature (much higher than both Venus and Jupiter, for which this particular phase function was derived) and a consequently much different atmospheric chemistry. Also, cloud properties could play a role, since HAT-P-7b most likely has some form of silicate or iron clouds (see, e.g., comparison of Kepler-7b and Jupiter in \citealp{webber2015}). Even for Jupiter and Saturn, cloud properties are thought to be responsible for the difference in observed phase functions (e.g., \citealp{dyudina2005}).

Such an impact of the choice of the phase function on mass estimates has also been discussed by \citet{mislis2012}.

\subsection{Rayleigh scattering}

To investigate the influence of Rayleigh scattering, we incorporated H$_2$ and He Rayleigh scattering in the model. These two species are thought to form the major constituents of gas-giant atmospheres.

The Rayleigh scattering cross sections of H$_2$ and He are calculated as 

\begin{equation}\label{rayleigh}
    \sigma_{\rm{ray,i}}(\lambda)= \left (\frac{\lambda_{\rm{0,i}}}{\lambda}\right)^4 \cdot,
    \sigma_{\rm{0,i}}
\end{equation}

where $\sigma_{\rm{ray,i}}$ of species $i$ is given in cm$^2$ per molecule, $\lambda$ in $\mu$m and $\lambda_{\rm{0,i}}$ is a reference wavelength where $\sigma_{\rm{0,i}}$ has been measured.

This approach is used in many approximative treatments of Rayleigh scattering (see, e.g., \citealp{lecav2008}). For the values of
$\lambda_0$ and $\sigma_{\rm{0,i}}$ in eq. \ref{rayleigh}, measurements from \citet{shardanand1977} were used in this work, as
tabulated in Table \ref{rayleigh_shard}.

\begin{table}[h]
  \centering
  \caption{Rayleigh scattering parameters for use in  eq. \ref{rayleigh}}\label{rayleigh_shard}
\begin{tabular}{ccc}
  \hline\hline
   Molecule & $\lambda_{\rm{0,i}}$ [$\mu$m]  & $\sigma_{\rm{0,i}}$ [cm$^2$] \\
  \hline\hline
   H$_2$    &   0.5145                       &  1.17 $\times$ 10$^{-27}$\\
   He       &   0.5145                       &  8.6 $\times$ 10$^{-29}$\\

\end{tabular}
\end{table}

The optical depth in an atmospheric layer $j$ due to Rayleigh scattering, $\tau_{\rm{ray,j}}$,  is obtained with the following equation:

\begin{equation}
\label{raylayer}
\tau_{\rm{ray,j}}=\sum_k\sigma_k C_{k,j},
\end{equation}

where $\sigma_k$, $C_{k,j}$ are the Rayleigh cross section and the column density of species $k$ respectively. We calculate the column density as

\begin{equation}
\label{columndens}
C_{k,j}=c_{k,j} \cdot \frac{P_k-P_{k+1}}{\mu_{\rm{atm}} g_P},
\end{equation}

where $c_{k,j}$ is the volume mixing ratio of species $k$ in layer $j$, $g_P$ planetary gravity, $\mu_{\rm{atm}}$ the mean molecular weight of the atmosphere ($\approx$2 for H$_2$-dominated atmospheres) and $P_k$ is the layer pressure. The atmospheric layers are approximately spaced evenly in $\log P$ from the "surface" pressure $P_S$ to 10$^{-4}$\,bar.

From there, the total optical depth $\tau_{\rm{ray}}$ for use in eq. \ref{transmission} is obtained by summing the optical depths of each layer from the surface to the model lid:

\begin{equation}
\label{raytautotal}
\tau_{\rm{ray}}=\sum_j \tau_{\rm{ray,j}}.
\end{equation}

\begin{figure}[h]
\includegraphics[width=250pt]{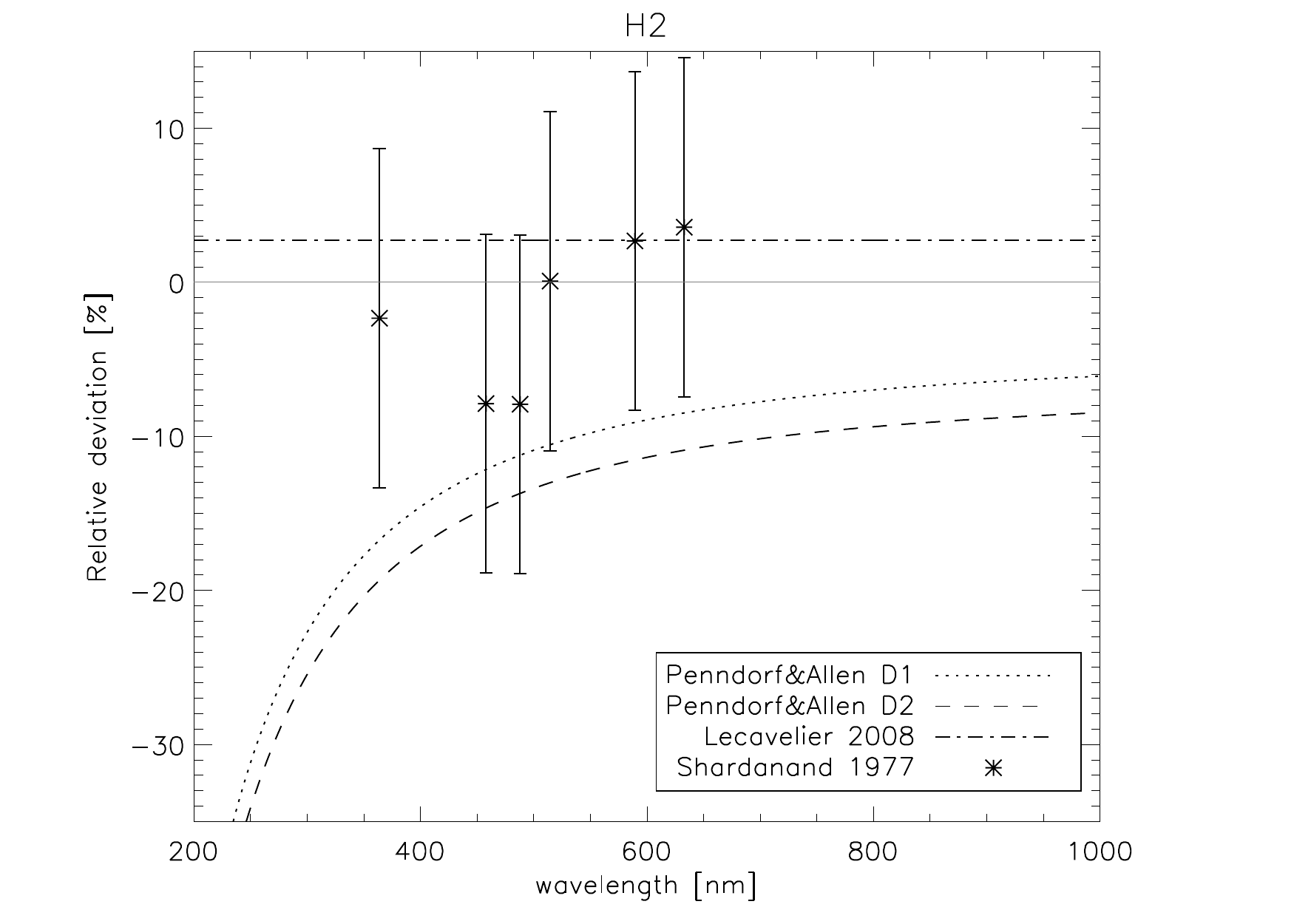}\\
  \caption{Comparison of H$_2$ Rayleigh scattering cross sections. Relative deviations $\frac{\sigma_{\rm{model}}-\sigma_{\rm{data}}}{\sigma_{\rm{model}}}$   in \% between model and data sources (as indicated). Vertical lines show measurement uncertainties. Grey horizontal line indicates 0\% deviation.}
  \label{h2_relative}
\end{figure}

In Fig. \ref{h2_relative}, the used Rayleigh scattering cross sections of H$_2$ are compared to measurements reported in the literature (\citealp{shardanand1977}) as well as different approximations used in various models (Table II of \citealp{penndorf1957}, \citealp{lecav2008}).

For H$_2$, the agreement with measurements is very good, again to within the stated error bars of \citet{shardanand1977}. Also, the agreement with the approximation of \citet{lecav2008} is very good.  The comparison with the parametrization of H$_2$ Rayleigh scattering using \citet{penndorf1957} data is less good.

At the "bottom" of the atmosphere, at a prescribed "surface" pressure $P_S$, we impose a Lambertian surface with scattering albedo $A_S$. Hence, eq. \ref{cellflux} is modified,

\begin{eqnarray}
\label{raycell}
F_R  &=&   T_{\rm{ray}}\cdot F_l+ \\
  \nonumber& & (1-T_{\rm{ray}}) \cdot \Phi_{\rm{ray}} \cdot \frac{\omega}{\cos z_s+\cos z_o} \\
\nonumber  && \cdot \cos z_s \cdot \frac{S}{r(t)^2}\cdot  \cos z_o \cdot \Delta \Omega \left(\frac{R_p}{d}\right)^2 ,
\end{eqnarray}

where $\omega$ is the single-scattering albedo (set to unity in the all-scattering, zero-absorption approximation used here), $T_{\rm{ray}}$ is the transmission along the optical path calculated as

\begin{equation}
\label{transmission}
T_{\rm{ray}}=e^{-\tau_{\rm{ray}}(\frac{1}{\cos z_s}+\frac{1}{\cos z_o})},
\end{equation}

with $\tau_{\rm{ray}}$ the (zenith) optical depth due to Rayleigh scattering (calculated at the mid-point of the spectral interval considered). The value of $\tau_{\rm{ray}}$ will depend critically on the choice of $P_S$ (see above, eqs. \ref{raylayer} and \ref{raytautotal}). $\Phi_{\rm{ray}}$ is the phase function of Rayleigh scattering

\begin{equation}
\label{rayphase}
\Phi_{\rm{ray}}=\frac{3}{16\pi}\left(1+(\cos\phi_o)^2\right),
\end{equation}

with $\phi_o$ the angle between observer and incoming stellar light, i.e., $\cos\phi_o=$\textbf{s}$\cdot$\textbf{o}. 

\begin{figure}[h]
\begin{center}
\includegraphics[width=250pt]{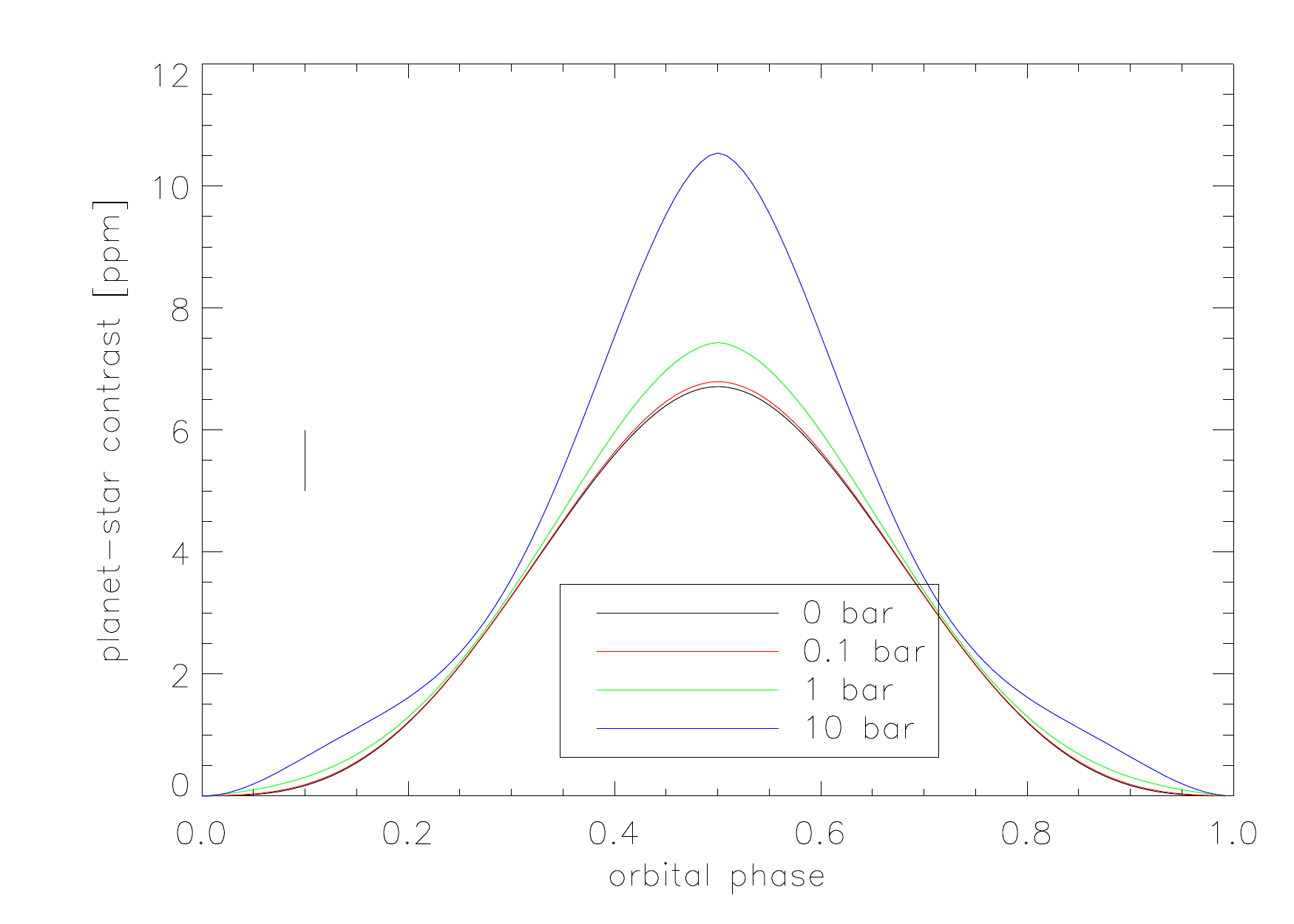}\\
\caption{Effect of Rayleigh scattering on the phase curve, for different values of $P_S$. 1\,ppm error bar to the left. See text for discussion.}
\label{rayeffect}
\end{center}
\end{figure}

Figure \ref{rayeffect} shows the effect of Rayleigh scattering on the phase curve. Together with the standard Lambertian scattering approximation of eq. \ref{cellflux}, we show phase curves with varying values of $P_S$. As expected, with increasing $P_S$, hence increasing contribution of Rayleigh scattering to the reflected light, the phase curve changes. For $P_S$=1\,bar, the atmosphere starts to become visible, and at $P_S$=10\,bar, dominates over the "surface" contribution.

\begin{figure}[h]
\begin{center}
\includegraphics[width=250pt]{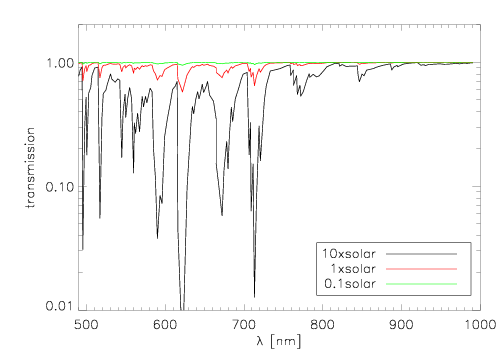}\\
\caption{Transmission due to TiO and VO, as a function of wavelength. See text for details.}
\label{tio}
\end{center}
\end{figure}

Atmospheric modeling of hot Jupiters predicts the formation of clouds around the 10$^{-2}$\,bar layer or even at lower pressures (e.g., \citealp{parmentier2013}, \citealp{webber2015}). This implies that the reflecting "surface" is at $P_S$<10$^{-2}$\,bar. Furthermore, optical absorption by, e.g., alkali metals or TiO/VO, greatly increases with pressure. Based on cross sections and solar abundances presented by \citet{desert2008}, Fig. \ref{tio} shows the transmission due to TiO and VO absorption assuming $P_S$=10$^{-4}$\,bar. Figure \ref{tio} suggests that not much radiation is expected to penetrate to levels where Rayleigh scattering becomes important.

Hence, we would expect that Rayleigh scattering does not play a large role, and we neglect it in our phase-curve studies (equivalent to setting $P_S$=0). Therefore, phase curves are calculated with the Lambert approximation.

\section{MCMC results}
\label{mcmcresults}
\clearpage

\begin{figure*}[t]
\centering
\includegraphics[width=250pt]{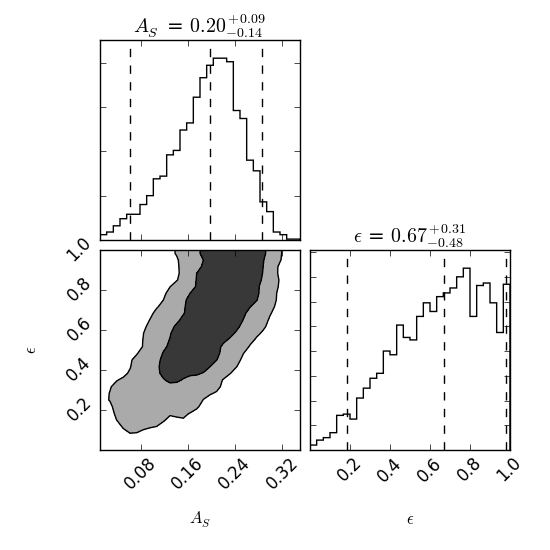}
\includegraphics[width=250pt]{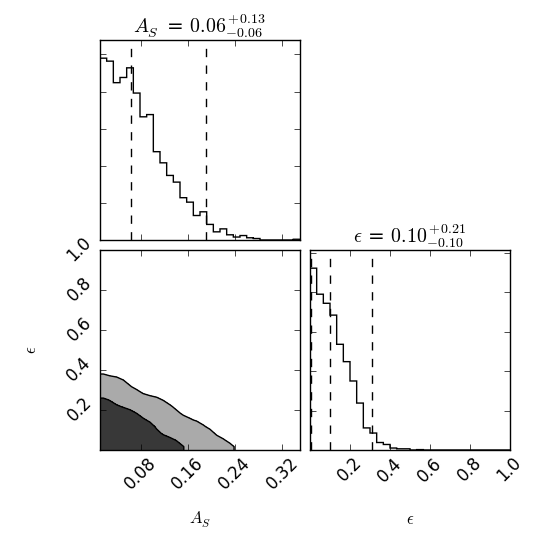}\\
\caption{CoRoT-1b phase curve constraints: Posterior projections for the standard model (left) and no-scattering model (right). Dashed vertical lines represent marginalized 95\,\% credibility regions. Smoothed 68\,\% and 95\,\% credibility regions in dark and light grey shade, respectively. Note that the results of both models are almost contrary to each other. }
\label{corot1triangle_2}
\end{figure*}

\begin{figure*}
\begin{center}
\includegraphics[width=250pt]{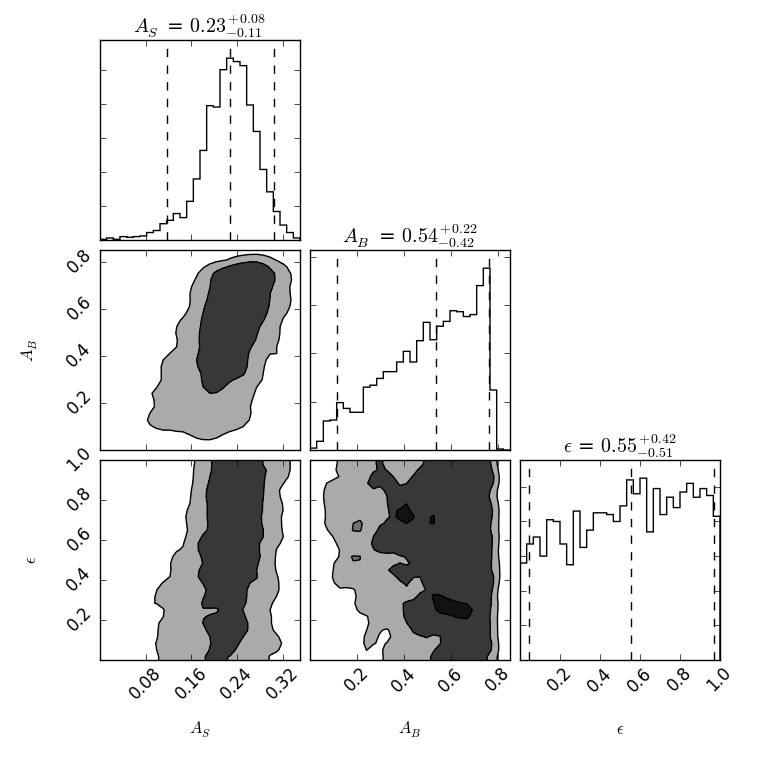}
\includegraphics[width=250pt]{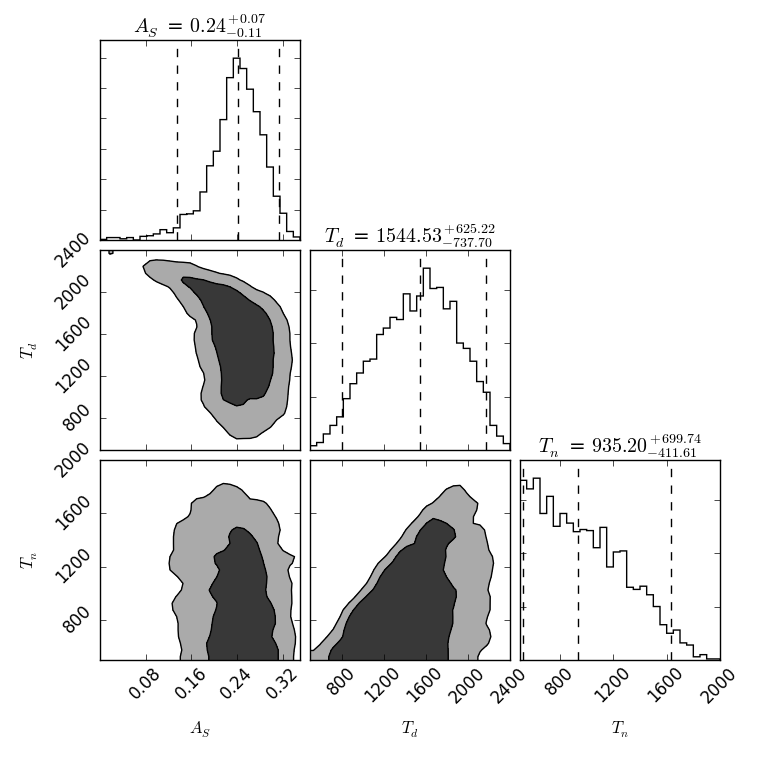}\\
\caption{CoRoT-1b phase curve constraints: Posterior projections for the free-albedo model (left) and free-temperature model (right). Dashed vertical lines represent marginalized 95\,\% credibility regions. Smoothed 68\,\% and 95\,\% credibility regions in dark and light grey shade, respectively. }
\label{corot1triangle_3}
\end{center}
\end{figure*}

\begin{table*}
  \centering 
  \caption{CoRoT-1b 95\,\% credibility regions, parameters of the maximum a-posteriori models and associated goodness-of-fit criteria for scenarios from Table \ref{corot1mcmcsummary}. Models are ranked by BIC. Model probability ratios $p_M$ are stated (see eq. \ref{bicdiff}). For comparison, we also state goodness-of-fit criteria for the best-fit model of \citet{snellen2009}, their Figure 1.}\label{corot1mcmcresults}
  \begin{tabular}{l|llcc|lll}
\hline
\hline
   scenario&95\,\% credibility regions &best-fit $V_P$&$\chi^2_{\rm{min}}$ & $\chi^2_{\rm{red,min}}$ & BIC  & $\Delta$ BIC & $p_M$ \\
\hline
standard  &0.06\hspace{0.3cm}< $A_S$\hspace{0.15cm}< 0.28&$A_S$\hspace{0.15cm}= 0.22& 19.36&1.38 &24.91 & 0& 1\\
  &0.18\hspace{0.3cm}< $\epsilon$\hspace{0.4cm}<  0.98&$\epsilon$\hspace{0.4cm}= 0.86& & &\\
\hline
free T  &0.13\hspace{0.3cm}< $A_S$\hspace{0.15cm}< 0.31&$A_S$\hspace{0.15cm}= 0.25& 17.90 & 1.37  & 26.22 & 1.31 & 0.51\\
  &806\hspace{0.35cm}< $T_d$\hspace{0.2cm}<2169\,K&$T_d$\hspace{0.15cm}= 874\,&  &  &\\
  &523\hspace{0.35cm}< $T_n$\hspace{0.2cm}<1634\,K&$T_n$\hspace{0.15cm}= 727\,K&  &  &\\
\hline
free albedo  &0.11\hspace{0.3cm}< $A_S$\hspace{0.15cm}< 0.30&$A_S$\hspace{0.15cm}= 0.25&17.93& 1.37 & 26.25 &1.34 &0.51\\
     &0.11\hspace{0.3cm}< $A_S$\hspace{0.15cm}< 0.76&$A_B$\hspace{0.15cm}= 0.77&& & \\
    &0.04\hspace{0.3cm}< $\epsilon$\hspace{0.4cm}<  0.96&$\epsilon$\hspace{0.4cm}= 0.77&&  & \\
\hline
no scattering  &0\hspace{0.7cm}< $A_B$\hspace{0.15cm}< 0.18&$A_B$\hspace{0.15cm}= 0.09& 24.40 & 1.74& 29.95& 5.04 & 0.08\\
  &0\hspace{0.7cm}< $\epsilon$\hspace{0.4cm}<  0.31&$\epsilon$\hspace{0.4cm}= 0.01& & &\\
\hline
\hline
\citet{snellen2009} &- &-&22.24& 1.58 &27.79&2.88&0.23\\
\end{tabular}
\end{table*}

\begin{figure*}
\begin{center}
\includegraphics[width=250pt]{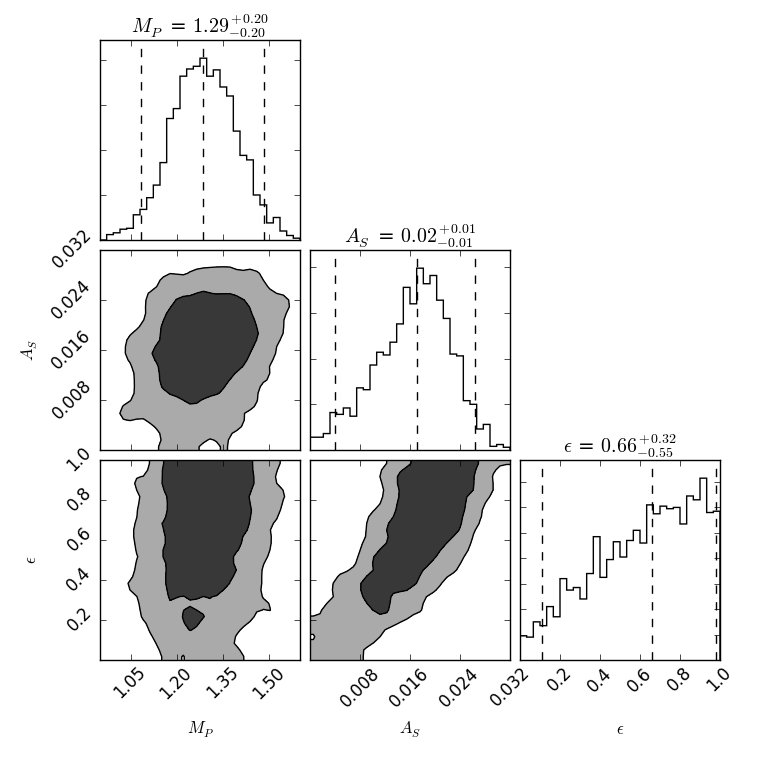}
\includegraphics[width=250pt]{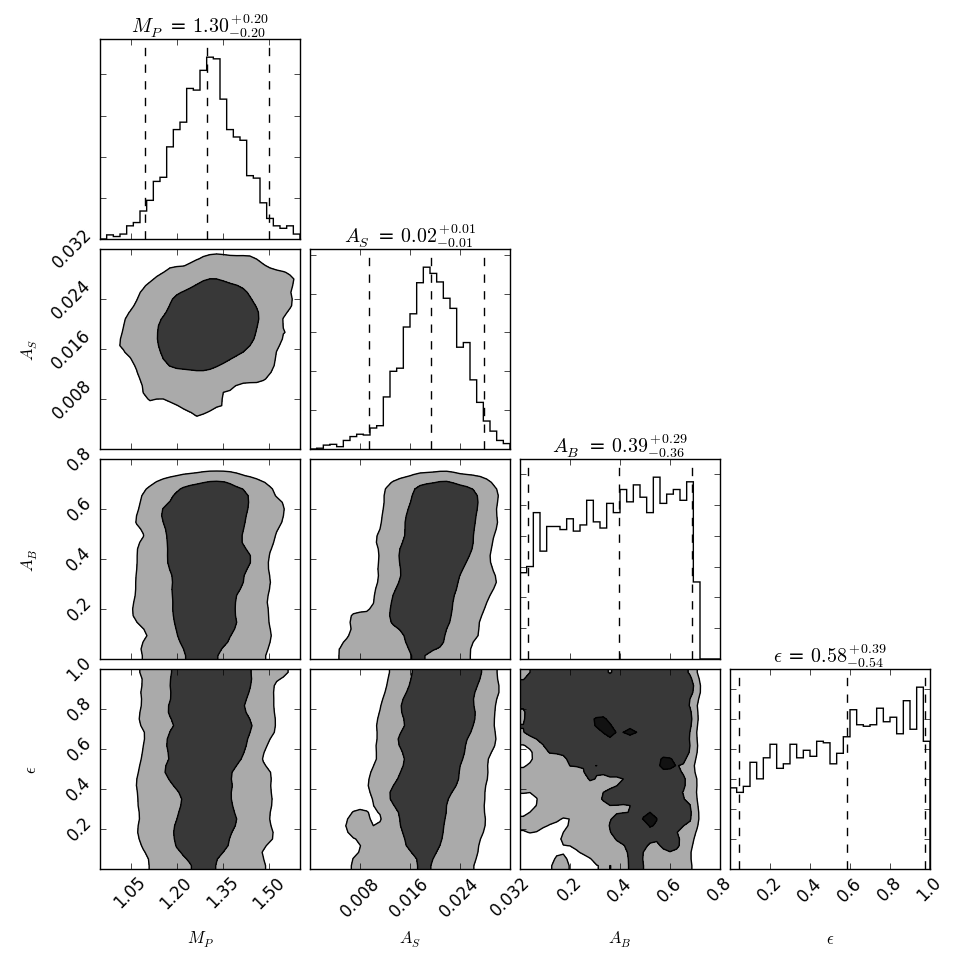}\\
\caption{TrES-2b phase curve constraints: Posterior projections for the "standard" model (left) and "free A" model (right). Dashed vertical lines represent marginalized 95\,\% credibility regions. Smoothed 68\,\% and 95\,\% credibility regions in dark and light grey shade, respectively. }
\label{tres2triangle_1}
\end{center}
\end{figure*}

\begin{figure*}
\begin{center}
\includegraphics[width=250pt]{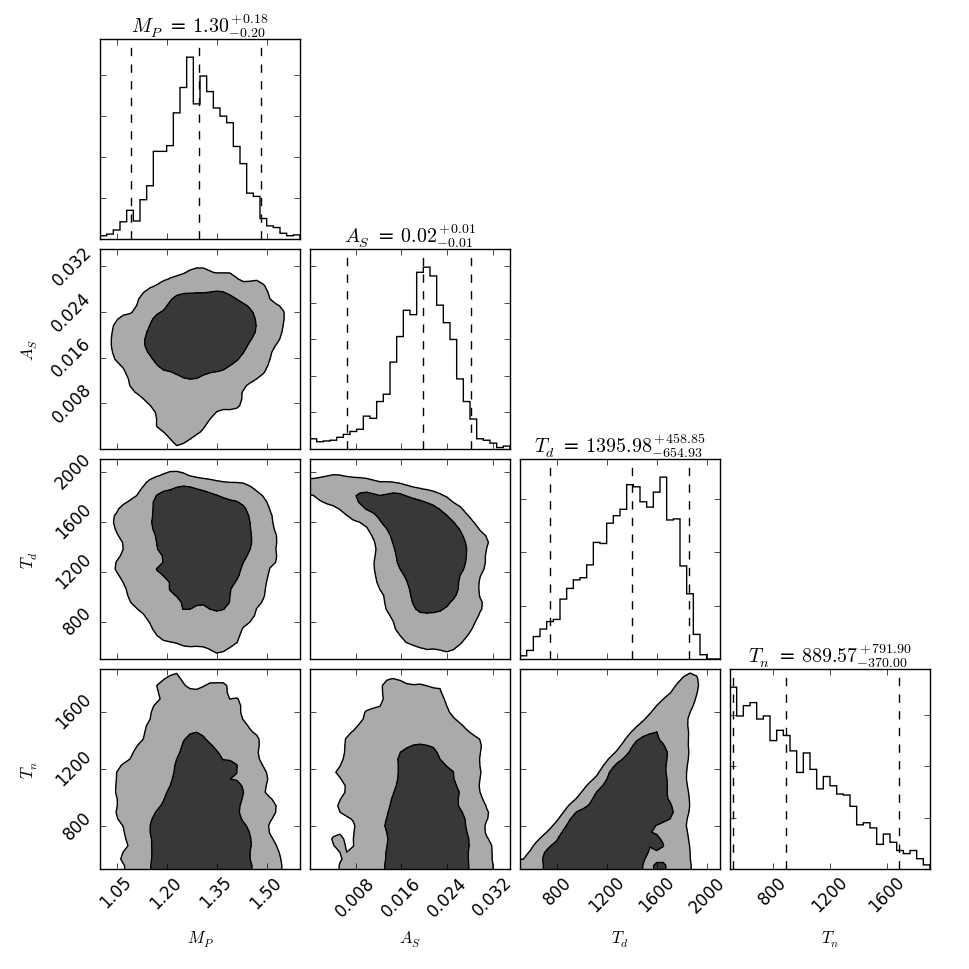}
\includegraphics[width=250pt]{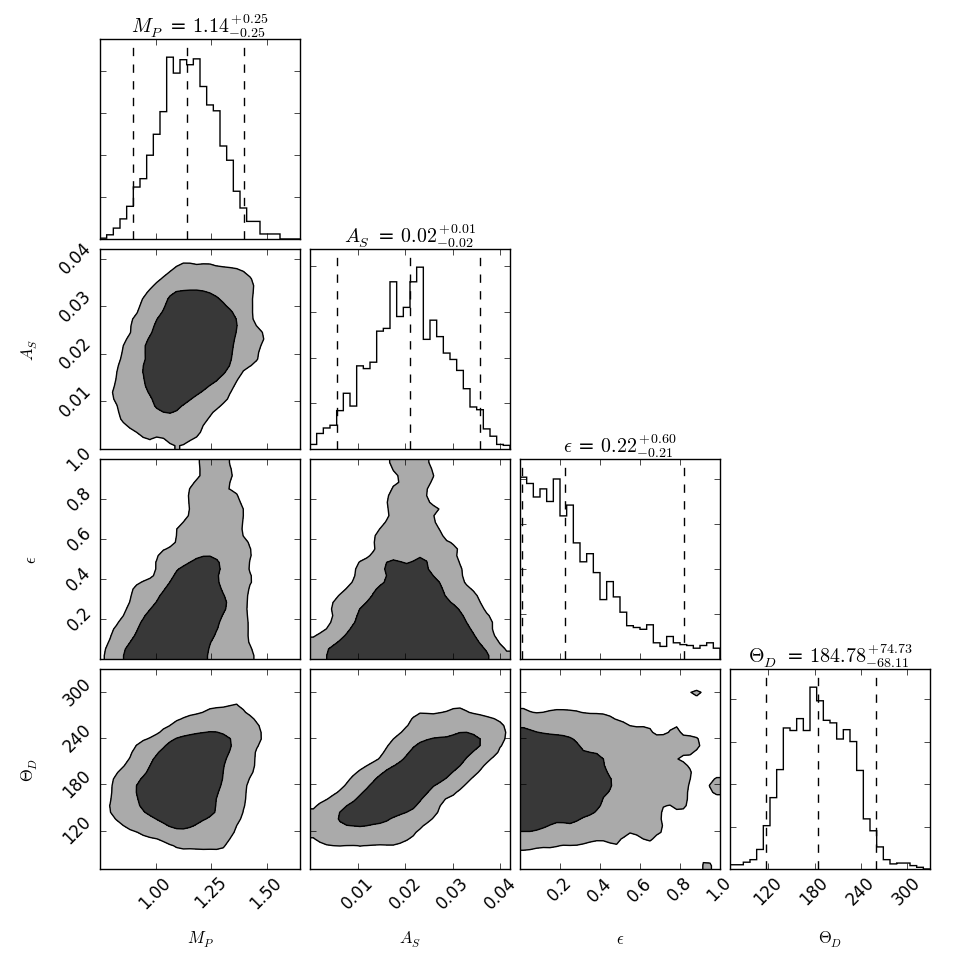}\\
\caption{TrES-2b phase curve constraints: Posterior projections for the "free T" model (left) and "standard + off" model (right). Dashed vertical lines represent marginalized 95\,\% credibility regions. Smoothed 68\,\% and 95\,\% credibility regions in dark and light grey shade, respectively.}
\label{tres2triangle_2}
\end{center}
\end{figure*}

\begin{figure*}
\begin{center}
\includegraphics[width=250pt]{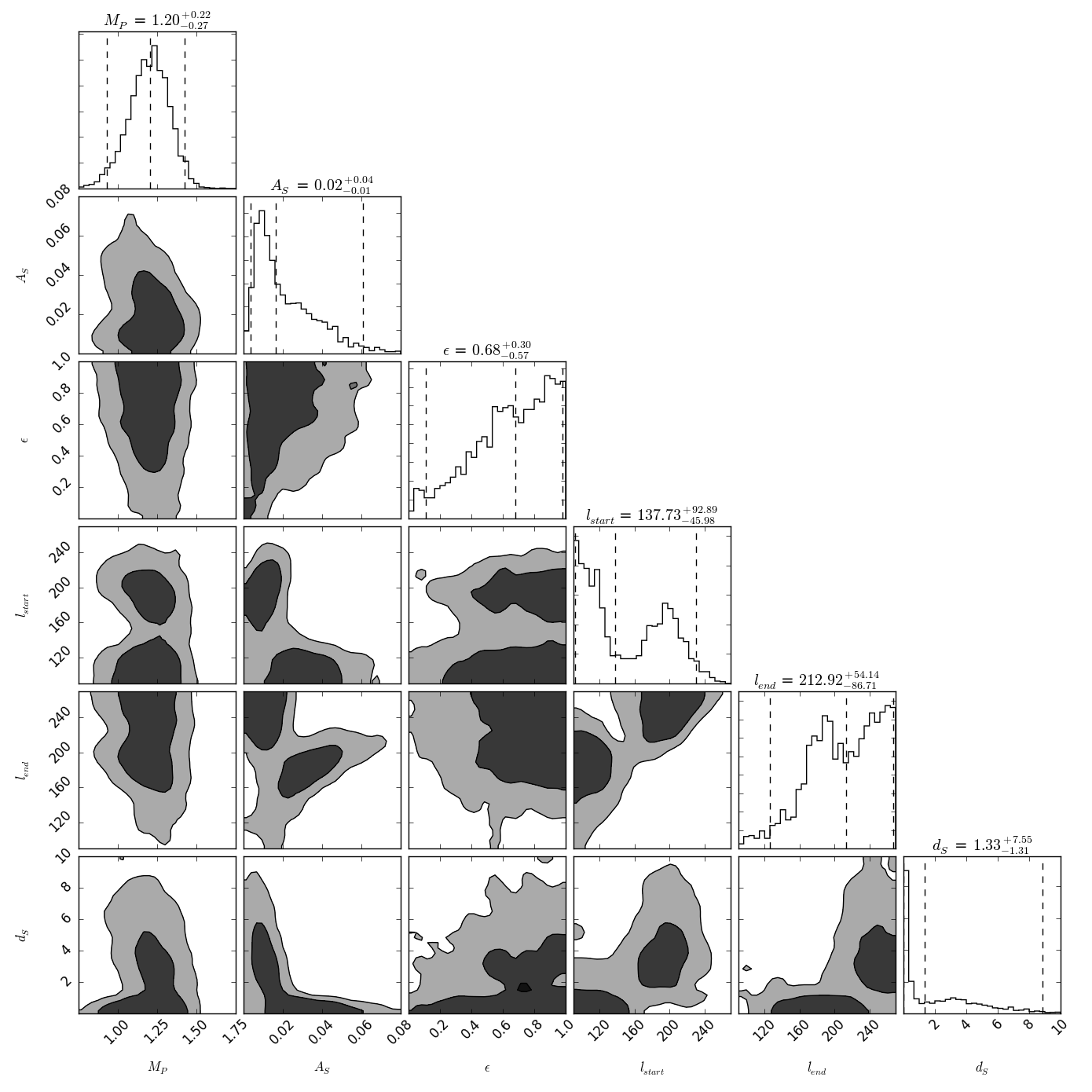}
\includegraphics[width=250pt]{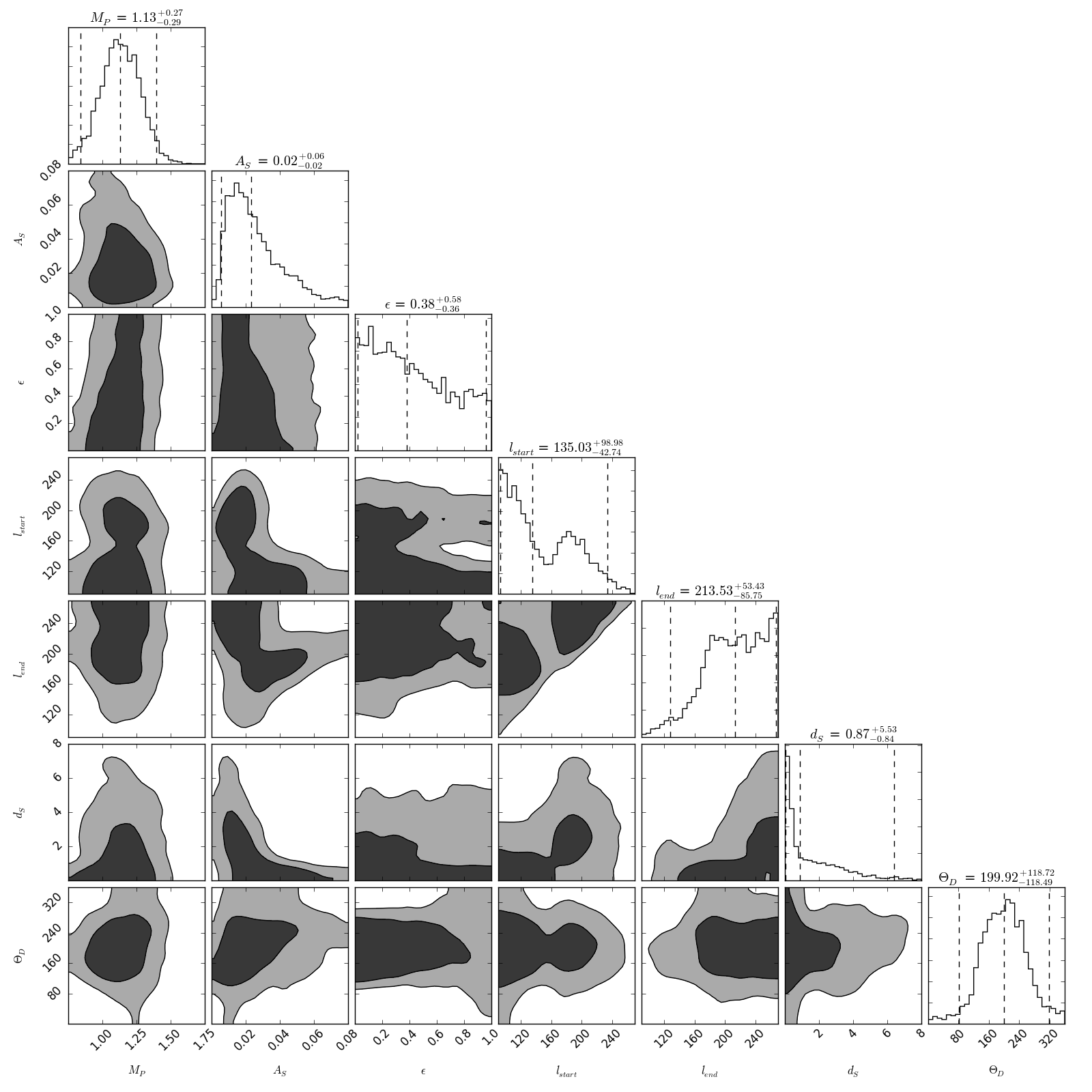}\\
\caption{TrES-2b phase curve constraints: Posterior projections for the "standard + asy" model (left) and "standard + both" model (right). Dashed vertical lines represent marginalized 95\,\% credibility regions. Smoothed 68\,\% and 95\,\% credibility regions in dark and light grey shade, respectively. }
\label{tres2triangle_3}
\end{center}
\end{figure*}

\begin{table*}
  \centering 
  \caption{TrES-2b 95\,\% credibility regions, parameters of the maximum a-posteriori models and associated goodness-of-fit criteria for scenarios from Table \ref{corot1mcmcsummary}. Models are ranked by BIC. Model probability ratios $p_M$ are stated (see eq. \ref{bicdiff}).  For comparison, we also state goodness-of-fit criteria for the best-fit model of \citet{barclay2012}, their Figure 5.}\label{tres2mcmcresults}
  \begin{tabular}{l|llcc|lll}
\hline
\hline
   scenario&95\,\% credibility regions &best-fit $V_P$&$\chi^2_{\rm{min}}$ & $\chi^2_{\rm{red,min}}$ & BIC & $\Delta$ BIC & $p_M$ \\
\hline
  standard + off  &0 \hspace{0.6cm}< $A_S$\hspace{0.17cm}<  0.035&$A_S$\hspace{0.15cm}= 0.028& 128.48&2.06 &149.42&0&1\\
  &0.01\hspace{0.3cm}< $\epsilon$\hspace{0.4cm}<  0.82&$\epsilon$\hspace{0.4cm}= 0.07& & &\\
  &0.89\hspace{0.3cm}< $M_P$ < 1.39 $M_{\rm{jup}}$&$M_P$ = 1.09\,$M_{\rm{jup}}$& & &\\
  &116\hspace{0.35cm}< $\theta_{\rm{day}}$\hspace{0.1cm}< 259$^{\circ}$& $\theta_{\rm{day}}$\hspace{0.05cm}= 217$^{\circ}$&&&\\
\hline
standard  &0 \hspace{0.6cm}< $A_S$\hspace{0.17cm}<  0.026&$A_S$\hspace{0.15cm}= 0.022& 136.19&2.19 &152.95 & 3.53 & 0.17\\
  &0.11\hspace{0.3cm}< $\epsilon$\hspace{0.4cm}<  0.98&$\epsilon$\hspace{0.4cm}= 0.94& & &\\
  &1.08\hspace{0.3cm}< $M_P$ < 1.48 $M_{\rm{jup}}$&$M_P$ = 1.31\,$M_{\rm{jup}}$& & &\\
\hline
standard + asy &0 \hspace{0.6cm}< $A_S$\hspace{0.17cm}<  0.06&$A_S$\hspace{0.15cm}= 0.046& 126.65&2.14 &155.97 &6.55 &0.04\\
  &0.10\hspace{0.3cm}< $\epsilon$\hspace{0.4cm}<  0.98&$\epsilon$\hspace{0.4cm}= 0.94& & &\\
  &0.93\hspace{0.3cm}< $M_P$ < 1.42 $M_{\rm{jup}}$&$M_P$ = 1.06\,$M_{\rm{jup}}$& & &\\
  &0.02\hspace{0.3cm}< $d_S$\hspace{0.25cm}< 8.87 &$d_S$\hspace{0.2cm}= 0.14&&&\\
  &91\hspace{0.5cm}< $l_{\rm{start}}$\hspace{0.16cm}< 230$^{\circ}$&$l_{\rm{start}}$\hspace{0.05cm}= 98$^{\circ}$&&&\\
  &126\hspace{0.35cm}< $l_{\rm{end}}$\hspace{0.2cm}< 267$^{\circ}$&$l_{\rm{end}}$\hspace{0.1cm}= 192$^{\circ}$&&&\\
\hline
  free A  &0 \hspace{0.6cm}< $A_S$\hspace{0.17cm}<  0.027&$A_S$\hspace{0.15cm}= 0.02& 135.97&2.22 &156.92&7.5& 0.02\\
  &0.03\hspace{0.3cm}< $A_B$\hspace{0.15cm}<  0.68&$A_B$\hspace{0.15cm}= 0.68& & &\\
  &0.04\hspace{0.3cm}< $\epsilon$\hspace{0.4cm}<  0.97&$\epsilon$\hspace{0.4cm}= 0.86& & &\\
  &1.09\hspace{0.3cm}< $M_P$ < 1.49 $M_{\rm{jup}}$&$M_P$ = 1.31\,$M_{\rm{jup}}$& & &\\
\hline
free T  &0 \hspace{0.6cm}< $A_S$\hspace{0.17cm}<  0.028&$A_S$\hspace{0.15cm}= 0.020& 135.98 & 2.22  & 156.93&7.51 &0.02\\
  &741\hspace{0.35cm}< $T_d$\hspace{0.2cm}<1854\,K&$T_d$\hspace{0.2cm}= 1100\,K&  &  &\\
  &519\hspace{0.35cm}< $T_n$\hspace{0.2cm}<1681\,K&$T_n$\hspace{0.2cm}= 890\,K&  &  &\\
  &1.09\hspace{0.3cm}< $M_P$ < 1.48 $M_{\rm{jup}}$&$M_P$ = 1.29\,$M_{\rm{jup}}$& & &\\
\hline
  standard + both  &0 \hspace{0.65cm}< $A_S$\hspace{0.17cm}<  0.085&$A_S$\hspace{0.15cm}= 0.018& 125.52&2.16 &159.03 & 9.61 & 8$\cdot$ 10$^{-3}$\\
  &0.01\hspace{0.3cm}< $\epsilon$\hspace{0.4cm}<  0.95&$\epsilon$\hspace{0.4cm}= 0.05& & &\\
  &0.84\hspace{0.3cm}< $M_P$ < 1.39 $M_{\rm{jup}}$&$M_P$ = 1.21\,$M_{\rm{jup}}$& & &\\
  &0.03\hspace{0.3cm}< $d_S$\hspace{0.25cm}< 6.4 &$d_S$\hspace{0.2cm}= 3.7&&&\\
  &92\hspace{0.5cm}< $l_{\rm{start}}$\hspace{0.16cm}< 234$^{\circ}$&$l_{\rm{start}}$\hspace{0.05cm}= 185$^{\circ}$&&&\\
  &127\hspace{0.35cm}< $l_{\rm{end}}$\hspace{0.2cm}< 266$^{\circ}$&$l_{\rm{end}}$\hspace{0.1cm}= 220$^{\circ}$&&&\\
  &81\hspace{0.5cm}< $\theta_{\rm{day}}$\hspace{0.15cm}< 318$^{\circ}$& $\theta_{\rm{day}}$\hspace{0.05cm}= 253$^{\circ}$&&&\\
\hline
\hline
\citet{barclay2012} &- &-&125.87& 2.06 &146.82 &-2.6 & 3.66\\
\end{tabular}
\end{table*}

\begin{figure*}
\begin{center}
\includegraphics[width=250pt]{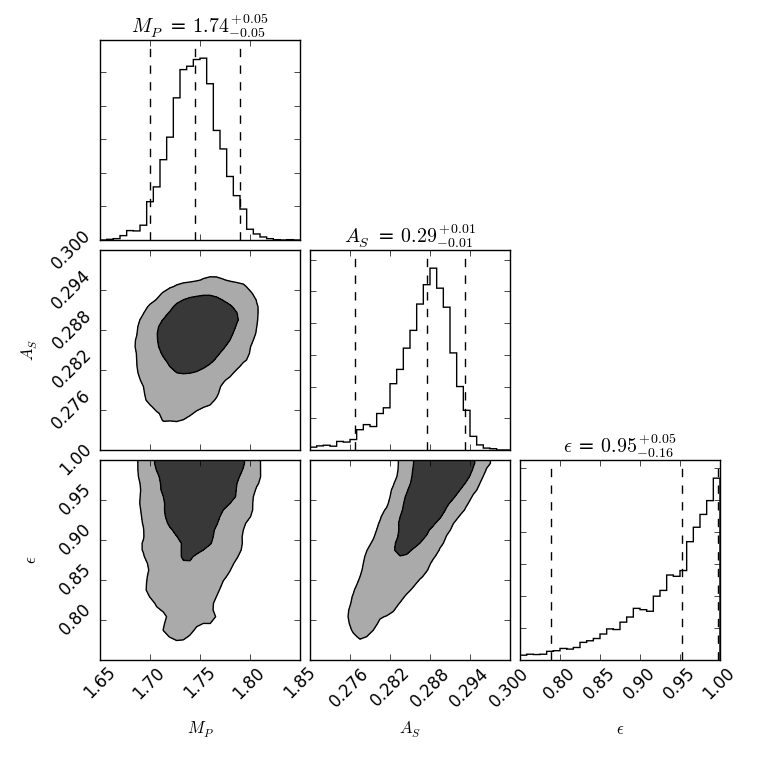}
\includegraphics[width=250pt]{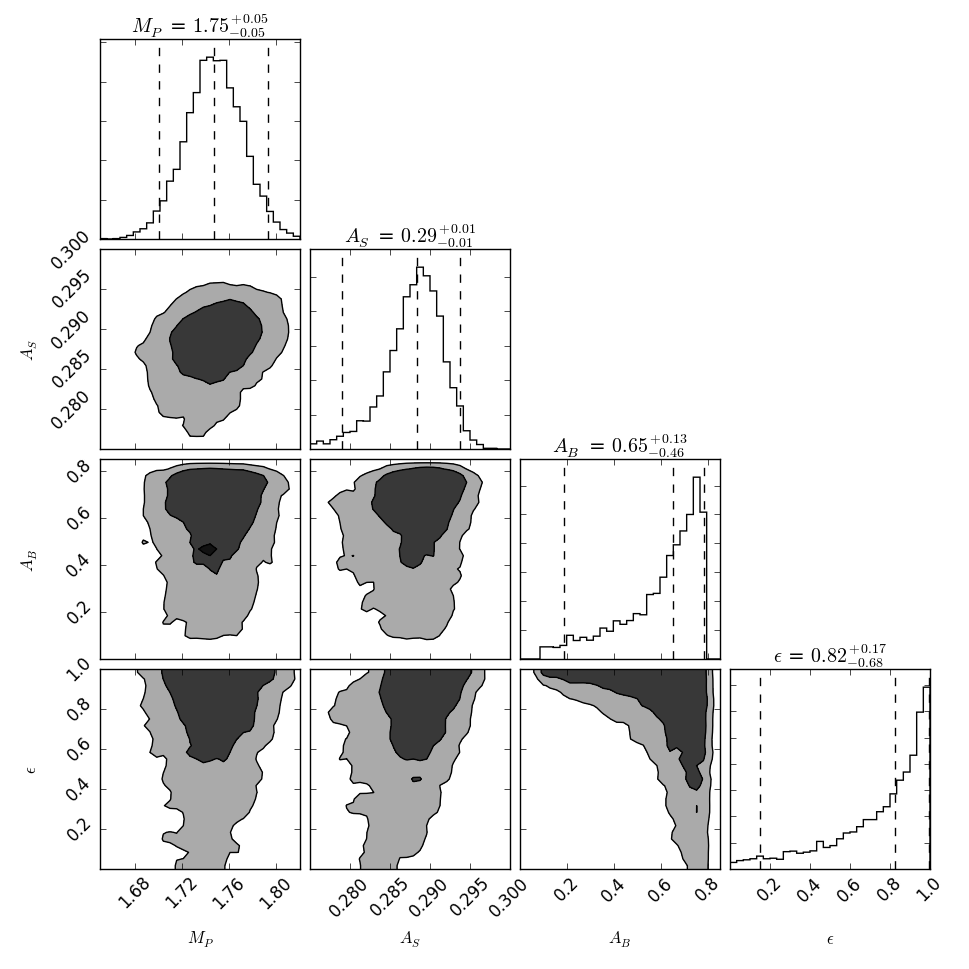}\\
\caption{HAT-P-7b phase curve constraints: Posterior projections for the "standard" model (left) and "free A" model (right). Dashed vertical lines represent marginalized 95\,\% credibility regions. Smoothed 68\,\% and 95\,\% credibility regions in dark and light grey shade, respectively.}
\label{hat7triangle_1}
\end{center}
\end{figure*}

\begin{figure*}
\begin{center}
\includegraphics[width=250pt]{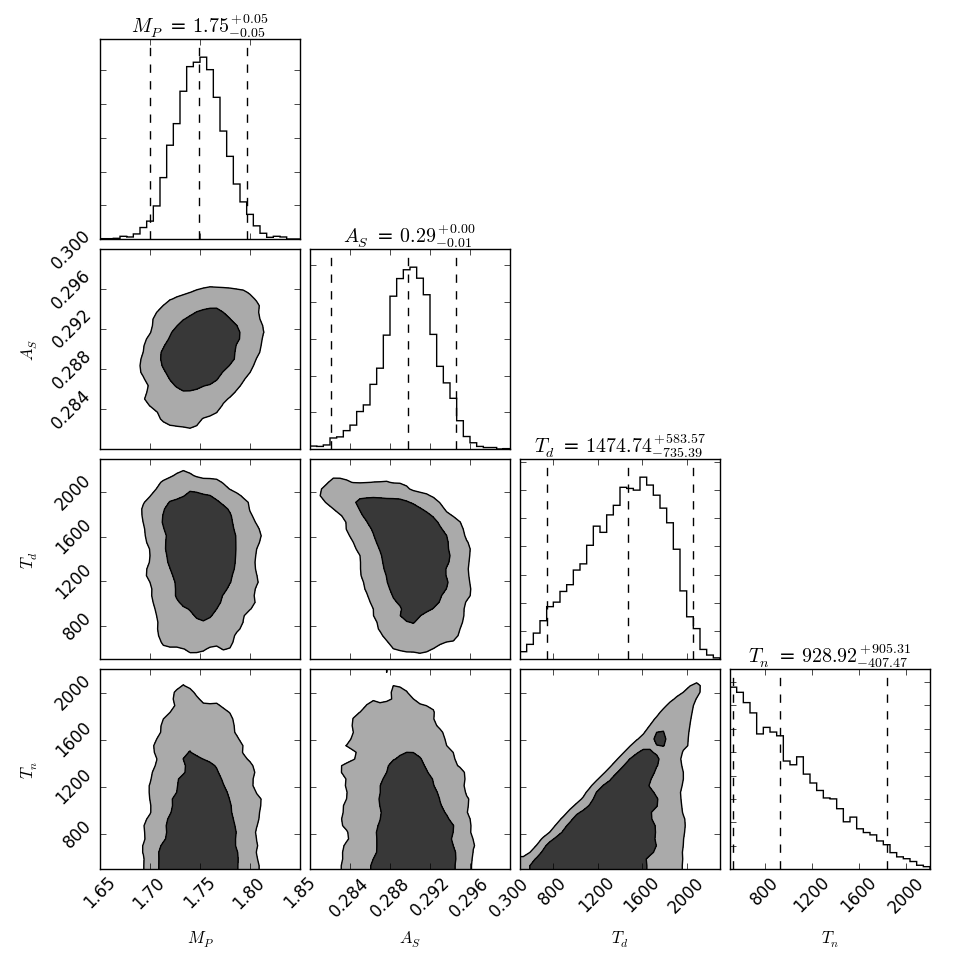}
\includegraphics[width=250pt]{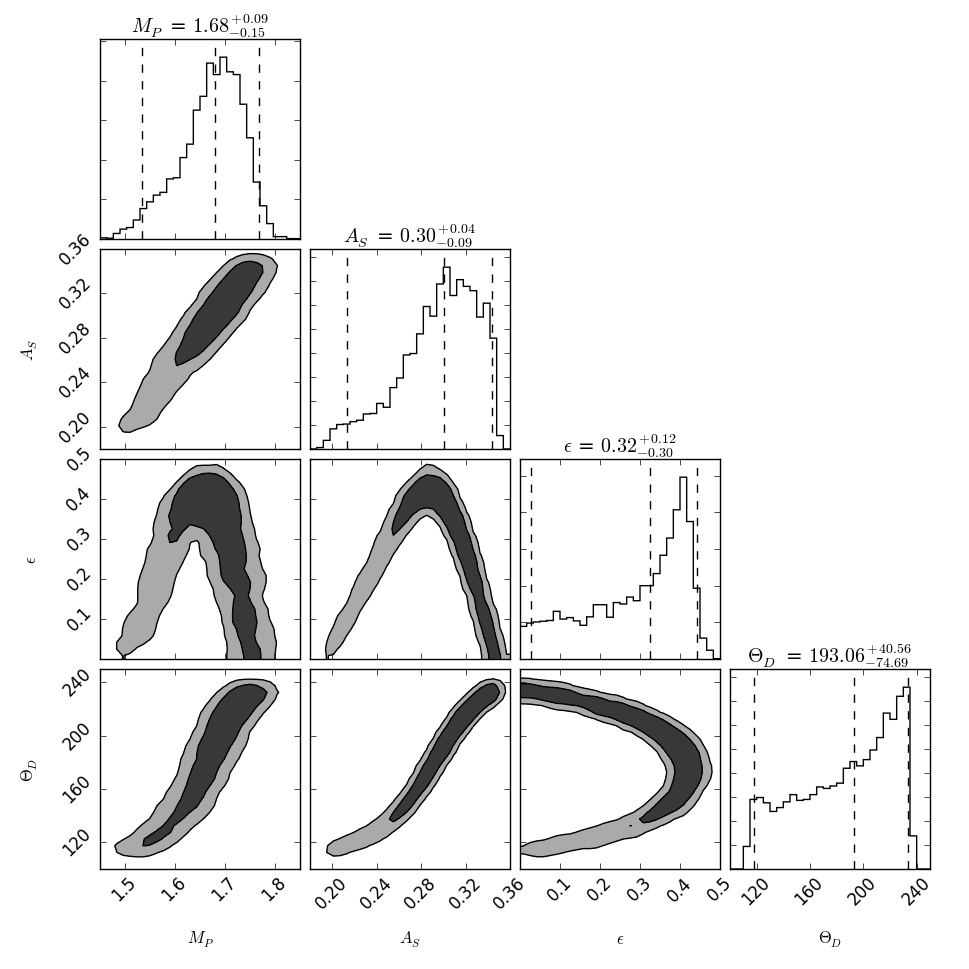}\\
\caption{HAT-P-7b phase curve constraints: Posterior projections for the "free T" model (left) and "standard + off" model (right). Dashed vertical lines represent marginalized 95\,\% credibility regions. Smoothed 68\,\% and 95\,\% credibility regions in dark and light grey shade, respectively.}
\label{hat7triangle_2}
\end{center}
\end{figure*}

\begin{figure*}
\begin{center}
\includegraphics[width=250pt]{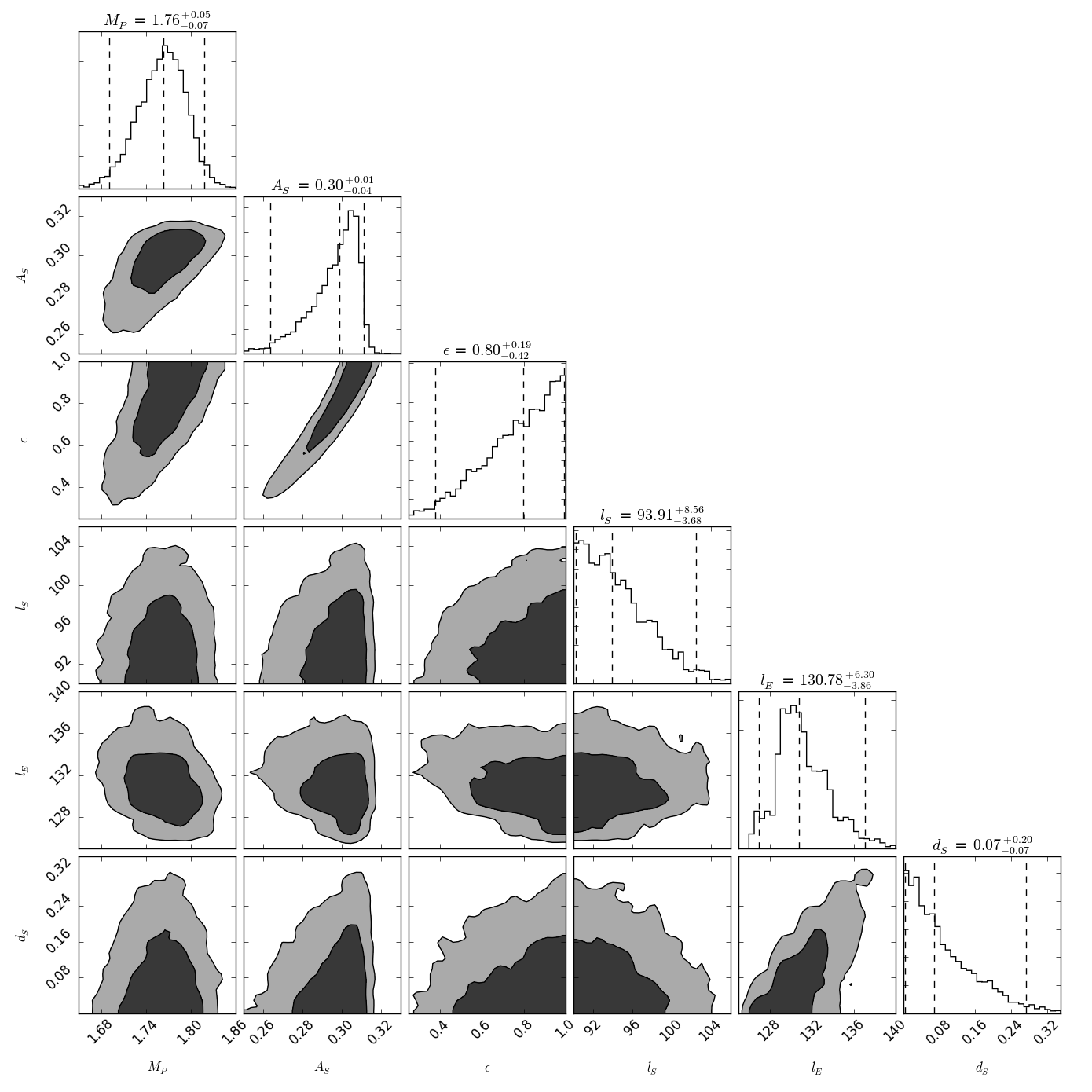}
\includegraphics[width=250pt]{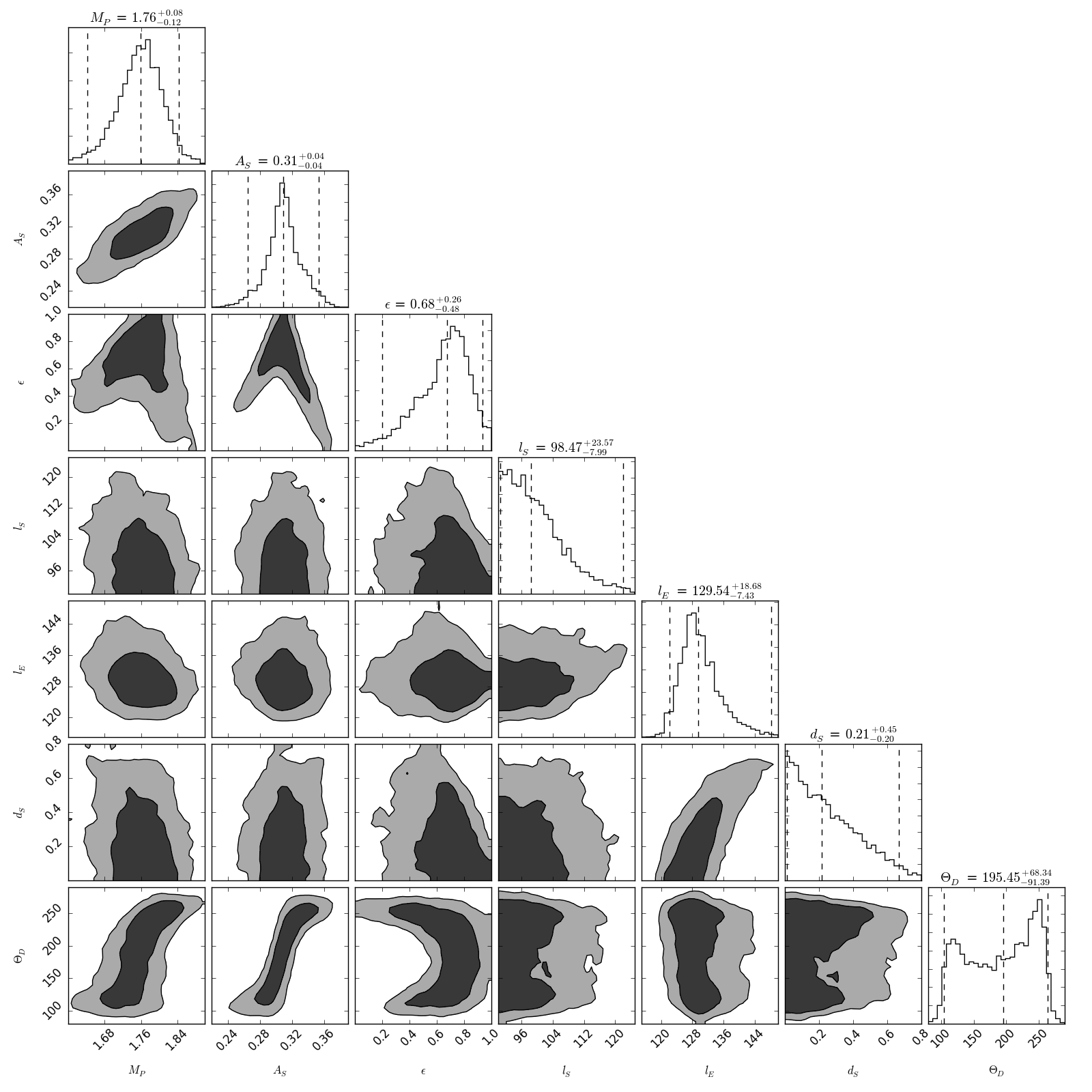}\\
\caption{HAT-P-7b phase curve constraints: Posterior projections for the "standard + asy" model (left) and "standard + both" model (right). Dashed vertical lines represent marginalized 95\,\% credibility regions. Smoothed 68\,\% and 95\,\% credibility regions in dark and light grey shade, respectively.}
\label{hat7triangle_3}
\end{center}
\end{figure*}

\begin{table*}
  \centering 
  \caption{HAT-P-7b 95\,\% credibility regions, parameters of the maximum a-posteriori models and associated goodness-of-fit criteria for scenarios from Table \ref{corot1mcmcsummary}. Models are ranked by BIC. Model probability ratios $p_M$ are stated (see eq. \ref{bicdiff}). For comparison, we also state goodness-of-fit criteria for the best-fit model of \citet{esteves2013}, their Figure 3.}\label{hat7mcmcresults}
  \begin{tabular}{l|llcc|lll}
\hline
\hline
   scenario&95\,\% credibility regions &best-fit $V_P$&$\chi^2_{\rm{min}}$ & $\chi^2_{\rm{red,min}}$ & BIC & $\Delta$ BIC & $p_M$\\
\hline
standard + asy &0.26\hspace{0.3cm}< $A_S$\hspace{0.15cm}<  0.31&$A_S$\hspace{0.15cm}= 0.30& 217.23&3.68 &246.56 & 0 & 1\\
  &0.37\hspace{0.3cm}< $\epsilon$\hspace{0.4cm}<  0.99&$\epsilon$\hspace{0.4cm}= 0.93& & &\\
  &1.69\hspace{0.3cm}< $M_P$ < 1.81 $M_{\rm{jup}}$&$M_P$ = 1.79\,$M_{\rm{jup}}$& & &\\
  &0.003 \hspace{0.05cm}< $d_S$\hspace{0.3cm}< 0.27 &$d_S$\hspace{0.2cm}= 0.002&&&\\
  &90\hspace{0.5cm}< $l_{\rm{start}}$\hspace{0.16cm}< 102$^{\circ}$&$l_{\rm{start}}$\hspace{0.05cm}= 93$^{\circ}$&&&\\
  &126\hspace{0.35cm}< $l_{\rm{end}}$\hspace{0.2cm}< 137$^{\circ}$&$l_{\rm{end}}$\hspace{0.1cm}= 126$^{\circ}$&&&\\
\hline
  standard + both  &0.26\hspace{0.3cm}< $A_S$\hspace{0.15cm}< 0.35&$A_S$\hspace{0.15cm}= 0.34& 214.74&3.70 & 248.26 & 1.7 & 0.42\\
  &0.19\hspace{0.3cm}< $\epsilon$\hspace{0.4cm}< 0.93&$\epsilon$\hspace{0.4cm}= 0.34& & &\\
  &1.64\hspace{0.3cm}< $M_P$ < 1.84 $M_{\rm{jup}}$&$M_P$ = 1.81\,$M_{\rm{jup}}$& & &\\
  &0.01 \hspace{0.2cm}< $d_S$\hspace{0.3cm}< 0.66 &$d_S$\hspace{0.2cm}= 0.005&&&\\
  &90\hspace{0.5cm}< $l_{\rm{start}}$\hspace{0.16cm}< 122$^{\circ}$&$l_{\rm{start}}$\hspace{0.05cm}= 91$^{\circ}$&&&\\
  &122\hspace{0.35cm}< $l_{\rm{end}}$\hspace{0.2cm}< 148$^{\circ}$&$l_{\rm{end}}$\hspace{0.1cm}= 121$^{\circ}$&&&\\
  &104\hspace{0.35cm}< $\theta_{\rm{day}}$\hspace{0.15cm}< 263$^{\circ}$& $\theta_{\rm{day}}$\hspace{0.05cm}= 259$^{\circ}$&&&\\
\hline
  standard + off  &0.21\hspace{0.3cm}< $A_S$\hspace{0.15cm}<  0.34&$A_S$\hspace{0.15cm}= 0.34& 247.34&4.05 &268.29 & 21.73 & 1.9$\cdot$10$^{-5}$\\
  &0.02\hspace{0.3cm}< $\epsilon$\hspace{0.4cm}<  0.44&$\epsilon$\hspace{0.4cm}= 0.007& & &\\
  &1.53\hspace{0.3cm}< $M_P$ < 1.76 $M_{\rm{jup}}$&$M_P$ = 1.75\,$M_{\rm{jup}}$& & &\\
  &118\hspace{0.35cm}< $\theta_{\rm{day}}$\hspace{0.1cm}< 233$^{\circ}$& $\theta_{\rm{day}}$\hspace{0.05cm}= 234$^{\circ}$&&&\\
\hline
standard  &0.27\hspace{0.3cm}< $A_S$\hspace{0.15cm}<  0.29&$A_S$\hspace{0.15cm}= 0.29& 669.74&10.80 &686.50 & 439.94 & 0\\
  &0.78\hspace{0.3cm}< $\epsilon$\hspace{0.4cm}<  0.99&$\epsilon$\hspace{0.4cm}= 0.99& & &\\
  &1.69\hspace{0.3cm}< $M_P$ < 1.79 $M_{\rm{jup}}$&$M_P$ = 1.75\,$M_{\rm{jup}}$& & &\\
\hline
  free A  &0.27\hspace{0.3cm}< $A_S$\hspace{0.15cm}<  0.29&$A_S$\hspace{0.15cm}= 0.29& 669.64&10.97 &690.59 & 444.03 & 0\\
  &0.18\hspace{0.3cm}< $A_B$\hspace{0.15cm}<  0.78&$A_B$\hspace{0.15cm}= 0.75& & &\\
  &0.14\hspace{0.3cm}< $\epsilon$\hspace{0.4cm}<  0.99&$\epsilon$\hspace{0.4cm}= 0.96& & &\\
  &1.69\hspace{0.3cm}< $M_P$ < 1.79 $M_{\rm{jup}}$&$M_P$ = 1.74\,$M_{\rm{jup}}$& & &\\
\hline
free T  &0.28\hspace{0.3cm}< $A_S$\hspace{0.15cm}<  0.29&$A_S$\hspace{0.15cm}= 0.29& 669.68 & 10.97  & 690.62 & 444.06 & 0\\
  &739\hspace{0.35cm}< $T_d$\hspace{0.2cm}<2058\,K&$T_d$\hspace{0.15cm}= 1204\,K&  &  &\\
  &521\hspace{0.35cm}< $T_n$\hspace{0.2cm}< 1834\,K&$T_n$\hspace{0.15cm}= 512\,K&  &  &\\
  &1.70\hspace{0.3cm}< $M_P$ < 1.79 $M_{\rm{jup}}$&$M_P$ = 1.75\,$M_{\rm{jup}}$& & &\\
\hline
\hline
\citet{esteves2013} &- &-&276.41& 4.53 &297.36 & 50.8 &  9.3$\cdot$10$^{-12}$\\
\end{tabular}
\end{table*}

\newpage

\section{MCMC convergence diagnostics}
\label{convergence_res}
\clearpage

\begin{figure*}
\begin{center}
\includegraphics[width=250pt]{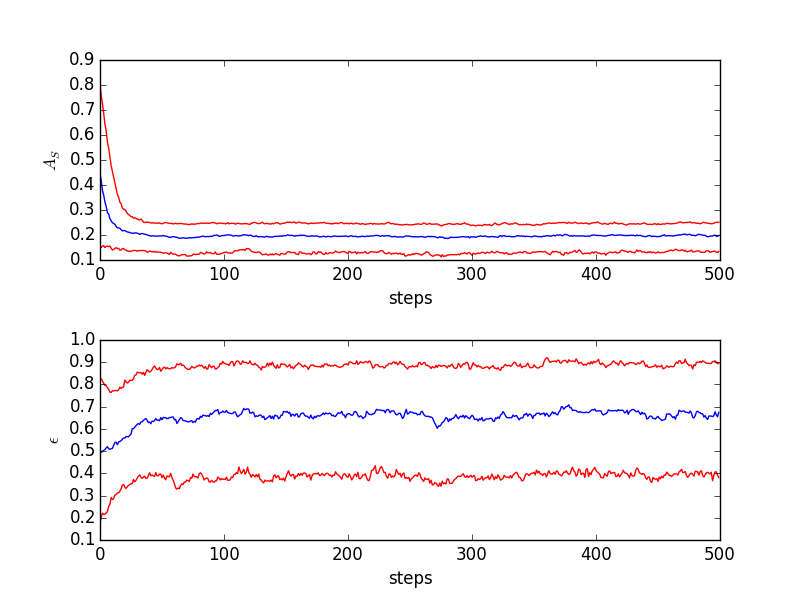}
\includegraphics[width=250pt]{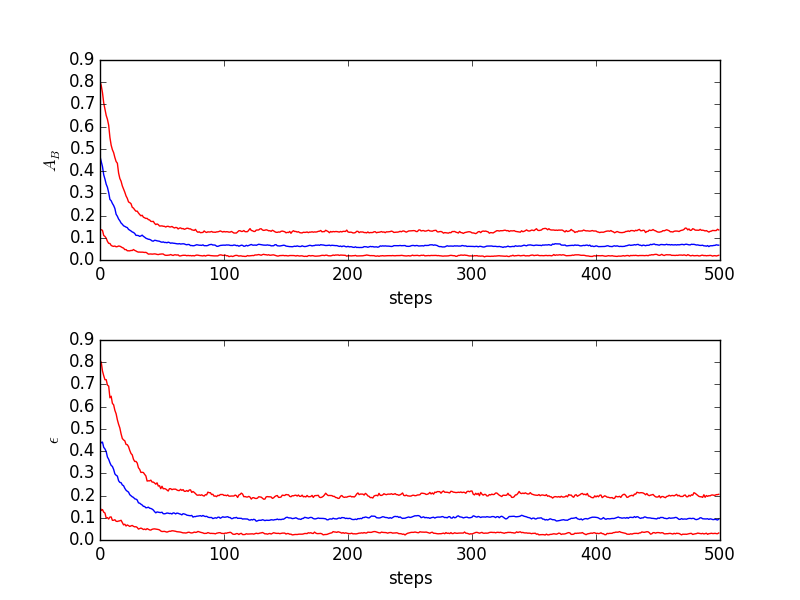}\\
\caption{Trace plots of model parameters for the CoRoT-1b "standard" (left) and "no scattering" (right) scenarios. Blue line traces the ensemble median, red lines correspond to the [0.16,0.84] percentiles.}
\label{corot1_conv_1}
\end{center}
\end{figure*}

\begin{figure*}
\begin{center}
\includegraphics[width=250pt]{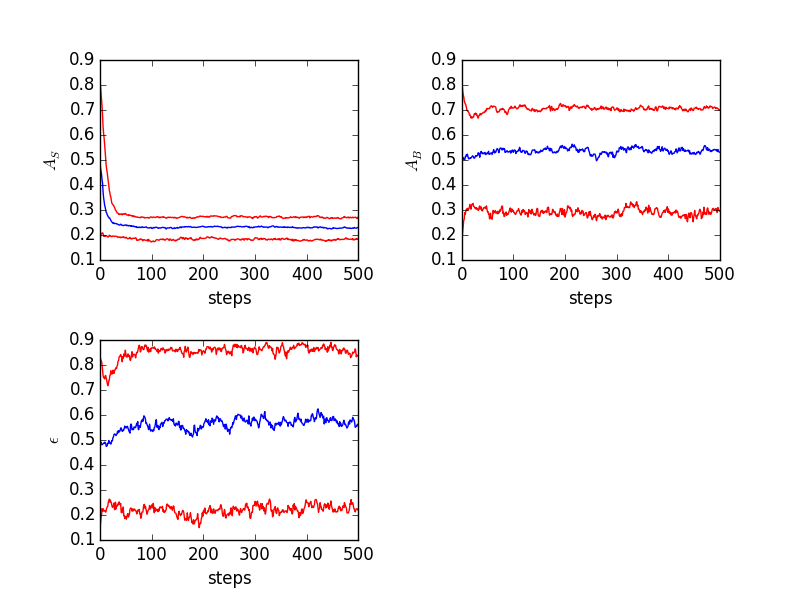}
\includegraphics[width=250pt]{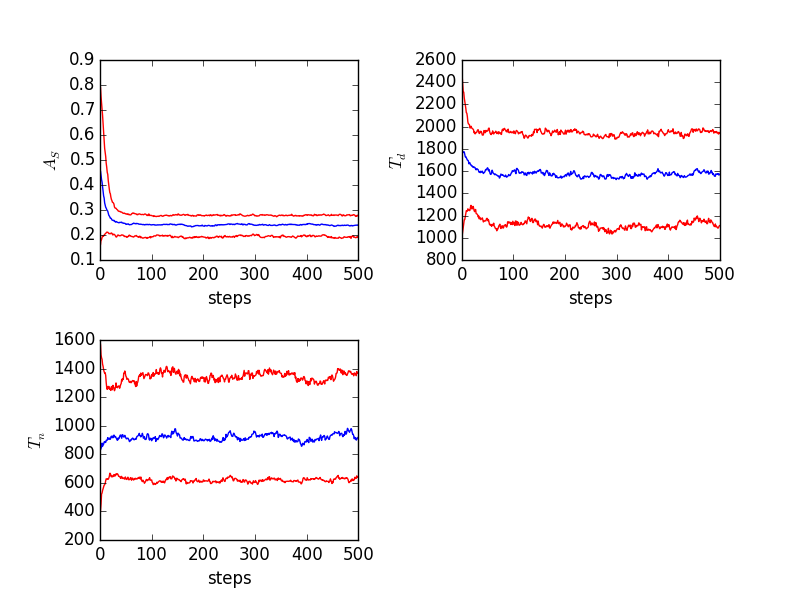}\\
\caption{Trace plots of model parameters for the CoRoT-1b "free alb" (left) and "free temp" (right) scenarios. Blue line traces the ensemble median, red lines correspond to the [0.16,0.84] percentiles.}
\label{corot1_conv_2}
\end{center}
\end{figure*}

\begin{figure*}
\begin{center}
\includegraphics[width=250pt]{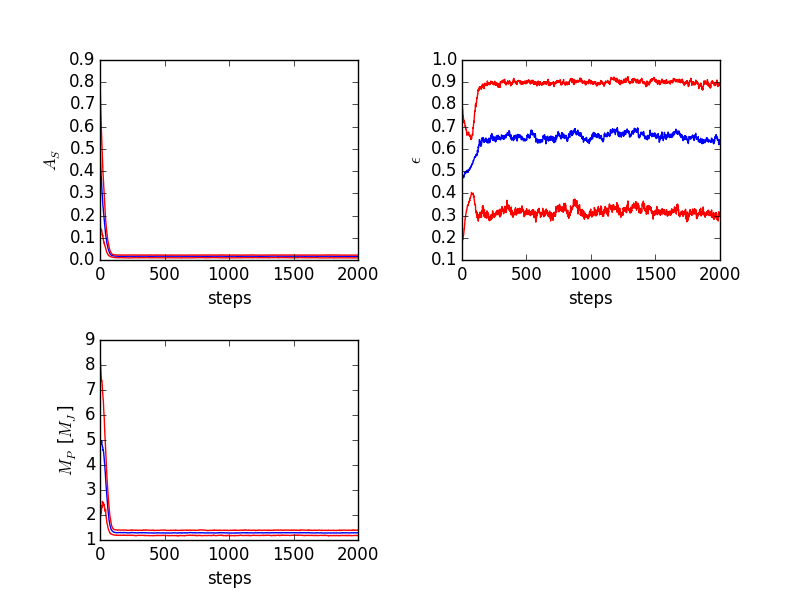}
\includegraphics[width=250pt]{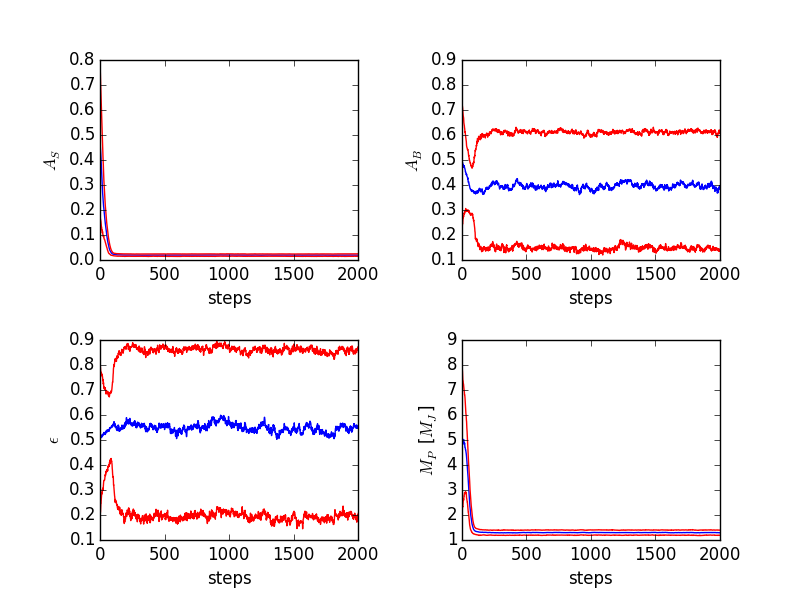}\\
\caption{Trace plots of model parameters for the TrES-2b "standard" (left) and "free alb" (right) scenarios. Blue line traces the ensemble median, red lines correspond to the [0.16,0.84] percentiles.}
\label{tres2_conv_1}
\end{center}
\end{figure*}

\begin{figure*}
\begin{center}
\includegraphics[width=250pt]{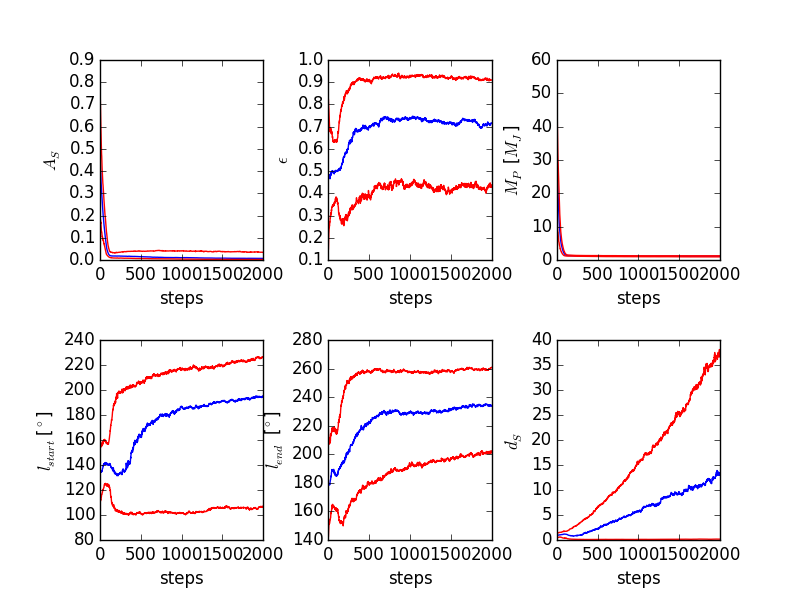}
\includegraphics[width=250pt]{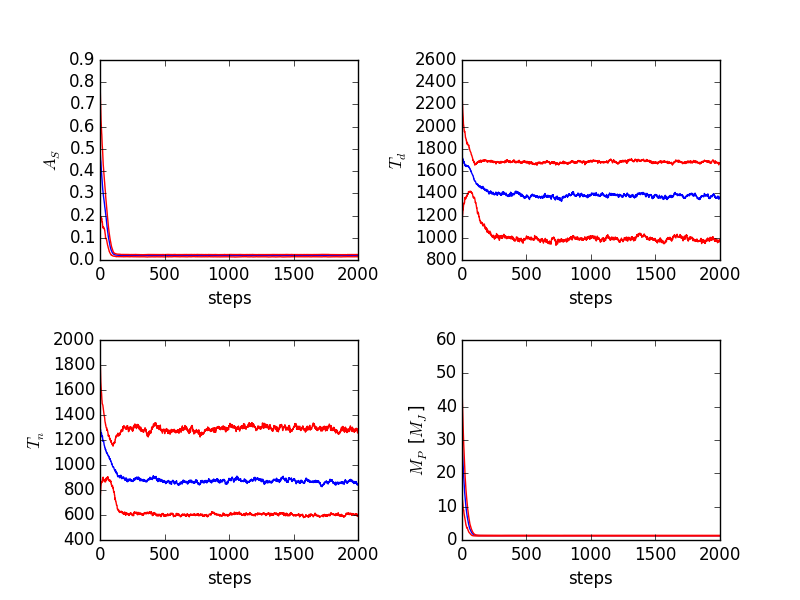}\\
\caption{Trace plots of model parameters for the TrES-2b "asymmetric" (left) and "free temp" (right) scenarios. Blue line traces the ensemble median, red lines correspond to the [0.16,0.84] percentiles.}
\label{tres2_conv_2}
\end{center}
\end{figure*}

\begin{figure*}
\begin{center}
\includegraphics[width=250pt]{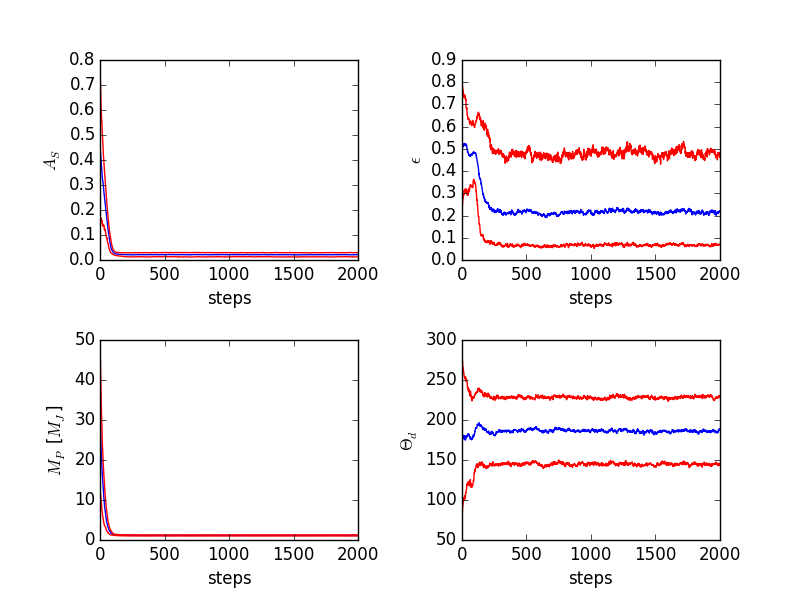}
\includegraphics[width=250pt]{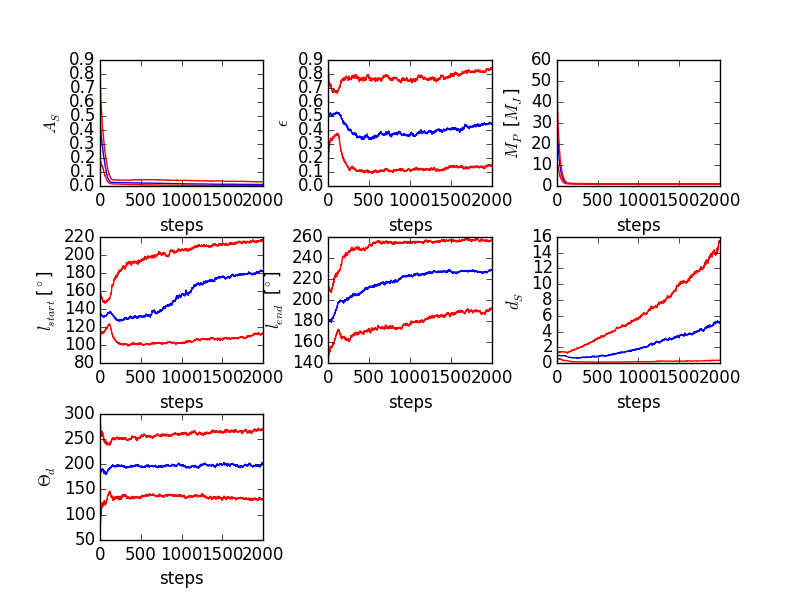}\\
\caption{Trace plots of model parameters for the TrES-2b "offset" (left) and "both" (right) scenarios. Blue line traces the ensemble median, red lines correspond to the [0.16,0.84] percentiles.}
\label{tres2_conv_3}
\end{center}
\end{figure*}

\begin{figure*}
\begin{center}
\includegraphics[width=250pt]{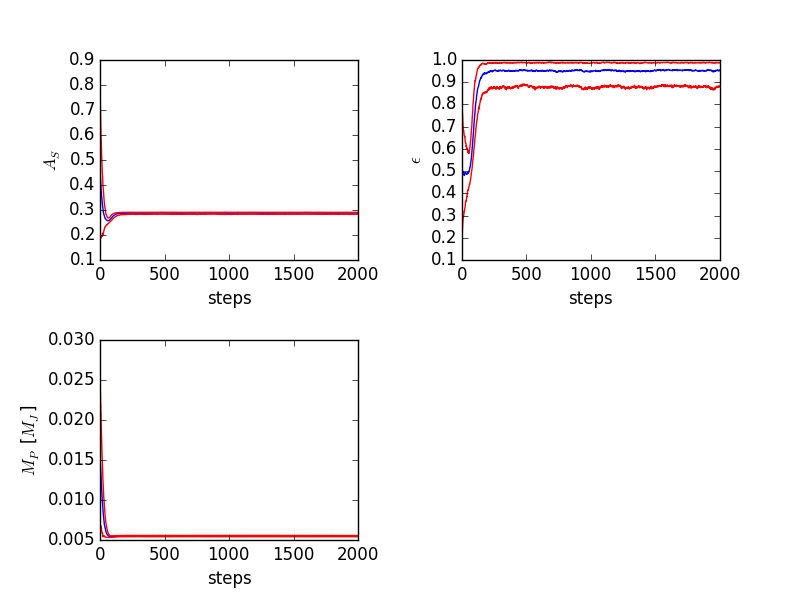}
\includegraphics[width=250pt]{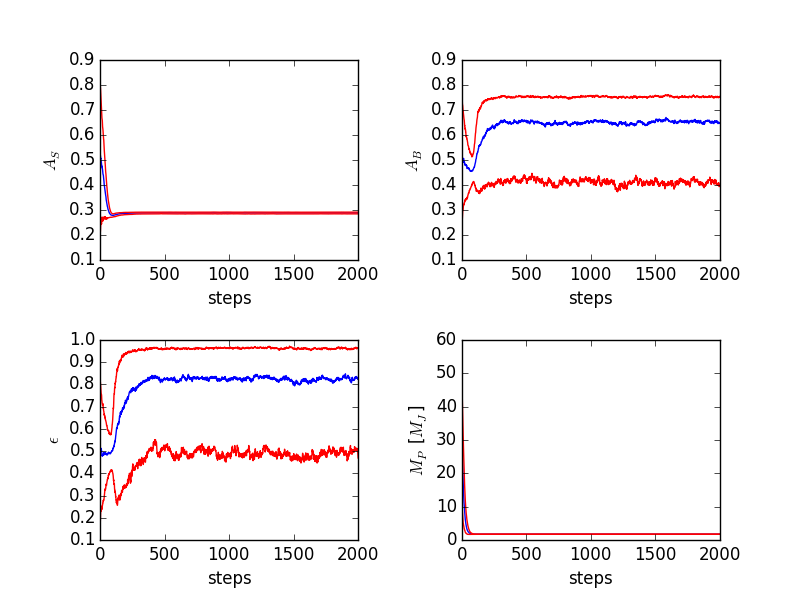}\\
\caption{Trace plots of model parameters for the HAT-P-7b "standard" (left) and "free alb" (right) scenarios. Blue line traces the ensemble median, red lines correspond to the [0.16,0.84] percentiles.}
\label{that7_conv_1}
\end{center}
\end{figure*}

\begin{figure*}
\begin{center}
\includegraphics[width=250pt]{asym_convergence}
\includegraphics[width=250pt]{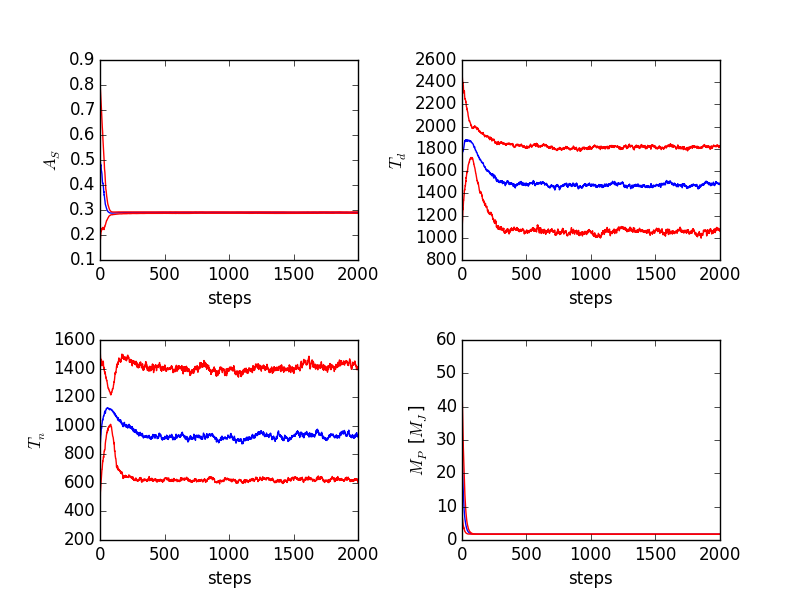}\\
\caption{Trace plots of model parameters for the HAT-P-7b "asymmetric" (left) and "free temp" (right) scenarios. Blue line traces the ensemble median, red lines correspond to the [0.16,0.84] percentiles.}
\label{hat7_conv_2}
\end{center}
\end{figure*}

\begin{figure*}
\begin{center}
\includegraphics[width=250pt]{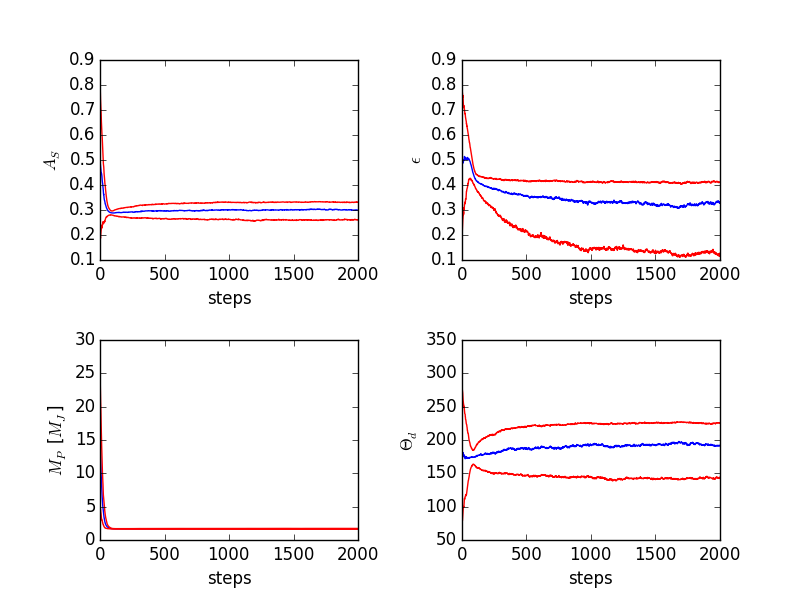}
\includegraphics[width=250pt]{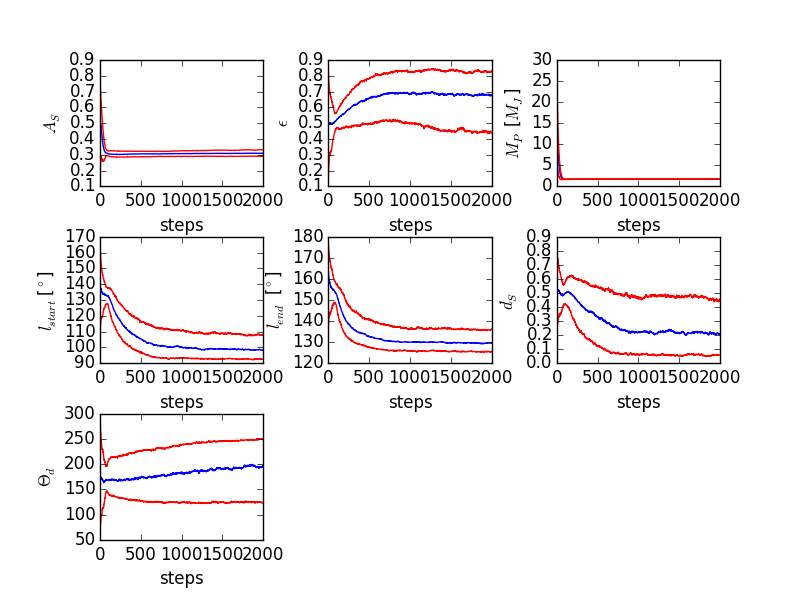}\\
\caption{Trace plots of model parameters for the HAT-P-7b "offset" (left) and "both" (right) scenarios. Blue line traces the ensemble median, red lines correspond to the [0.16,0.84] percentiles.}
\label{hat7_conv_3}
\end{center}
\end{figure*}

\begin{table*}
  \centering 
  \caption{Gelman-Rubin statistics for the CoRoT-1b scenarios.}\label{corot1gelman}
  \begin{tabular}{l|ccccc}
\hline
\hline
   scenario& $A_S$&$A_B$&$\epsilon$& $T_d$ & $T_n$ \\
\hline
standard  &1.026 & - & 1.036 &-&-\\
free T  &1.037 &-&-&1.061&1.055\\
free albedo  & 1.029 & 1.054 &1.059 &-&-\\
no scattering  &-&1.024 &1.025&-&-\\
\end{tabular}
\end{table*}

\begin{table*}
  \centering 
  \caption{Gelman-Rubin statistics for the TrES-2b scenarios.}\label{tres2gelman}
  \begin{tabular}{l|cccccccccc}
\hline
\hline
   scenario& $A_S$&$A_B$&$\epsilon$& $T_d$ & $T_n$ &  $M_P$& $\theta_{\rm{day}}$ & $l_{\rm{start}}$& $l_{\rm{end}}$ & $d_S$\\
\hline
standard & 1.002 & -& 1.017 &-&-&1.002&-&-&-&-\\
 free T & 1.031 &-&-&1.059 & 1.09 & 1.002 &-&-&-&-\\
 free albedo & 1.002 & 1.025 & 1.025&-&-&1.002&-&-&-&-\\
asymmetric & 1.056&-&1.0952&-&-&1.001&-&1.534 & 1.251 &1.2\\
offset & 1.002 &-&1.022 &-&-&1.001&1.022&-&-&-\\
both & 1.041 &-&1.164 & -&-&1.001&1.148&1.319 & 1.193 &1.218\\
 \end{tabular}
\end{table*}

\begin{table*}
  \centering 
  \caption{Gelman-Rubin statistics for the HAT-P-7b scenarios.}\label{hat7gelman}
  \begin{tabular}{l|cccccccccc}
\hline
\hline
   scenario& $A_S$&$A_B$&$\epsilon$& $T_d$ & $T_n$ &  $M_P$& $\theta_{\rm{day}}$ & $l_{\rm{start}}$& $l_{\rm{end}}$ & $d_S$\\
\hline
standard & 1.006 & -& 1.013 &-&-&1.002&-&-&-&-\\
 free T & 1.093 &-&-&1.137 & 1.124 & 1.002 &-&-&-&-\\
 free albedo & 1.01 & 1.061 & 1.037&-&-&1.001&-&-&-&-\\
asymmetric & 1.025&-&1.112&-&-&1.002&-&1.009 & 1.011 &1.021\\
offset & 1.099 &-&1.063 &-&-&1.002&1.167&-&-&-\\
both & 1.089 &-&1.256 & -&-&1.002&1.332&1.167 & 1.257 &1.186\\
 \end{tabular}
\end{table*}
  
\end{document}